\documentclass[%
 preprint
 amsmath,amssymb,
 aps, physrev,
]{revtex4-2}

\usepackage{graphicx}
\usepackage{dcolumn}
\usepackage{bm}
\usepackage[dvipsnames]{xcolor}

\usepackage{placeins} 
\usepackage{float}
\usepackage{amsmath}
\usepackage{amssymb}
\usepackage{esint}
\usepackage{cancel}
\usepackage{subcaption}
\usepackage{mwe}
\usepackage{array}
\usepackage{tikz}
\usepackage{xcolor}
\usepackage{stackengine}
\usepackage{wasysym}
\usepackage{comment}
\usepackage{relsize}
\usepackage[percent]{overpic}
\usepackage{mathtools}
\usepackage{mdframed}
\usepackage{booktabs}
\usepackage{enumitem}
\usepackage{siunitx}
\usepackage{gensymb}
\usepackage{stfloats} 
\usepackage{epigraph}
\usepackage[normalem]{ulem}

\newcommand{\RomanNumeralCaps}[1]
    {\MakeUppercase{\romannumeral #1}}

\usepackage[absolute,overlay]{textpos}  
\usepackage[export]{adjustbox} 
\usepackage{gensymb} 

\usepackage[colorlinks=true, allcolors=blue]{hyperref}

\begin{document}


\title{\textbf{Drafting-Kissing-Tumbling Dynamics of Two Particles\\Subjected to Horizontal Oscillations}
}%

\author{Fabian Kleischmann}
  \email{fabian.kleischmann@tu-dresden.de}
\author{Bernhard Vowinckel} 
\email{bernhard.vowinckel@tu-dresden.de}
\affiliation{%
 Institute of Urban and Industrial Water Management, Technische Universit\"{a}t Dresden, 01062 Dresden, Germany
}%

\date{\today}

\begin{abstract}
    We investigate the effects of horizontal oscillations on the drafting–kissing–tumbling (DKT) dynamics of two monodisperse spherical particles settling under gravity in a viscous fluid. 
    Applying particle-resolved direct numerical simulations, we systematically vary the oscillation frequency and amplitude to assess their impact on the behavior of individual particles, their mutual interaction, and the orientation of the particle arrangement.
    The results demonstrate that the oscillatory effects on DKT become significant only when the particle Reynolds number $Re_p$, defined as the ratio of oscillation-induced inertial to viscous forces, exceeds unity. 
    In this regime, oscillations alter the temporal characteristics of the DKT process with moderate amplitudes tending to prolong and larger amplitudes to reduce the kissing phase. 
    Moreover, oscillations affect particle reorientation.
    At low $Re_p$, the particles maintain their initial orientation throughout the interaction, whereas an increasing $Re_p$ promotes a preferential alignment perpendicular to the direction of oscillation.
    We explain these findings by analyzing the oscillation-induced pressure fields surrounding the individual particles, which develops increasingly pronounced lateral anisotropy with increasing $Re_p$. 
    The corresponding lateral hydrodynamic forcing likewise becomes increasingly anisotropic, providing a consistent physical basis for the observed modification of particle interactions and reorientation.
    These findings provide a physical framework for understanding how horizontal oscillations govern binary particle–particle interactions and orientation during gravitational settling.
    \\

\end{abstract}


\maketitle

\section{Introduction} \label{sec:Introduction}

The sedimentation of particles suspended in a viscous fluid refers to their settling under the influence of gravity, which occurs when the weight force of the solid phase exceeds the buoyant and drag forces exerted by the surrounding fluid phase.
This process has been observed since ancient times and has played a fundamental role in early human survival, for example, by enabling water purification to obtain a drinkable resource \cite{2005_Shammas_etal, 2022_Hirom_Devi}.
To this day, sedimentation represents a vital part in the treatment of drinking water and wastewater \cite{2022_Juraev_etal, 2025_Bandh_Mushtaq}.
When particles settle in a suspension, hydrodynamic interactions lead to collisions and contacts with neighboring particles. 
This can either result in sustained particle–particle contact or in separation after collision \cite{1979_Fitch}.
In both cases, the particle interactions modify the surrounding flow field and thus alter the effective settling velocity and trajectory of the particles \cite{2012_Kourki_Famili, 2015_Zaidi_etal, 2019_Hu_Guo}.
To isolate and analyze the governing mechanisms, this inherently complex phenomenon is often reduced to the binary interaction of two particles.
This simplified scenario provides fundamental insights into the hydrodynamic coupling and collision dynamics that underlie more complex collective settling behavior, forming the conceptual basis for the well-established Drafting–Kissing–Tumbling (DKT) mechanism.

DKT describes the settling dynamics of a sequence of interactions of two particles that are initially vertically aligned as they settle due to gravity \cite{1987_Fortes_etal}.
The leading (lower) particle generates a wake characterized by a pressure drop that causes a relative acceleration of the trailing (upper) particle \cite{2005_Xue-Ming_etal}.
Consequently, the trailing particle catches up with the leading one during the drafting phase until they come into contact, which introduces the kissing phase.
During contact, the two particles align almost perpendicular to the settling direction, which is an unstable configuration and causes them to drift apart \cite{1987_Fortes_etal}.
The phenomenon of DKT is known since the work of \textcite{1987_Fortes_etal}, who conducted experiments on spheres settling due to gravity in a viscous fluid.
Since then, numerous experimental and numerical studies have been performed to analyze how the DKT process is affected by various effects such as particle arrangements \cite{2015_Cao_etal, 2021_Nie_etal, 2022_Li_etal}, different particle sizes \cite{1994_Feng_etal, 2005_Xue-Ming_etal, 2014_Wang_etal, 2020_Nie_Lin, 2020_Ghosh_Kumar}, varying particle densities \cite{2020_Nie_Lin, 2022_Li_etal}, and particle properties such as cohesion \cite{2019_Vowinckel_etal}.
These studies collectively provide a comprehensive understanding of pairwise interactions in otherwise quiescent environments.

However, in both natural and engineered systems, the surrounding fluid is rarely quiescent.
Fluid motion arises from a multitude of driving forces and can modify fluid-particle and particle-particle interactions \cite{2016_Fornari_etal}.
Among the various flow regimes, oscillatory flows, which are characterized by periodic accelerations and decelerations, have recently gained increasing attention.
Several studies, primarily on cohesive particles, have shown that oscillatory flows can enhance particle interactions and promote aggregate formation through oscillatory shear \cite{2019a_Halfi_etal, 2019b_Halfi_etal, 2020_Halfi_etal, 2025_Bendory_Friedler} or particle inertia \cite{2024_Kleischmann_etal, 2025_Kleischmann_etal}.
Such changes in particle interactions can, in turn, affect the settling dynamics.
However, the influence of oscillatory forcing on the dynamics of particles that settle under gravity remains largely unexplored.
Elucidating this coupling between gravitational settling and oscillatory dynamics is essential to advance our understanding of particle interactions in such flow environments.
Furthermore, a characterization of the impact of oscillations on the particle contact time may contribute fundamental insights into interaction durations and associated information exchange  between two entities, such as the transfer of pathogens for biologically active particles \cite{2025_Rodriguez-Grande_etal}.

With the present work, we aim to address this research gap  of how horizontal oscillations modify the interactions of two monodisperse particles during settling.
We provide new insights into the interplay between gravitational and oscillation-induced effects in particle pair dynamics through highly resolved numerical simulations by conducting a systematic simulation campaign that varies oscillation frequency and amplitude. 
The remainder of the article is structured as follows.
We introduce the numerical model in~\S\ref{sec:NumericalModel} and present the results in \S\ref{sec:DKT_results}. 
We first discuss the influence of the oscillations on the particle dynamics (\S\ref{sec:ParticleDynamics}), followed by an analysis of the impact of the individual oscillation parameters (\S\ref{sec:ImpactOscParameters}).
We then examine the dynamics during the kissing phase and the associated oscillation-induced pressure field (\S\ref{sec:InteractionPhases}), before addressing the particle orientation relative to the settling and oscillation directions (\S\ref{sec:ParticleOrientation}). 
Finally, we elucidate the mechanism underlying the observed lateral reorientation through an analysis of the relative particle velocities and hydrodynamic forces induced by the oscillatory flow (\S\ref{sec:SingleParticle}), followed by our conclusions (\S\ref{sec:Conclusion}).

\section{Numerical model}\label{sec:NumericalModel}

\subsection{Computational setup}\label{sec:DKT_setup}

\begin{figure}[h!]
    \centering
    \includegraphics[width=0.6\linewidth]{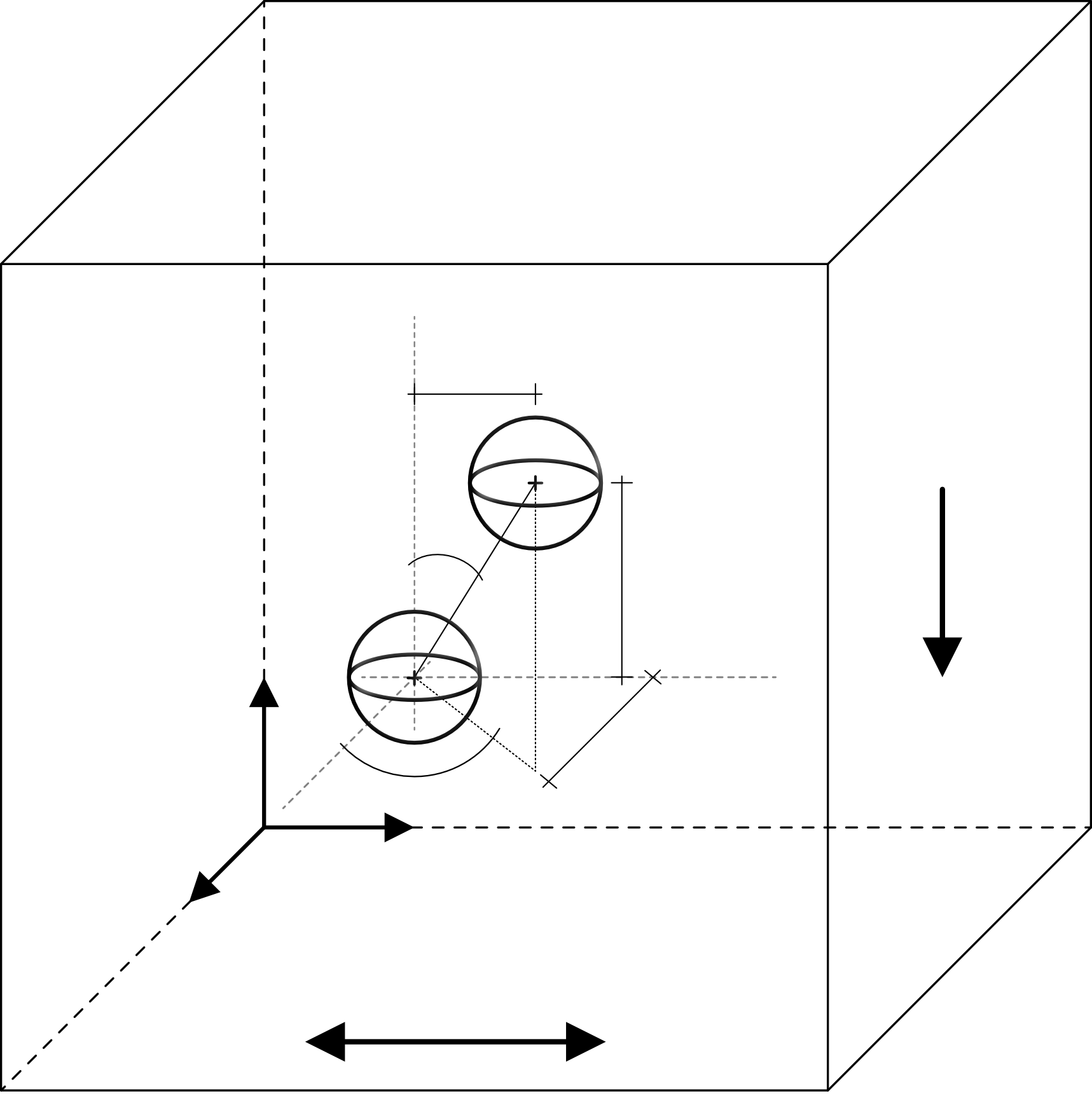}
    \put(-224,100){\footnotesize $x$}
    \put(-185,60){\footnotesize $y$}
    \put(-224,50){\footnotesize $z$}
    \put(-120,130){\footnotesize $\Delta x$}
    \put(-165,185){\footnotesize $\Delta y$}
    \put(-125,90){\footnotesize $\Delta z$}
    \put(-170,131){\footnotesize $\vartheta$}
    \put(-162,124){\footnotesize $\zeta_{c,i}$}
    \put(-170,77){\footnotesize $\varphi$}
    \put(-198,122){\footnotesize $P_1$}
    \put(-131,172){\footnotesize $P_2$}
    \put(-169,19){\footnotesize $u_f$}
    \put(-34,140){\footnotesize $g$}
    \caption{\label{fig:NumericalSetup} Schematic representation of the particle arrangement with the leading particle $P_1$ and the trailing particle $P_2$. 
    The imposed oscillation is directed along the $y$-axis (indicated by the double arrow) and gravity in the negative $x$-direction.
    Particle positions are defined relative to the origin of the Cartesian coordinate system located at the lower back left corner of the numerical domain. 
    Orientations are evaluated in a relative framework, with the origin of the spherical coordinate system placed in the center of $P_1$.}
\end{figure}
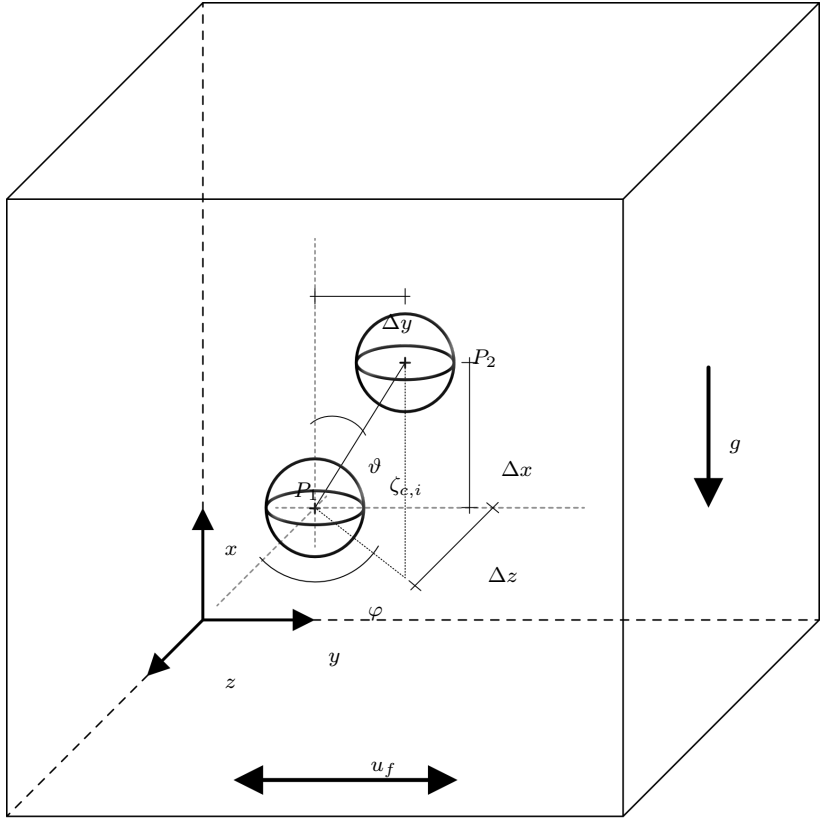

We conduct numerical simulations for a cubical domain of size $L_{x,y,z} = 15 d_p$, with $L_{x,y,z}$ the size in all Cartesian coordinate directions and $d_p$ the diameter of the spherical particles.
The container is filled with viscous fluid and is initially at rest. 
As illustrated in Fig.~\ref{fig:NumericalSetup}, we define the positions of the particles in Cartesian coordinates relative to the origin located in the lower back left corner of the domain with gravity $g$ pointing in the negative $x$-direction. 
The particles of size $d_p$ are placed in the center of the domain with zero initial velocity, where the leading particle $P_1$ is initially positioned at $(7.5 d_p,7.5 d_p,7.5 d_p)^T$ and the trailing particle $P_2$ at $(9.5 d_p,7.525 d_p, 7.525 d_p)^T$.
This yields an initial distance between the particle surfaces of $\zeta_i = \zeta_{c,i} - 2 R_p \approx d_p$.
Here, $\zeta_{c,i} = \sqrt{{\Delta_x}^2 + {\Delta_y}^2 + {\Delta_z}^2}$ represents the initial distance between the centers of the particles, $\Delta_{i}$ the distance along the respective coordinate direction indicated by the subscript $i=x,y,z$, and $R_p$ the particle radius.
Applying small offsets of $\Delta_{y} = \Delta_z =  d_p/40=0.025$, we dislocate $P_2$ from the centerline to promote and control the DKT dynamics, which is common practice for these types of simulations \cite{2015_Dash_Lee, 2019_Vowinckel_etal, 2021_Chen_etal, 2022_Li_etal}.
We evaluate the orientations of the particles in a relative framework, with the origin of the spherical coordinate system placed at the center of $P_1$.
The applied offsets yield an initial orientation of $\vartheta = 1.01 \degree$ and $\varphi = 45 \degree$.
Here, $\vartheta$ represents the polar angle that describes the particle orientation relative to the settling direction and $\varphi$ the azimuthal angle that states the orientation relative to the oscillation direction.

For our computational setup, we employ two non-inertial reference frames.
The first frame is a moving frame that follows the settling motion of the leading particle $P_1$, where its center defines the reference point of this frame \cite{2025_Metelkin_Vowinckel}.
The application of such a moving frame allows for smaller domain sizes and, consequently, enhances the computational efficiency of the simulation campaign \cite{2017_Panah_etal}.
For the implementation, we define the bottom boundary as an inflow  condition, in which the velocity is set to $\textbf{u}=(u_1, 0, 0)$ at $x=0$ to match the settling velocity $u_1$ of the leading particle $P_1$.
Consequently, we specify the boundary condition at the top as a convective outflow ($\textit{D} \textbf{u} / \textit{D} t = 0$ at $x=L_x$).
The remaining horizontal boundaries left, right, front and back are specified as stress-free walls with $d u_t / d n=0$ and $u_n=0$.
Here, $u_t$ and $u_n$ are the components of tangential and normal fluid velocity relative to the wall and $n$~denotes the normal direction on the same wall. 
The particle surfaces are characterized by no-slip boundary conditions.
Within the moving frame, the vertical position of $P_1$ remains at $x = 7.5 d_p$ at all times, while it is free to move in the lateral $y$- and $z$-directions.
The trailing particle $P_2$, on the other hand, is free to move in all spatial directions.
Its settling velocity $u_2$ is considered relative to that of the leading particle $u_2' = u_2 - u_1$.
The second non-inertial reference frame accounts for the external oscillations.
As a result, the particle dynamics are described relative to the motion of the domain \cite{2024_Kleischmann_etal, 2025_Kleischmann_etal}, which is further elaborated in \S \ref{sec:DKT_governingEqs}.

We establish a non-oscillating setup that considers DKT solely due to gravity and designate this simulation as a reference case.
In all remaining simulations, we apply a horizontal oscillation perpendicular to $g$ in the $y$-direction, as indicated by the horizontal double arrow in Fig.~\ref{fig:NumericalSetup}.
These oscillations are monochromatic and unidirectional, which allows a convenient characterization by applying $u_f = -A_f \Omega \sin \Omega t$, where $A_f$ denotes the distance amplitude, $\Omega = 2 \pi f$ the angular frequency, $f$ the frequency and $t$ the time.
The maximum oscillation velocity is defined by the velocity amplitude $A_f \Omega$.

\subsection{Governing equations}\label{sec:DKT_governingEqs}

Simulations are conducted by means of a well established computational code for particle-resolved direct numerical simulations (pr-DNS) \citep[e.g.][]{2017_Biegert_etal, 2019_Vowinckel_etal, 2024_Kleischmann_etal, 2025_Kleischmann_etal} that solve the continuity equations and Navier-Stokes equations for an incompressible Newtonian fluid, given by
\begin{equation}\label{eq:continuity_DKT}
    \nabla\cdot\textbf{u}=0 \qquad , 
\end{equation}
\begin{equation} \label{eq:navier_stokes_DKT}
    \frac{\partial{\textbf{u}}}{\partial{t}}+\nabla\cdot(\textbf{u}\textbf{u}) = -\frac{1}{\rho_f}\:\nabla p + \nu_f \nabla^2 \textbf{u} + \textbf{f}_\text{IBM} + \frac{\partial \textbf{u}_1}{\partial t} \textbf{e}_x \qquad . 
\end{equation}

Here, $\textbf{u}=(u,v,w)^{T}$ denotes the fluid velocity vector in Cartesian components, $\rho_f$~the fluid density, $p$ the fluid pressure, $\nu_f$ the kinematic fluid viscosity, and $\textbf{f}_\text{IBM}$ the volume force based on the immersed boundary method (IBM).
The IBM accounts for the fluid-particle interactions by geometrically resolving the flow around the particles. 
Hence, $\textbf{f}_\text{IBM}$ connects the motion of the particle to the fluid phase by imposing the no-slip condition on the particle surface via interpolation and the application of Dirac delta functions \cite{1999_Roma_etal, 2005_Uhlmann, 2012_Kempe_Froehlich, 2017_Biegert_etal, 2019_Vowinckel_etal}.
The last term on the right-hand side of \eqref{eq:navier_stokes_DKT} represents the inflow in the $x$-direction, as indicated by the unit vector $\textbf{e}_x$, and establishes the non-inertial moving reference frame that follows the settling motion of the leading particle $P_1$.
We apply a third-order low-storage Runge-Kutta scheme in time along a second order finite difference method in space to integrate \eqref{eq:continuity_DKT} and \eqref{eq:navier_stokes_DKT}.
Furthermore, we utilize a fast Fourier transform to calculate the pressure correction for continuity via a direct solver.
We discretize the domain by a uniform rectangular grid with a cell size of $h = L_{x,y,z} / 300$ that results in $d_p / h = 20$.

We account for the non-inertial reference frame of the external oscillations solely in the equations of the particle motion, as we exclude free surfaces and density variations within the fluid.
As detailed in \cite{2024_Kleischmann_etal}, we calculate the translational velocity of a particle relative to that of the fluid $\textbf{u}_p' = \textbf{u}_p - \textbf{u}_f$.
Here, $\textbf{u}_p =(u_p, v_p, w_p)^T$ is the vector of the translational particle velocity in the inertial reference frame.
This allows us to transform Newton's equation to the non-inertial reference frame:
\begin{equation}
    m_p \frac{\text{d}\textbf{u}_p'}{\text{d}t}=  \textbf{F}_\text{h} + \textbf{F}_\text{c} + \textbf{F}_\text{g} - \textbf{F}_\text{osc} 
    \label{eq:NewtonEq_nonInertial}
\end{equation}
%
Euler's equation for the rotational particle velocity remains unaffected by the transformation of the reference frame, as we do not apply an external rotation:
\begin{equation}
    I_p \frac{\text{d}\bm{\omega}_p}{\text{d}t}= \textbf{M}_h \: + \: \textbf{M}_c 
    \label{eq:EulerEq_DKT}
\end{equation}
Here, $m_p$ is the mass of a particle, $I_p = \pi \rho_p d_p ^ 5 / 60$ the moment of inertia with the particle density $\rho_p$, and $\bm\omega_p = \left( \omega_{p,x}, \omega_{p,y}, \omega_{p,z} \right)^{T}$ the angular velocity vector.
The hydrodynamic forces and torques are defined as $\textbf{F}_\text{h} = \oint\limits_{S_p}\, \bm\tau \cdot \textbf{n} \,dS$ and $\textbf{M}_\text{h} = \oint\limits_{S_p}\textbf{r}_p \times (\bm{\tau} \cdot \textbf{n}) \,dS$, respectively, where \mbox{$\bm\tau = -p \textbf{I} \: + \: \mu_f \left[ \nabla \textbf{u} + \left(\nabla \textbf{u} \right)^T \right]$} is the hydrodynamic stress tensor including the identity tensor $\textbf{I}$ and the dynamic fluid viscosity $\mu_f$. 
Furthermore, \mbox{$\textbf{n}$ denotes} the outward pointing normal vector, and $\textbf{r}_p$ the position vector pointing from the particle center of mass to a point on the particle surface $S_p$. 
The forces and torques $\textbf{F}_\text{c}$ and $\textbf{M}_\text{c}$ originate from collisions and contacts, and
$\textbf{F}_\text{g} = V_p \left(\rho_p - \rho_f \right) \textbf{g}$ as well as $\textbf{F}_\text{osc} = V_p \left(\rho_p - \rho_f \right) \frac{\text{d}\textbf{u}_f}{\text{d}t}$ constitute body forces due to external accelerations. In our case, those are gravitation and oscillation, with $V_p$ the particle volume and $\textbf{g}=(g,0,0)^T$ the gravitational acceleration.

The collision force $\textbf{F}_\text{c} = \textbf{F}_\text{l} + \textbf{F}_\text{n} + \textbf{F}_\text{t}$ comprises the lubrication force $\textbf{F}_\text{l}$ and the contact force with its normal $\textbf{F}_\text{n}$ and tangential components $\textbf{F}_\text{t}$.
In principle, particle-wall collisions are also included, but these are not relevant for the present configuration.
The methodology for evaluating these forces has been well established and validated by \textcite{2017_Biegert_etal} for various benchmark cases.
As soon as the distance between the surfaces of the two particles $\zeta_n$ reaches $\zeta_n \leq 2h$, we apply a lubrication force model for $\textbf{F}_\text{l}$ to account for fluid within the gap that is too small to be resolved by the IBM \cite{2017_Biegert_etal, 2019_Vowinckel_etal}.
In line with the study by Cox \& Brenner \citep{1967_Cox_Brenner}, we apply the lubrication force model
\begin{equation}
    \textbf{F}_\text{l} = - 6 \pi \rho_f \nu_f R_{eff}^2 \textbf{u}_{n,cp} \; / \, \text{max} \left( \zeta_n, \zeta_{min} \right) \; ,
\end{equation}
with  $R_{eff} = R_1 R_2 \, / \, (R_1 +R_2)$ the effective radius, where $R_1$ and $R_2$ represent the individual radii of $P_1$ and $P_2$.
$\textbf{u}_{n,cp} = (\textbf{u}_{cp} \cdot \textbf{n) \textbf{n}}$ denotes the normal component of the relative velocity of the surface contact point $\textbf{u}_{cp} = \textbf{u}_{r,p} + R_{1,cp} (\omega_{1} \times \textbf{n}) + R_{2,cp} (\omega_{2} \times \textbf{n})$.
Where $\textbf{u}_{r,p} = \textbf{u}_{2} - \textbf{u}_{1}$ is the relative velocity between the centers of mass of the two particles, $R_{1,cp} = || \textbf{x}_{cp} - \textbf{x}_{1} ||$ and $R_{2,cp} = || \textbf{x}_{cp} - \textbf{x}_{2} ||$ are the respective radii of the two particles with respect to the contact point, and $\omega_{1}$ as well as $\omega_{2}$ are the respective angular velocities.
$\zeta_{min}$ characterizes the surface roughness of the particle and serves as a limiter to prevent $\textbf{F}_\text{l} \rightarrow \infty$ with decreasing gap size.
We set $\zeta_{min} = 3 \cdot 10^{-3} \, R_m$, where $R_m = (R_1 +R_2) \, / \, 2$ denotes the mean radius of the two particles.
This choice follows the work of \textcite{2017_Biegert_etal}, who calibrated this parameter to reproduce the rebound trajectory of a particle colliding with a wall based on the experiments of \textcite{2002_Gondret_etal}.

As soon as the particles are in contact, that is, $\zeta_n \leq 0$, two models account for normal contact and tangential friction.
A nonlinear spring-dashpot system based on the work of Kempe \& Fr\"ohlich \citep{2012_Kempe_Froehlich} models the contact force in the normal direction: 
\begin{equation}
    \textbf{F}_\text{n} = - k_n |\zeta_n|^{3/2} \, \textbf{n} - d_n \, \textbf{u}_{n,cp} 
    \label{eq:ContactForce_Normal}
\end{equation}
Here, $k_n$ and $d_n$ represent the normal components of the stiffness and damping coefficients that are iteratively calibrated for each collision to achieve a restitution coefficient prescribed by $e_{dry} = -u_{out} / u_{in}$, with $u_{in}$ and $u_{out}$ stating the velocities before and after the impact of the particles \cite{2017_Biegert_etal, 2019_Vowinckel_etal}.
To incorporate the tangential component of the contact force, we apply the Coulomb friction criterion to limit the maximum force $\textbf{F}_\text{t} = min \left( ||\textbf{F}_\text{t,m}|| \, , \, ||\mu \textbf{F}_\text{n} ||  \right) \, \textbf{t}_n$, where $\textbf{F}_\text{t,m}$ represents a linear spring-dashpot model \cite{2013_Thornton_etal}:
\begin{equation}
    \textbf{F}_\text{t,m} = - k_t \bm{\zeta}_t  - d_t \, \textbf{u}_{t,cp} 
    \label{eq:ContactForce_Tangential}
\end{equation}
Here, $\mu$ is the friction coefficient that describes the friction between the surfaces of the particles, $\textbf{t}_n$ is the normal vector in the direction of the tangential force, and $k_t$ as well as $d_t$ are the tangential coefficients of stiffness and damping \cite{2013_Thornton_etal}.
$\bm{\zeta}_{t}$ represents the tangential spring displacement and $\textbf{u}_{t,cp}$ the tangential equivalent to $\textbf{u}_{n,cp}$ \cite{2017_Biegert_etal}.
For the present study, we choose $e_{dry} = 0.97$ and $\mu = 0.15$ for silicate materials \cite{2001_Joseph_etal, 2004_Joseph_Hunt, 2017_Biegert_etal, 2019_Vowinckel_etal}.

\subsection{Non-dimensionalization}\label{sec:DKT_nonDim}

We non-dimensionalize the governing equations by scaling all variables with characteristic quantities.
As such, we choose the variables that remain constant for all setups of the numerical simulations in the present study, namely $d_p$, $g$, and $\rho_f$.
The non-dimensional quantities are represented by the tilde symbol.

\begin{equation} \label{eq:nonDim_scales_DKT}
    \begin{split}
        \begin{gathered}
            \vspace{0.1cm}
            \ell = d_{p} \, \tilde{\ell}, \quad %
            t = \sqrt{d_p / g} \; \tilde{t}, \quad %
            \textbf{u} = \sqrt{d_p  g} \; \tilde{\textbf{u}}, \\
            \vspace{0.1cm}
            \rho = \rho_{f} \tilde{\rho}, \quad %
            p = \rho_f d_p g \Tilde{p}, \quad %
            g = g \, \Tilde{g}, \\
            \vspace{0.1cm}
            V = d_p^3 \: \tilde{V}, \quad %
            m = m_{f} \tilde{m} = \rho_f V_{p} \tilde{m}, \quad %
            \color{black}{\nu = d_p^2 \, / \sqrt{d_p / g} \; \tilde{\nu}} \\
            \vspace{0.1cm}
            F = m_f g \tilde{F}, \quad
            I = m_f \: d_p^2 \tilde{I}, \quad %
            M = m_f d_p g \Tilde{M},
       \end{gathered}
    \end{split}
\end{equation}
with $\ell$ representing any relevant length scale and $m_f$ the fluid mass.
We apply \eqref{eq:nonDim_scales_DKT} to \eqref{eq:continuity_DKT}-\eqref{eq:EulerEq_DKT} to obtain their non-dimensional forms, in which $Re_g = d_p \sqrt{d_p g} / \nu_f$ represents the gravitational Reynolds number.

\begin{equation}\label{eq:nondim_continuity_DKT}
    \tilde{\nabla}\cdot\tilde{\textbf{u}}=0 \quad ,
\end{equation}
\begin{equation} \label{eq:nondim_navier_stokes_DKT}
    \frac{\partial{\tilde{\textbf{u}}}}{\partial{\tilde{t}}} + \tilde{\nabla} \cdot (\tilde{\textbf{u}}\tilde{\textbf{u}}) = -\tilde{\nabla} \tilde{p} + \frac{1}{Re_g} \tilde{\nabla}^2 \tilde{\textbf{u}} + \tilde{\textbf{f}}_\text{IBM} + \frac{\partial \tilde{\textbf{u}}_1}{\partial \tilde{t}} \textbf{e}_x \quad ,
\end{equation}
\begin{equation}\label{eq:nondim_NewtonEq_nonInertial_DKT}
    \tilde{m}_p \frac{\text{d}\tilde{\textbf{u}}_p}{\text{d}\tilde{t}}=  \tilde{\textbf{F}}_\text{h} + \tilde{\textbf{F}}_\text{c} + \tilde{\textbf{F}}_\text{g} - \tilde{\textbf{F}}_\text{osc} \quad ,
\end{equation}
\begin{equation}\label{eq:nondim_EulerEq_DKT}
    \tilde{I}_p \frac{\text{d}\tilde{\bm{\omega}}_p}{\text{d}\tilde{t}}= \tilde{\textbf{M}}_h \: + \: \tilde{\textbf{M}}_c \qquad .
\end{equation}

The non-dimensional form of the density differences in $\tilde{\textbf{F}}_\text{g}$ and $\tilde{\textbf{F}}_\text{osc}$ (i.e. $\rho_p - \rho_f$) results in $(\rho_s - 1)$, with the density ratio $\rho_s = \rho_p / \rho_f$.
In addition, we introduce a particle Reynolds number $Re_p = d_p (1-\eta) A_f \Omega / \nu_f$, where the characteristic velocity, $(1-\eta) A_f \Omega$, corresponds to the maximum oscillatory velocity of the particle induced by the oscillating fluid.
Here, ${\eta = A_p / A_f}$ represents the ratio between the amplitudes of the particle and the fluid.
Details on the calculation of $A_p$ are presented in Appendix \ref{appendix:ParticleAmplitude}. 
We further introduce a non-dimensional frequency $S = d_p^2 \Omega \, / \, (36 \nu_f)$ \cite{2001_Coimbra_Rangel, 2004_Coimbra_etal, 2005_LEsperance_etal, 2024_Kleischmann_etal}.

For the numerical setup, we consider a dimensional system containing water of density $\rho_f = 1,000 \, [kg / m^3]$ and viscosity $\nu_f = 10^{-6} \, [m^2/s]$ with two submersed sediment particles of size  $d_p = 200 \, [\mu m]$ and density $\rho_p = 2,600 \, [kg / m^3]$.
The particles are subjected to gravitational acceleration, $g = 9.81 \, [m/s^2]$, which results in $Re_g = 8.86$.
Applying the non-dimensional scales from \eqref{eq:nonDim_scales_DKT}, we obtain dimensionless values of the density ratio $\rho_s = 2.6$, the fluid viscosity $\tilde{\nu_f} = 0.11$, the particle diameter $\tilde{d}_p = 1$ and the gravity $\tilde{g} = 1$.
This system is unaffected by oscillations and we define it as a reference case.

\subsection{Oscillation setups}\label{sec:OscillationSetups}

In addition to the non-oscillating reference case, we consider oscillatory configurations defined by varying combinations of amplitudes and frequencies.
We consider frequencies in the range from $f = 20$ to $200 \, [\text{Hz}]$ with increments of $10 \, [\text{Hz}]$ and five different amplitudes $A_f = 20$, $50$, $100$, $150$, and $200 \, [\mu m]$.
This yields the non-dimensional numbers $Re_p = [0.01, 17.79]$ and $S=[0.14,1.4]$. 
We select this parameter space based on a prior convergence analysis, which confirms that the results are independent of spatial and temporal resolution and remain unaffected by confinement or wall-induced effects on the particle dynamics.

The application of the non-dimensional time scale $t = \sqrt{d_p/g} \, \tilde{t}$ together with a numerical timestep of $\Delta \tilde{t} = 10^{-3}$ yields approximately $1100$ timesteps for $S=1.4$ and $22,147$ timesteps for $S=0.14$ for each oscillation period. 
Together with the spatial discretization introduced in \S\ref{sec:DKT_governingEqs} ($\Delta \tilde{x} = \tilde{d}_p / 20$), we obtain a maximum CFL number of $\text{CFL} = \tilde{A}_f \tilde{\Omega} \Delta \tilde{t} / \Delta \tilde{x} = 0.11$.
We deliberately chose this small CFL value to minimize the influence of numerical effects, thereby ensuring that the results remain effectively independent of the spatial and temporal discretizations.
Unless explicitly indicated, we express all subsequent equations, results, and quantities in dimensionless form and drop the tilde symbol for the remainder of the article. All simulation setups as well as the data used for the present analysis have been made publicly available as supplemental material \cite{supplemental_material}

\section{Results}\label{sec:DKT_results}

\subsection{Particle dynamics} \label{sec:ParticleDynamics}

We start our analysis by evaluating the effects of the horizontal oscillations on the particle-particle interaction. 
In this context, we systematically examine the differences in particle trajectories and settling velocities induced by the imposed oscillations.
To this end, we compare the non-oscillating reference case to an oscillating setup defined by $A_f=1.0$ and $S=1.12$ yielding $Re_p = 13.48$.

\subsubsection{Particle trajectories}\label{sec:ParticleTrajectories}

We present the particle trajectories for the non-oscillating reference case and the characteristic oscillating case in Figs.~\ref{fig:Trajectories_Settling}(a,c) and \ref{fig:Trajectories_Settling}(b,d), respectively.
We reconstruct the vertical trajectories of the particles by the temporal integration of their settling velocities, \mbox{$x(t) = x_0 + \int_{0}^{t} u_{1,2}(\tau)\, d\tau$}, where $x_0$ denotes the prescribed initial vertical position, chosen sufficiently large to allow for an adequately long settling distance, and $u_{1,2}$ represents the instantaneous settling velocities of particles $P_1$ and $P_2$, respectively.
Figs.~\ref{fig:Trajectories_Settling}(a,b) illustrate the settling trajectories with respect to the $y$-axis 
{and Figs.~\ref{fig:Trajectories_Settling}(c,d) with respect to the \mbox{$z$-axis}.
The leading particle $P_1$ is presented in black and the trailing particle $P_2$ in red.
The particles are initially aligned vertically, with $P_2$ slightly displaced by lateral offsets in the \mbox{$y$- and $z$-directions} (cf. \S\ref{sec:DKT_setup}). 
The horizontal dashed lines indicate the transitions from the drafting to the kissing phase and from the kissing to the tumbling phase, as labeled in Fig.~\ref{fig:Trajectories_Settling}(a).
We consider the particles to be in the kissing phase when the normalized surface-to-surface distance $\zeta_n$ is below the kissing threshold $\zeta_k$, i.e., $\zeta_n \leq \zeta_k$.
We define this threshold as $\zeta_k = 2 \, \zeta_{min}$, where $\zeta_{min}$ represents the prescribed surface roughness (cf.~\S\ref{sec:DKT_governingEqs}).
A detailed analysis presented in Appendix \ref{appendix:KissingDistance} demonstrates that the particle–particle interaction during the kissing period is not strongly affected by the choice of $\zeta_{min}$.

\begin{figure}[h!]
    \centering
    \captionsetup[subfigure]{labelformat=empty}
    \begin{subfigure}[b]{0.23\textwidth}
        \centering
        \topinset{{\footnotesize (a)}}{\includegraphics[trim=25cm 1cm 24cm 2cm, clip,width=\textwidth]{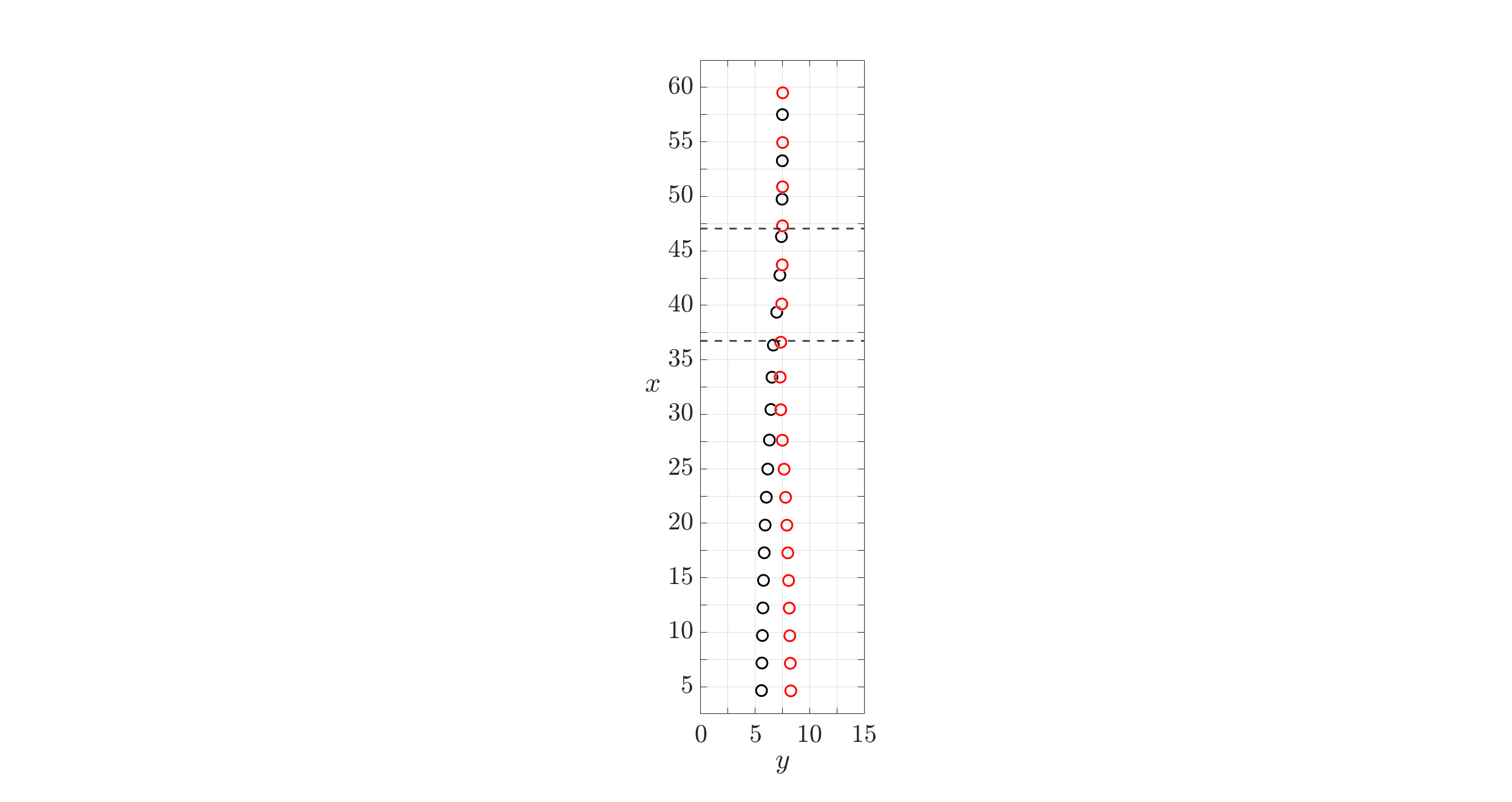}}{-0.1cm}{-1.5cm}
        \put(-28,287){\rotatebox{90}{\footnotesize Drafting}}
        \put(-28,218){\rotatebox{90}{\footnotesize Kissing}}
        \put(-28,155){\rotatebox{90}{\footnotesize Tumbling}}
    \end{subfigure}%
    \begin{subfigure}[b]{0.23\textwidth}
        \centering
        \topinset{{\footnotesize (b)}}{\includegraphics[trim=25cm 1cm 24cm 2cm, clip,width=\textwidth]{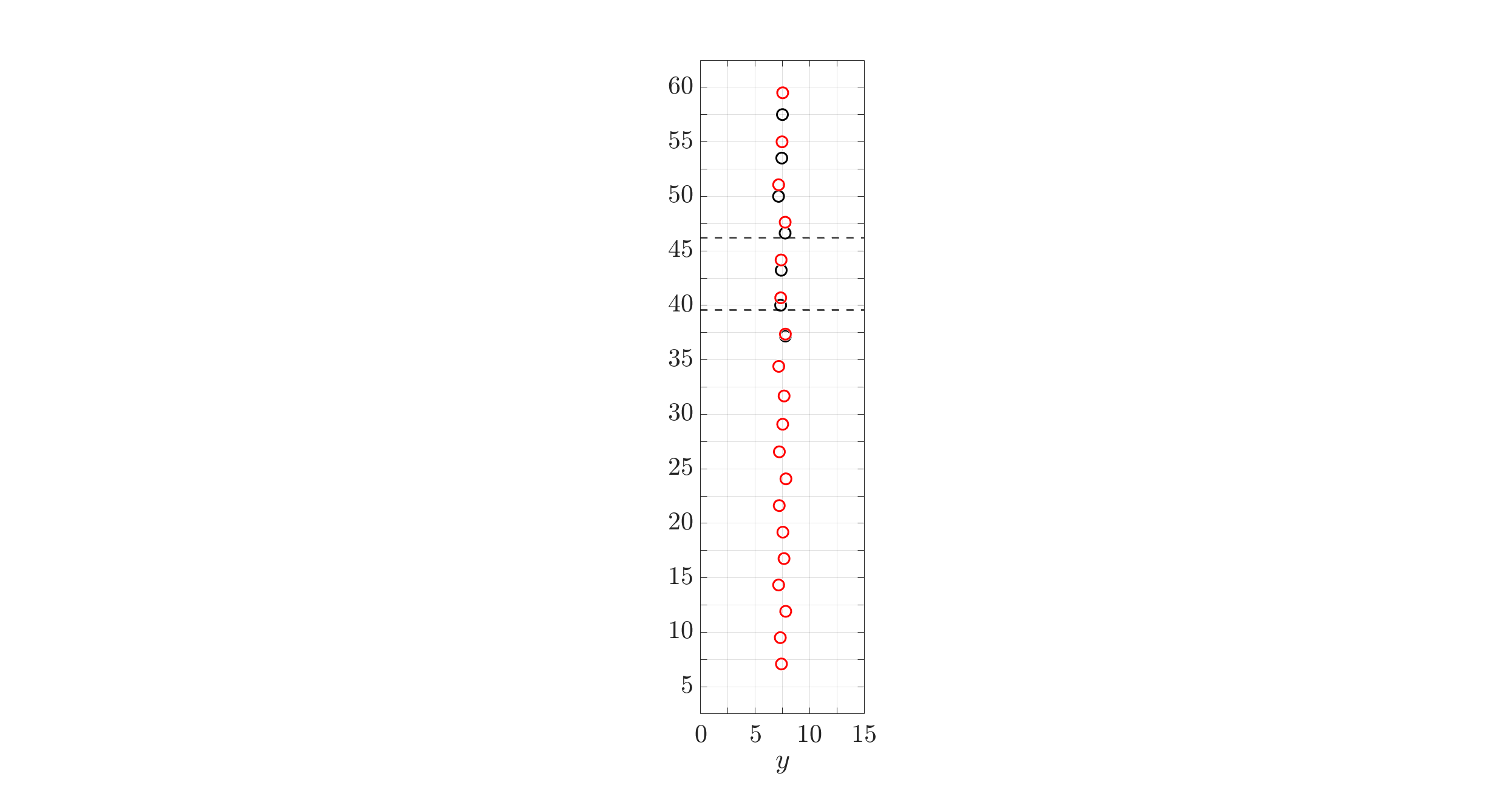}}{-0.1cm}{-1.5cm}
    \end{subfigure}%
    \hfill
    %
    \begin{subfigure}[b]{0.23\textwidth}
        \centering
        \topinset{{\footnotesize (c)}}{\includegraphics[trim=25cm 1cm 24cm 2cm, clip,width=\textwidth]{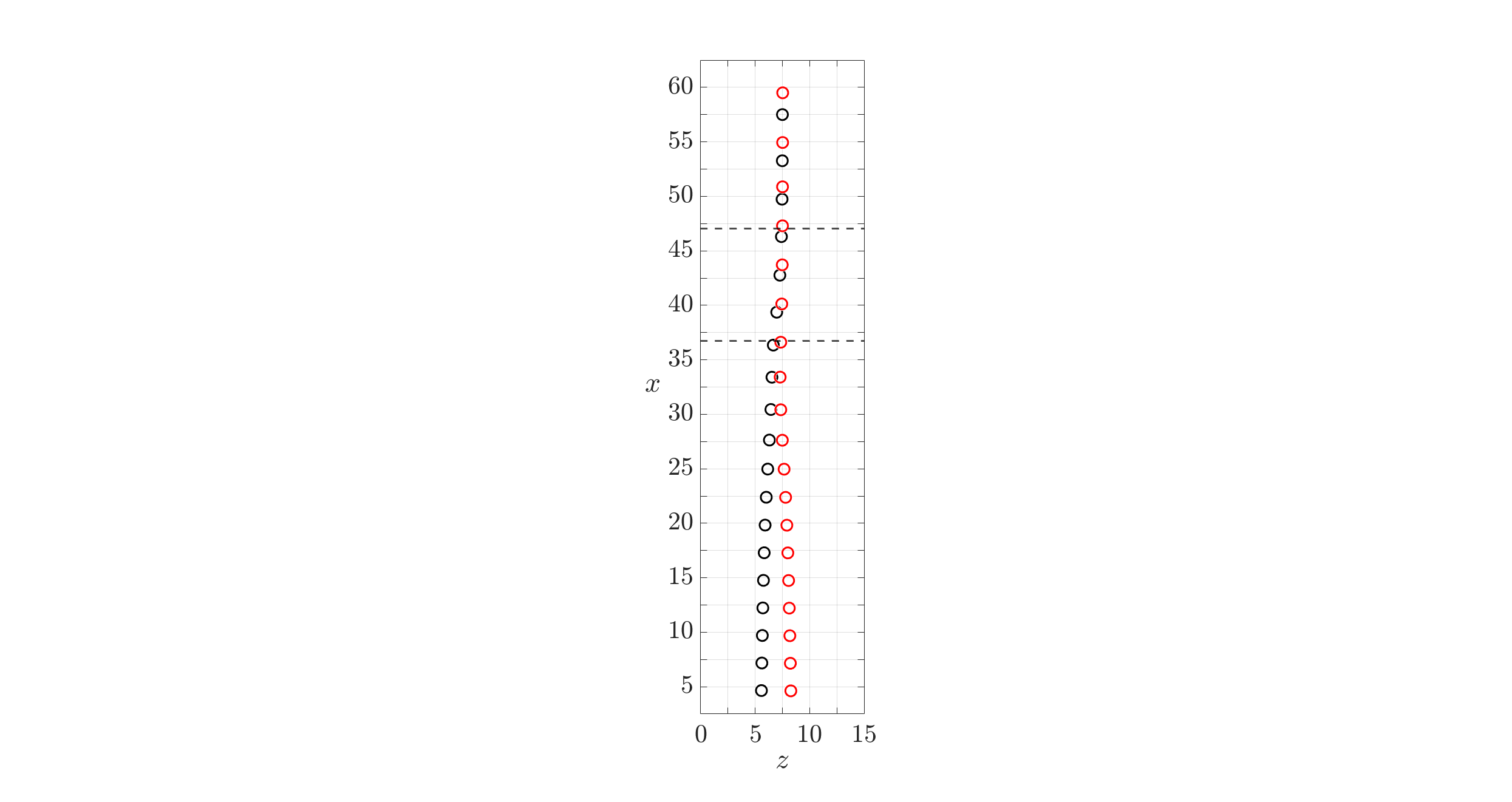}}{-0.1cm}{-1.5cm}
    \end{subfigure}%
    \begin{subfigure}[b]{0.23\textwidth}
        \centering
        \topinset{{\footnotesize (d)}}{\includegraphics[trim=25cm 1cm 24cm 2cm, clip,width=\textwidth]{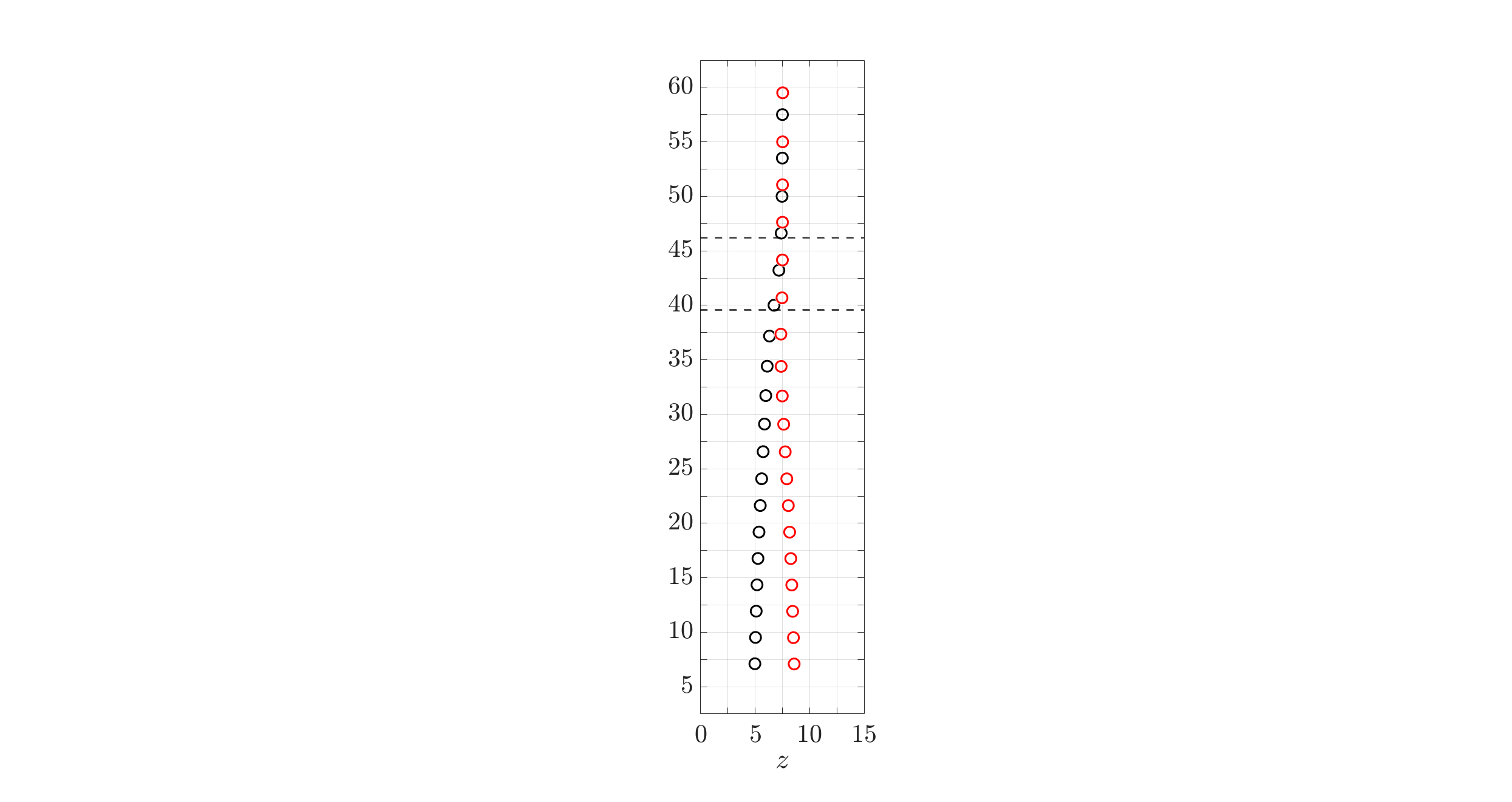}}{-0.1cm}{-1.5cm}
    \end{subfigure}
    \caption{\label{fig:Trajectories_Settling}%
    Comparison of visualized particle trajectories between the non-oscillating reference case (a, c) and the oscillating setup $A_f=1.0$, $S=1.12$ (b, d).
    The leading particle, $P_1$, is presented in black and the trailing particle, $P_2$, in red, with positions recorded at intervals of $5000 \Delta t$.
    The settling trajectory in $x$-direction is reconstructed by integrating the settling velocities over time and illustrated with respect to the $y$-axis (a, b) and the $z$-axis (c, d).
    The horizontal dashed lines indicate the kissing phase, which delineates the preceding drafting phase and the subsequent tumbling phase.
    The disappearance of $P_1$ in (b) is due to its perpendicular alignment relative to the oscillation direction.
    }
\end{figure}

The comparison of the particle trajectories between the non-oscillating reference setup and the oscillating configuration (Fig.~\ref{fig:Trajectories_Settling}) reveals three key observations:
(i) the kissing phase is altered in the presence of oscillations, exhibiting a slightly shifted onset and a reduced duration;
(ii) the settling distance decreases due to the impact of the oscillation;
and (iii) the oscillatory forcing leads to a modified particle orientation.
The latter effect is particularly evident in Fig.~\ref{fig:Trajectories_Settling}(b), where the particle representations overlap along the $y$-axis, indicating an alignment perpendicular to the oscillation direction.

To gain further insight on the influence of oscillations on the particle-particle interaction, we compare the evolution of the individual coordinate directions and the the total distance~$\zeta_n$ between the same setups, that are the non-oscillating reference case and the oscillating scenario ($A_f=1.0$, $S=1.12$, $Re_p = 13.48$). 
Fig.~\ref{fig:Trajectories_IndividualComparison} shows the temporal development of the individual trajectories for the reference case (left, Figs.~\ref{fig:Trajectories_IndividualComparison}a,c,e) and the oscillating setup (right, Figs.~\ref{fig:Trajectories_IndividualComparison}b,d,f), as well as for the total distance (Figs.~\ref{fig:Trajectories_IndividualComparison}g and \ref{fig:Trajectories_IndividualComparison}h, respectively).
For the individual trajectories, we keep the same color scheme to differentiate between the particles. 
The vertical dashed lines mark the beginning and end of the kissing phase. 
Along the $x$-direction, the particle trajectories exhibit similar behavior in both cases. 
The particles are initially separated by a center-to-center distance of $\zeta_{c,i} = 2$, corresponding to an initial surface-to-surface distance of $\zeta_i = \zeta_{c,i} - 2R_p = 1$. 
This separation decreases as the particles approach each other, culminating in the onset of the kissing phase. 
During this phase, the interparticle distance remains approximately constant over a substantial duration. 
Near the end of the kissing phase, however, the distance decreases again and ultimately vanishes, indicating that both particles attain the same position along the settling direction.
In contrast, pronounced differences appear along the $y$-axis (Figs.~\ref{fig:Trajectories_IndividualComparison}c,d).
In the non-oscillating configuration, the particles begin to diverge just prior to the onset of the kissing phase. 
Initially, $P_2$ moves in the direction of $P_1$ while the inter-particle distance increases, then reverses course and moves in the opposite direction, with the two particles continuing to separate progressively over time.
Conversely, in the oscillating setup, the particle trajectories overlap and exhibit a persistent back-and-forth motion along the direction of the imposed oscillations.
In the $z$-direction (Figs.~\ref{fig:Trajectories_IndividualComparison}e,f), the trajectories of the reference case mirror those observed along $y$. 
In the oscillating setting, the trajectories follow a trend similar to the non-oscillating reference, with a slightly increased final particle separation.

\begin{figure}[h!]
    \def\stackalignment{l}
    \centering
    \captionsetup[subfigure]{labelformat=empty}
    \begin{subfigure}[b]{\textwidth}
         \centering 
         \topinset{{\footnotesize (a)}}{\includegraphics[trim=7cm 1.2cm 4cm 2.9cm, clip,width=0.49\textwidth]{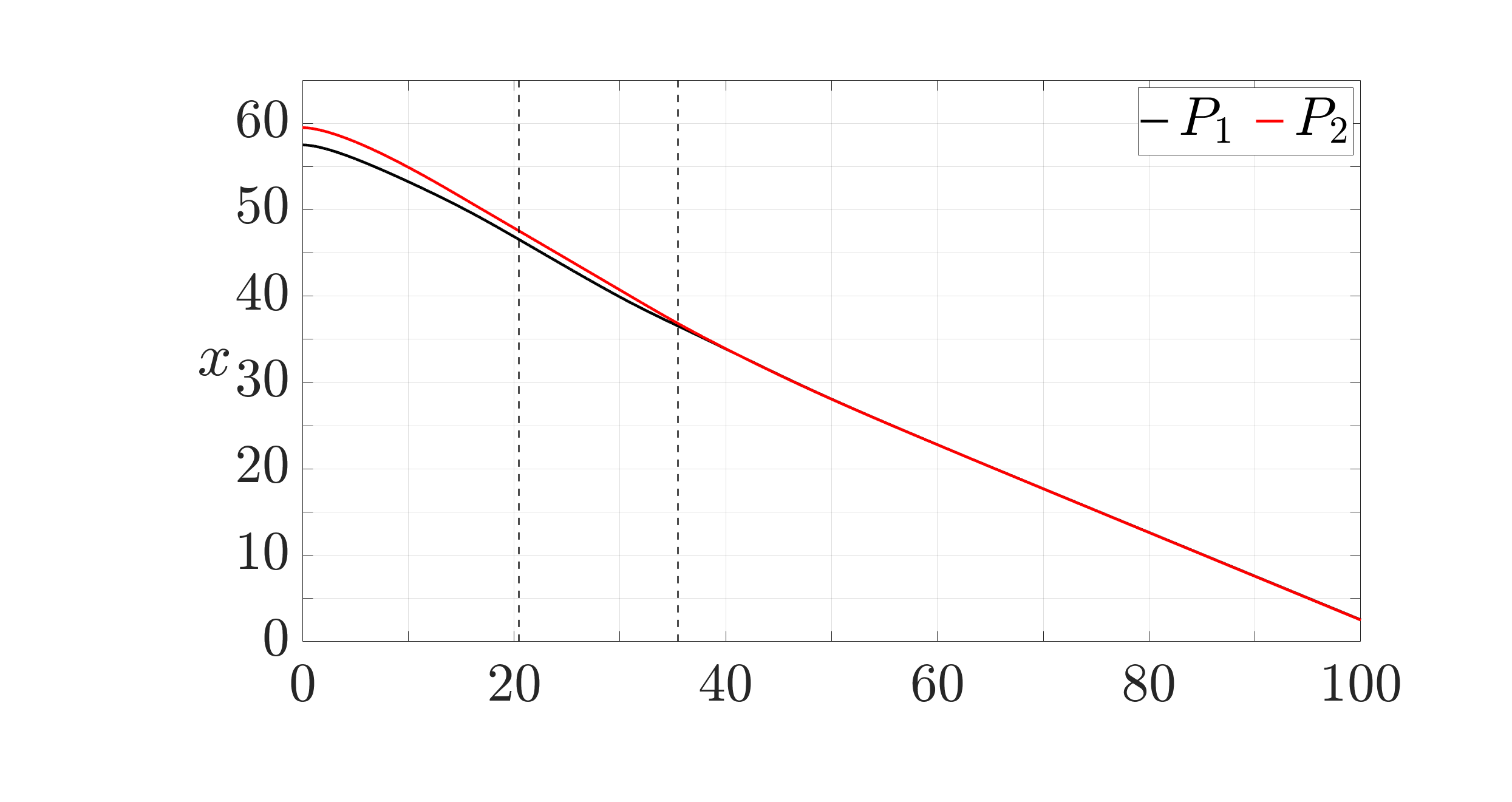}}{0cm}{-0.15cm}
         \topinset{{\footnotesize (b)}}{\includegraphics[trim=7cm 1.2cm 4cm 2.9cm, clip,width=0.49\textwidth]{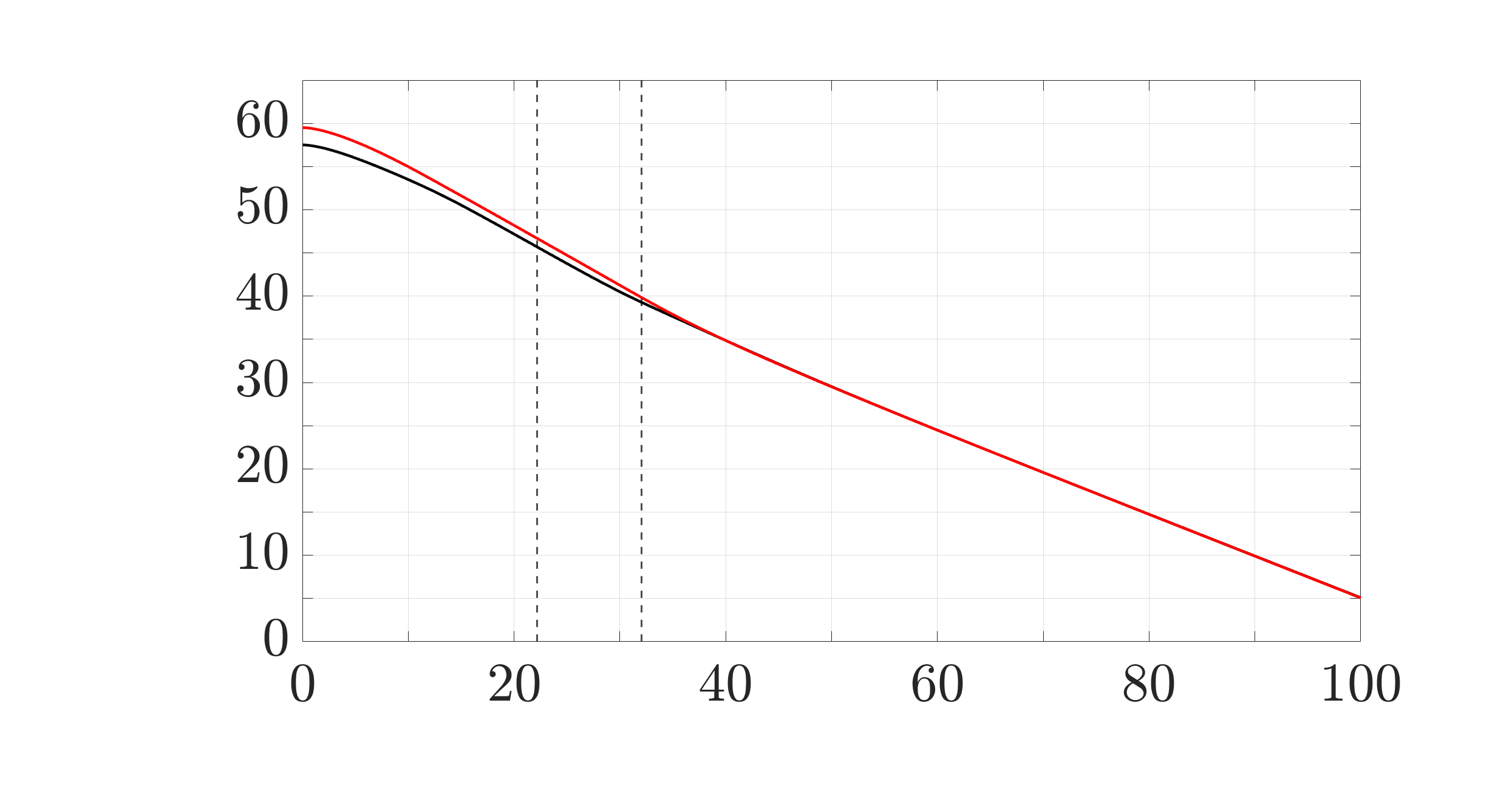}}{0cm}{-0.15cm}
    \end{subfigure}

    \begin{subfigure}[b]{\textwidth}
         \topinset{{\footnotesize (c)}}{\includegraphics[trim=7cm 1.2cm 4cm 2.9cm, clip,width=0.49\textwidth]{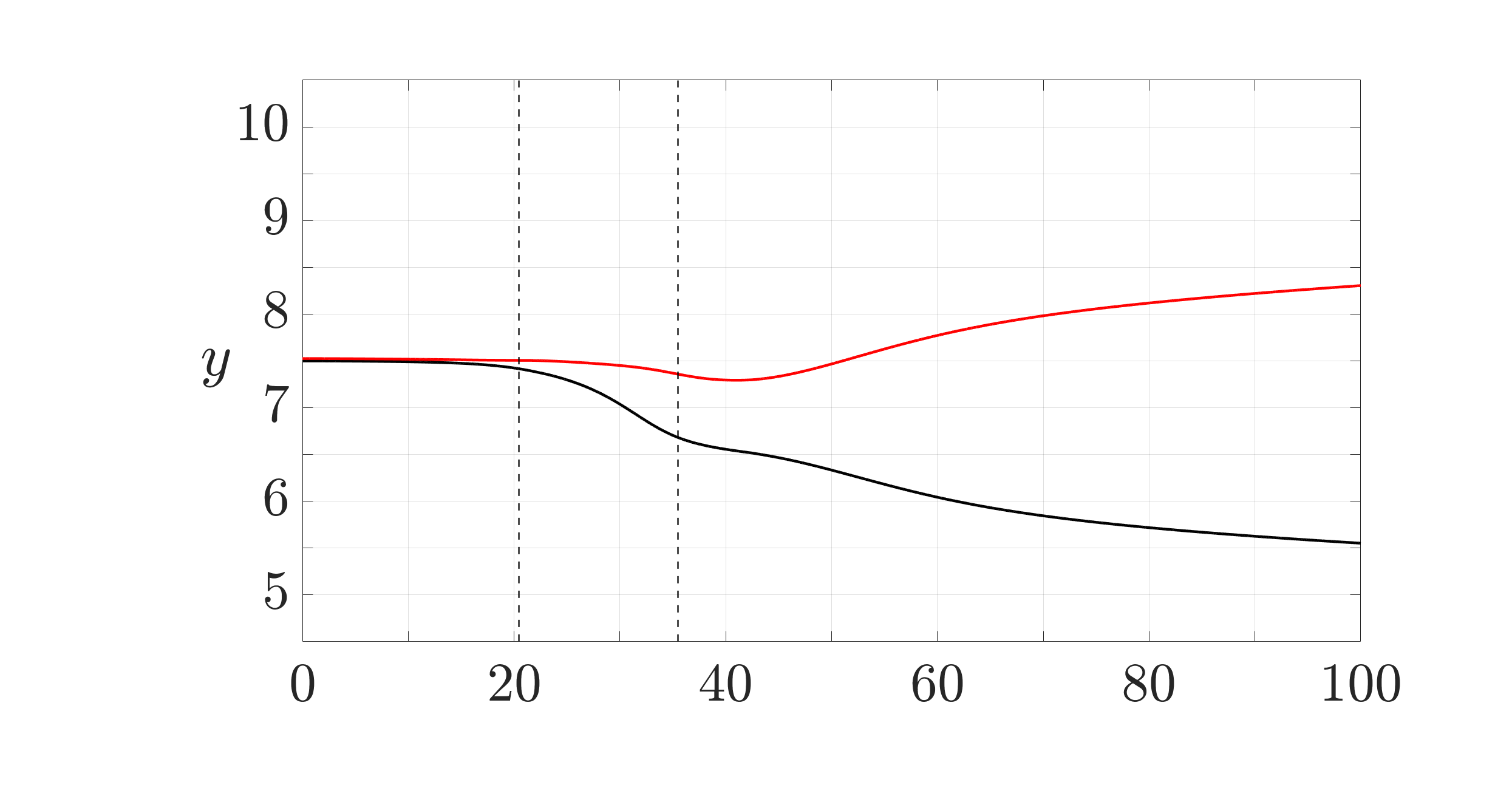}}{0cm}{-0.15cm}
         \topinset{{\footnotesize (d)}}{\includegraphics[trim=7cm 1.2cm 4cm 2.9cm, clip,width=0.49\textwidth]{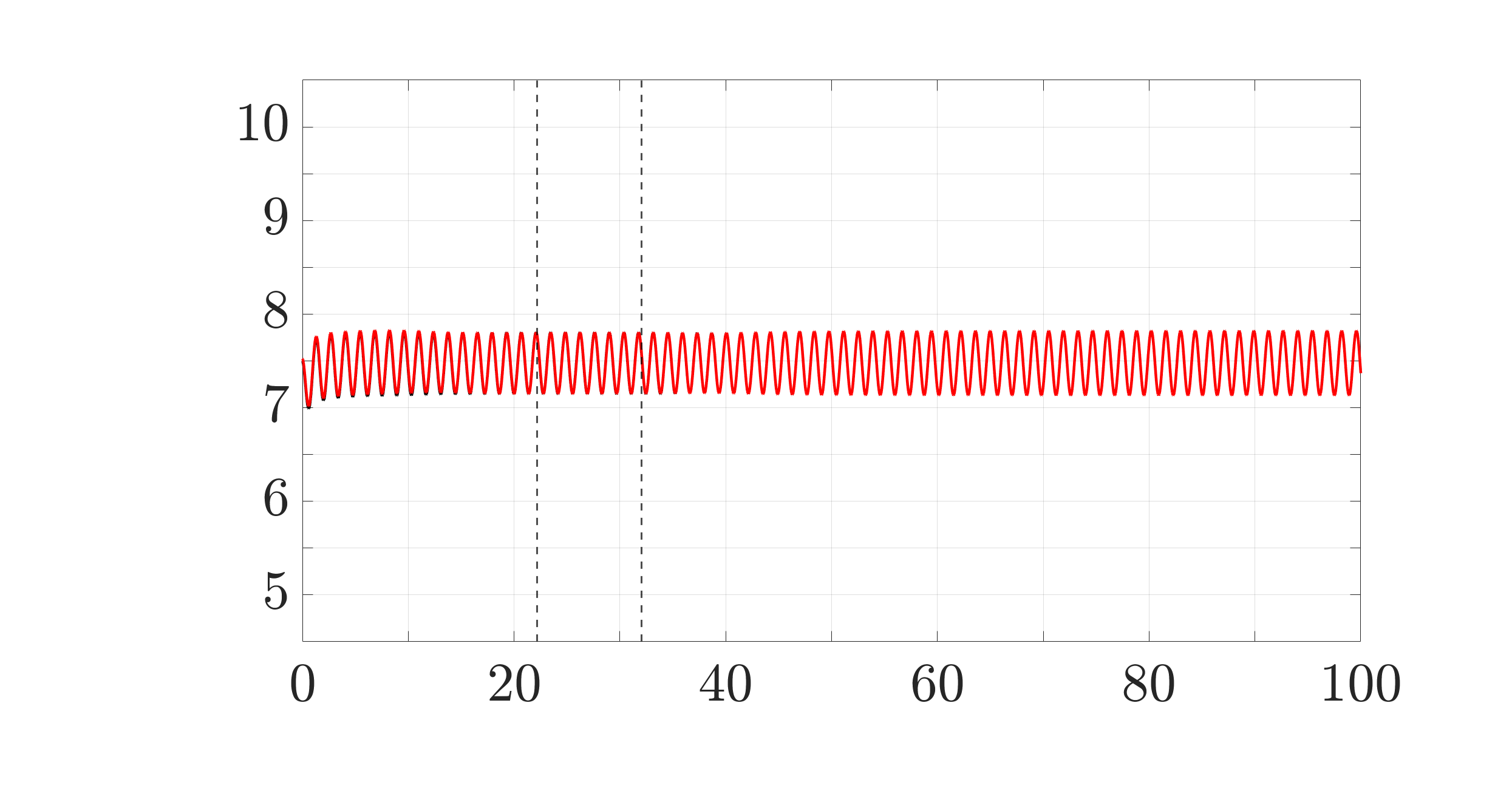}}{0cm}{-0.15cm}
    \end{subfigure}

    \begin{subfigure}[b]{\textwidth}
         \topinset{{\footnotesize (e)}}{\includegraphics[trim=7cm 1.2cm 4cm 2.9cm, clip,width=0.49\textwidth]{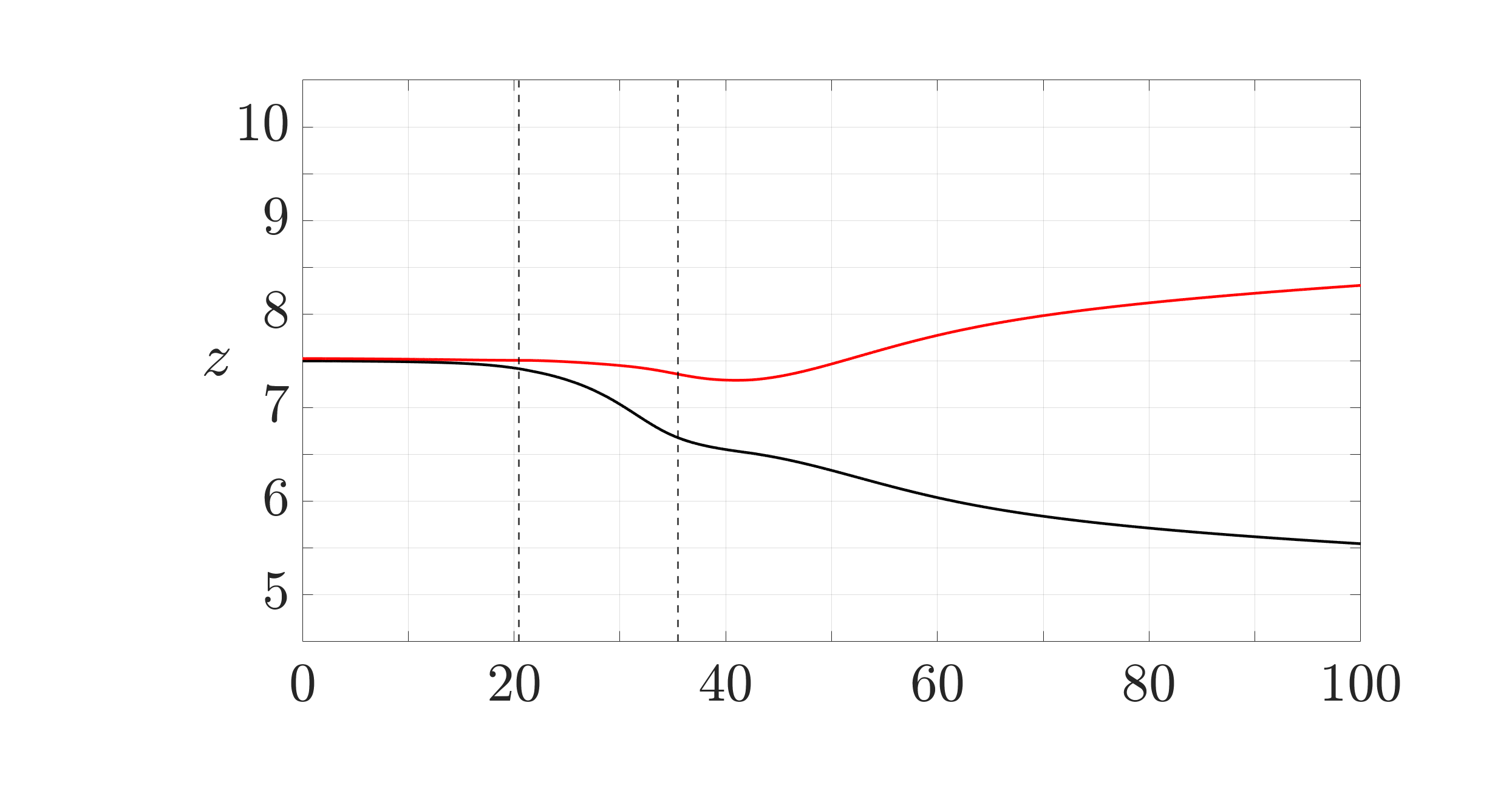}}{0cm}{-0.15cm}
         \topinset{{\footnotesize (f)}}{\includegraphics[trim=7cm 1.2cm 4cm 2.9cm, clip,width=0.49\textwidth]{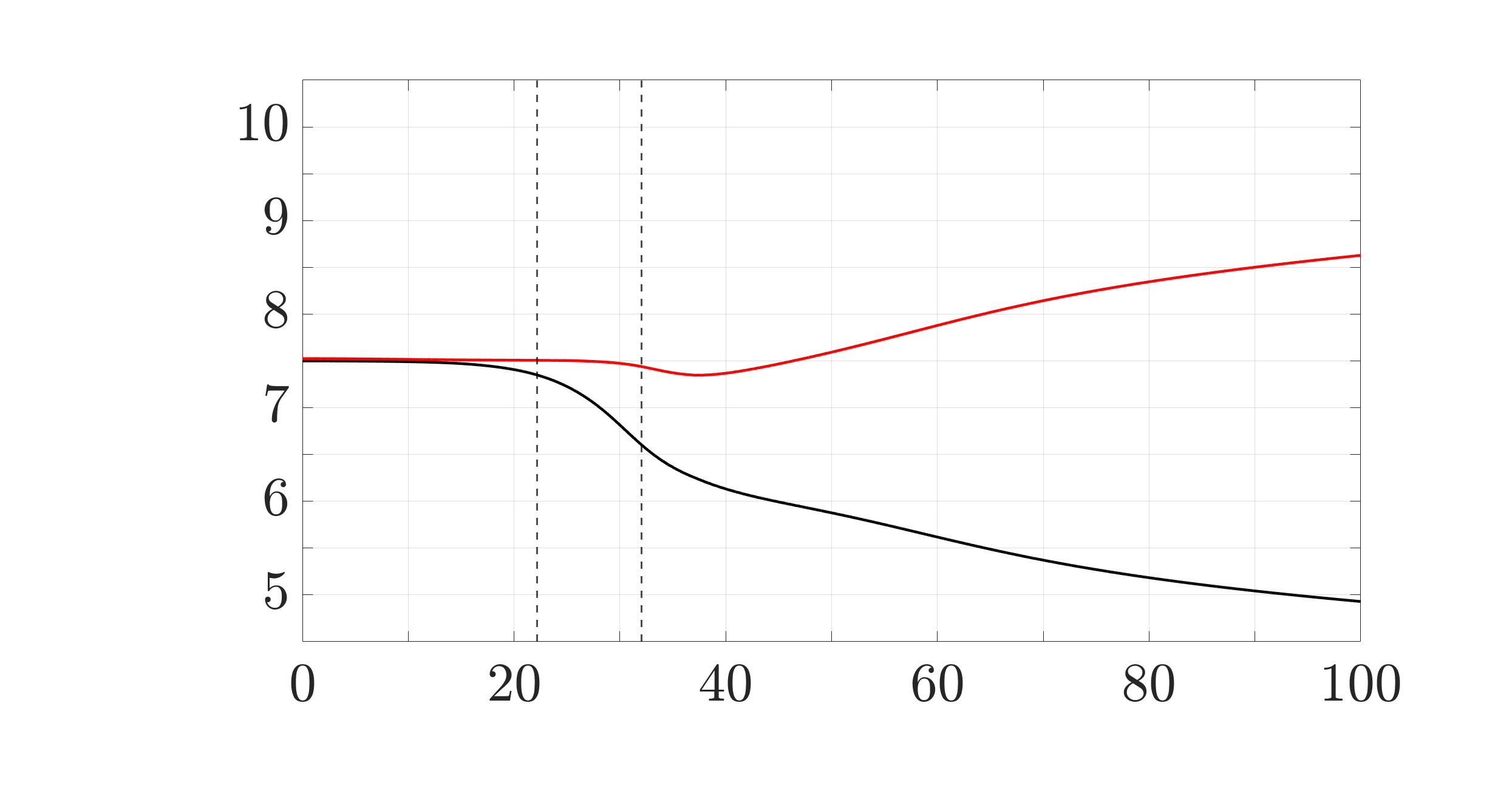}}{0cm}{-0.15cm}    
    \end{subfigure}
    
    \begin{subfigure}[b]{\textwidth}
         \topinset{{\footnotesize (g)}}{\includegraphics[trim=7cm 1.1cm 4cm 2.9cm, clip,width=0.49\textwidth]{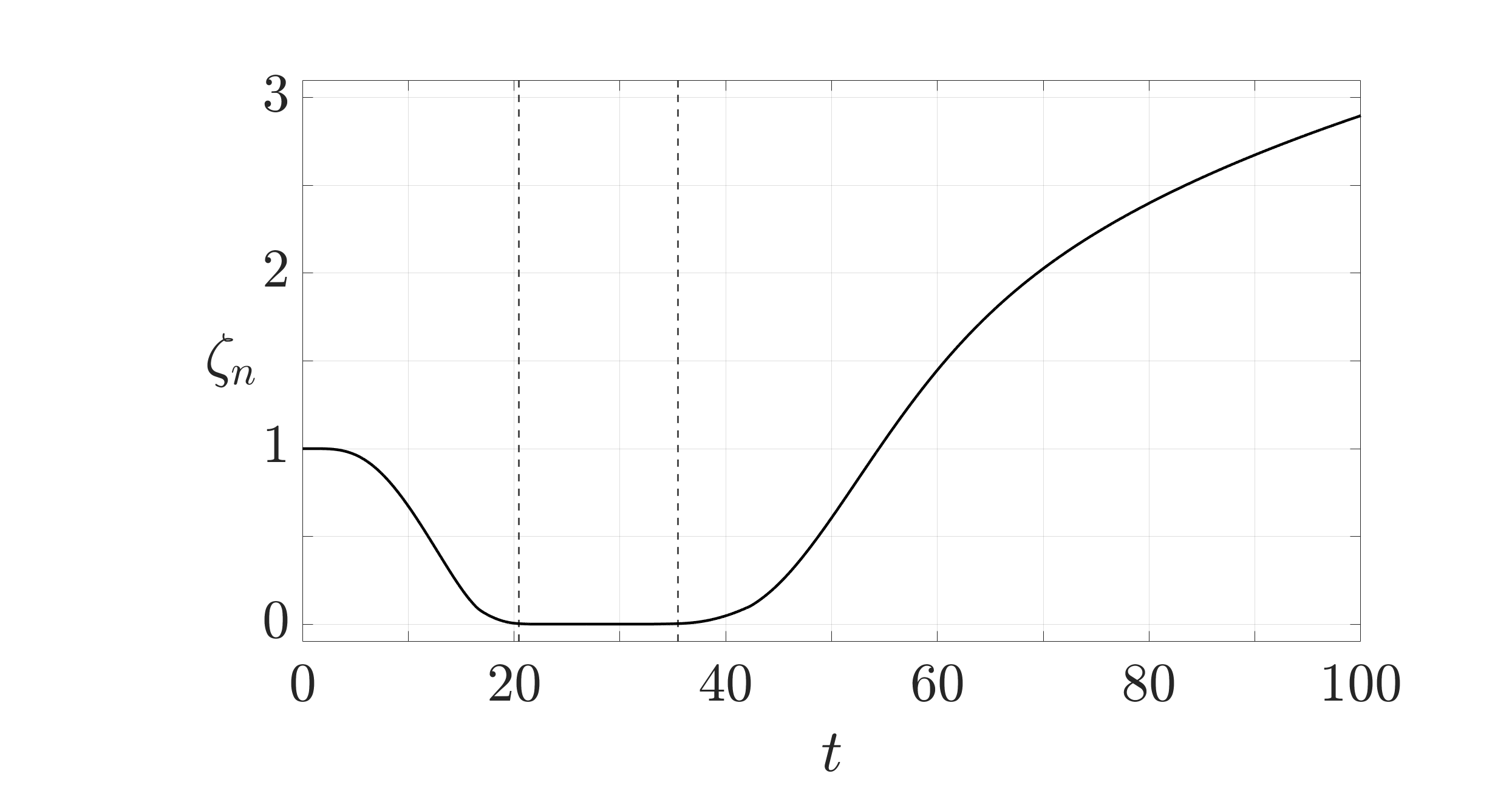}}{0cm}{-0.15cm}
         \topinset{{\footnotesize (h)}}{\includegraphics[trim=7cm 1.1cm 4cm 2.9cm, clip,width=0.49\textwidth]{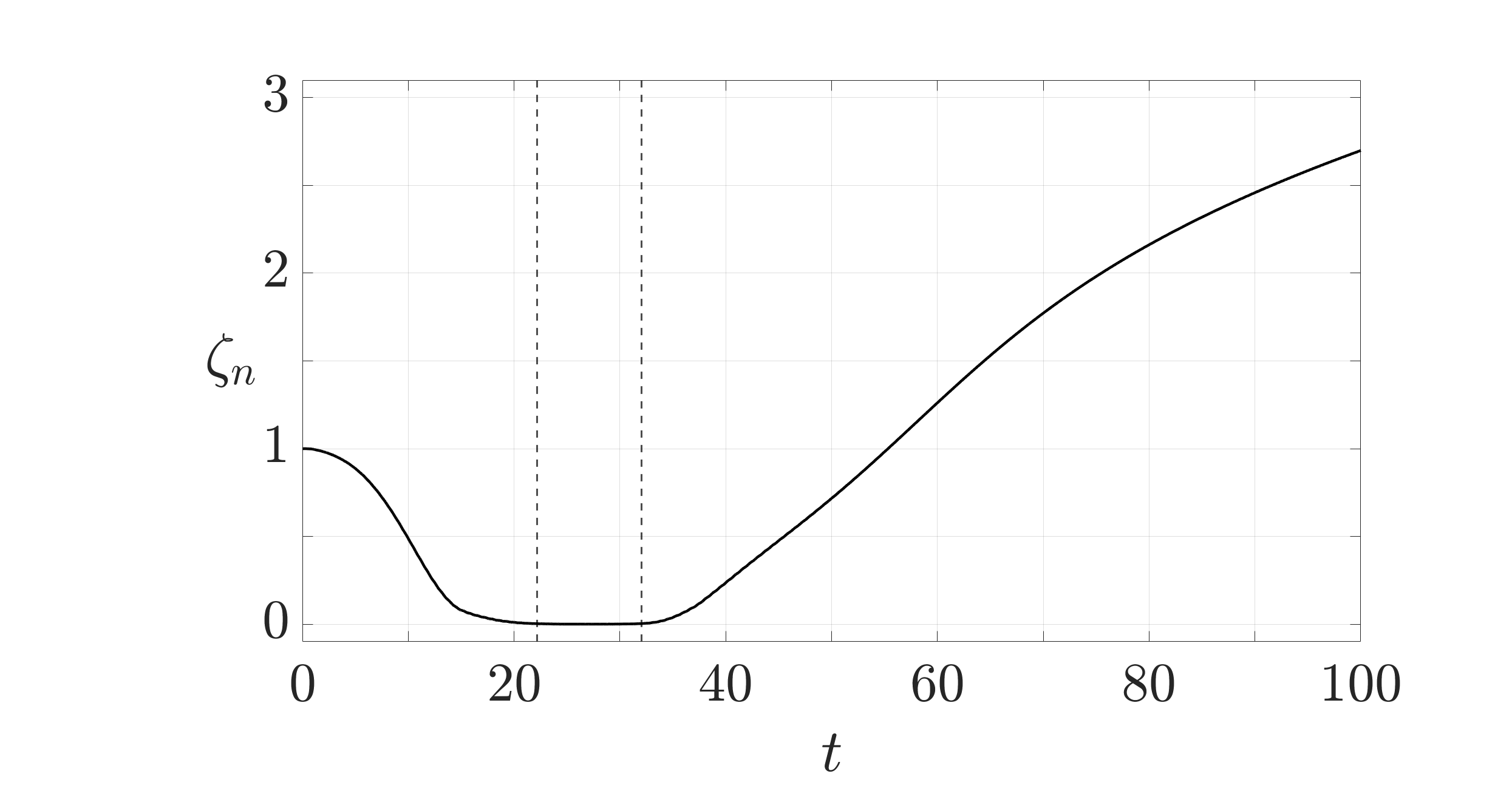}}{0cm}{-0.15cm}

    \end{subfigure}
    \caption{\label{fig:Trajectories_IndividualComparison}%
    Comparison of the individual trajectories and the total distance between the non-oscillating reference case (left: a, c, e, g) and the oscillating setup $A_f=1.0$, $S=1.12$ (right: b, d, f, h). 
    Trajectories along the $x$-, $y$-, and $z$-axes are shown in (a,b), (c,d), and (e,f), respectively.
    The leading particle $P_1$ is presented in black and the trailing particle $P_2$ in red.
    The total distances $\zeta_n$ are shown in (g) and (h).
    The vertical dashed lines indicate the kissing phase.}
\end{figure}

When the contributions of the individual components are combined to obtain the total distance $\zeta_n$ (Figs.~\ref{fig:Trajectories_IndividualComparison}g,h), this quantity decreases from initially $\zeta_n = 1$ after a brief start-up period.
Prior to the kissing phase, the decrease in $\zeta_n$ exhibits a temporary attenuation.
The oscillatory forcing appears to prolong this attenuation period, as Fig.~\ref{fig:Trajectories_IndividualComparison}(h) indicates an extended dampening time before the onset of kissing.
During the kissing phase, $\zeta_n$ remains approximately zero.
The duration of the kissing phase amounts to $15 t$ in the reference case and $9.9 t$ in the oscillating setup, corresponding to a reduction of $34 \%$.
After the end of the kissing phase, the particles tumble apart and $\zeta_n$ increases.

\subsubsection{Settling velocities}\label{sec:settling_velocities}

The representations of the particle trajectories along the settling direction in Figs.~\ref{fig:Trajectories_Settling} and \ref{fig:Trajectories_IndividualComparison}(a,b) exhibit noticeable differences in the total distance traveled over the considered time interval. This variation indicates that the imposed oscillations influence the vertical transport of the particles and, consequently, their effective settling velocity.
To quantify this effect, we analyze the temporal evolution of the settling velocities of both particles. 
We present the corresponding results for the non-oscillating reference case in Fig.~\ref{fig:Settling_Velocities}(a) and for the oscillating configuration in Fig.~\ref{fig:Settling_Velocities}(b), enabling a direct comparison of the oscillation-induced modifications.

In both configurations, the magnitudes of the settling velocities of the two particles initially increase at an approximately identical rate, reflecting a comparable acceleration.
As the particles approach one another, however, the velocity of $P_2$ begins to exceed that of $P_1$. 
This divergence is due to hydrodynamic interactions, in particular drafting effects, whereby the trailing particle $P_2$ experiences reduced fluid resistance within the wake of the leading particle $P_1$ and thus attains a higher settling velocity.
Shortly before contact, the velocities of both particles approach a plateau and remain nearly constant. 
As the system transitions toward the tumbling phase, the velocity of $P_1$ decreases first, followed by that of $P_2$. 
This sequential deceleration is associated with the reconfiguration of the particle pair and the breakdown of the drafting–kissing arrangement. 
During tumbling, both particles attain approximately constant settling velocities.

\begin{figure}[h!]
    \def\stackalignment{l}
    \centering
    \captionsetup[subfigure]{labelformat=empty}
    \begin{subfigure}[b]{\textwidth}
         \centering 
         \topinset{{\footnotesize (a)}}{\includegraphics[trim=6.5cm 1.0cm 4cm 2.2cm, clip,width=0.49\textwidth]{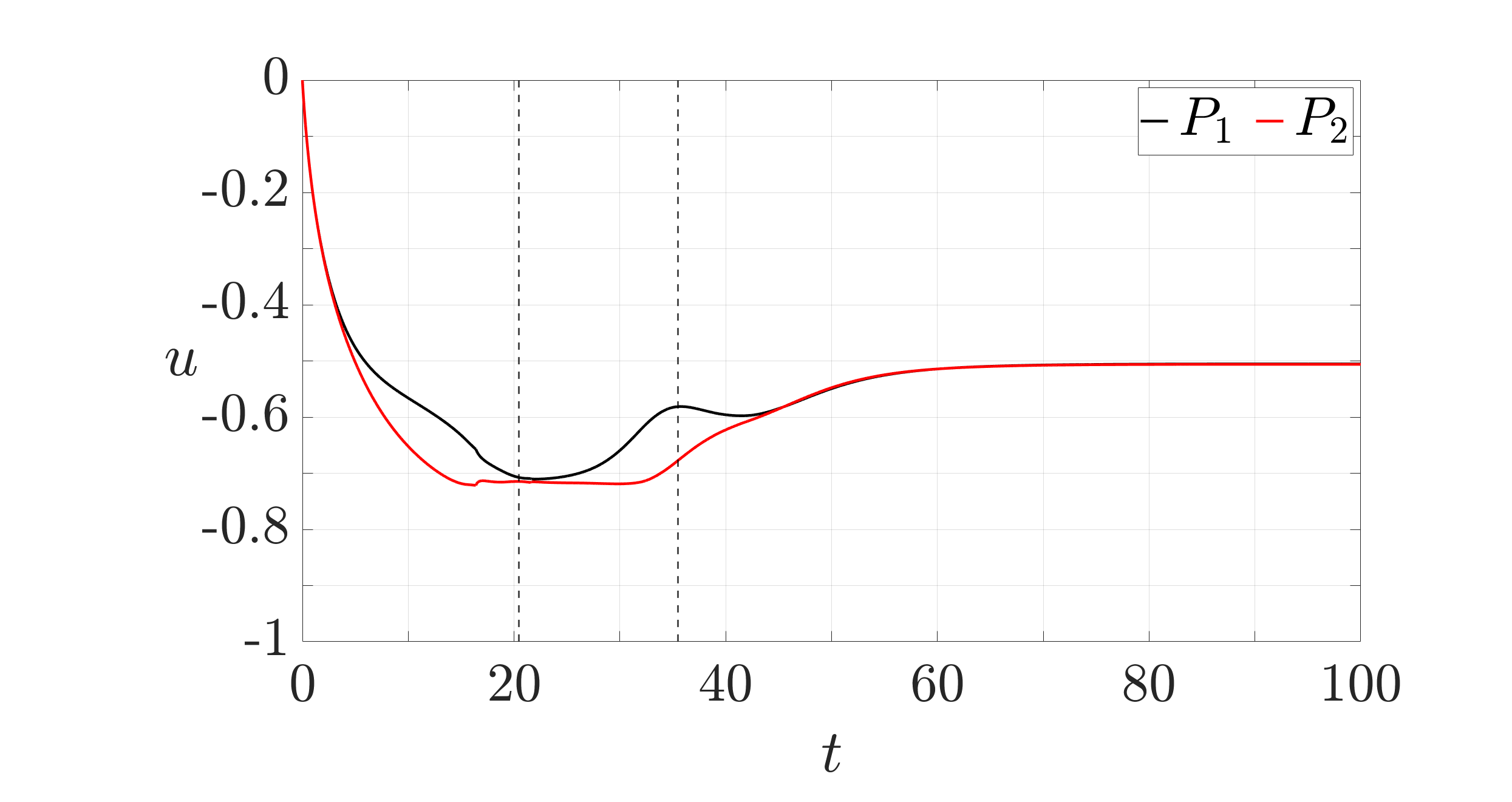}}{0cm}{-0.15cm}
         \topinset{{\footnotesize (b)}}{\includegraphics[trim=6.5cm 1.0cm 4cm 2.2cm, clip,width=0.49\textwidth]{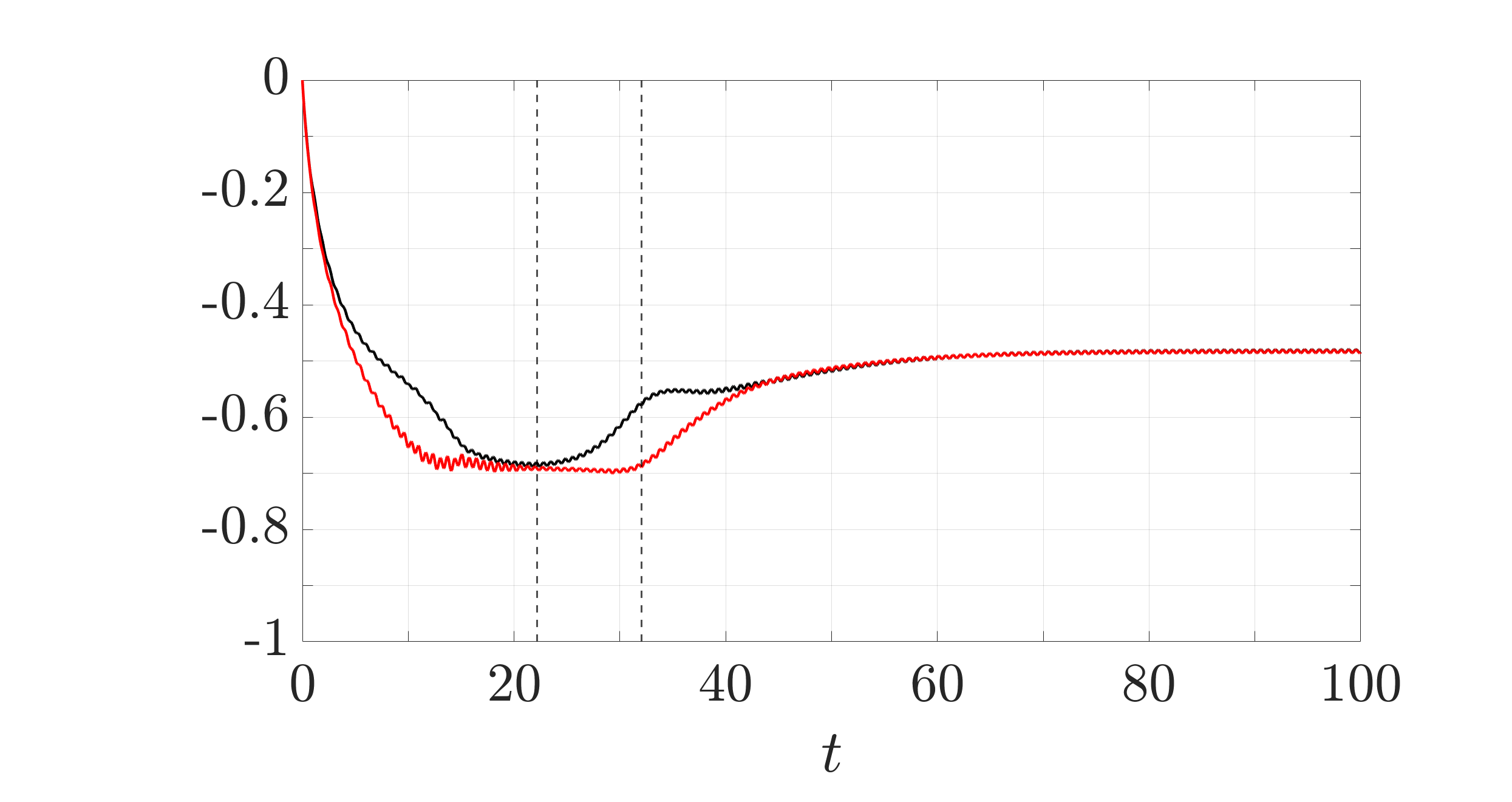}}{0cm}{-0.15cm}
    \end{subfigure}
    \caption{\label{fig:Settling_Velocities}%
    Comparison of the settling velocities between (a) the non-oscillating reference case and (b) the oscillating setup $A_f=1.0$, $S=1.12$. 
    $P_1$ is presented in black and $P_2$ in red.
    The vertical dashed lines indicate the kissing phase.}
\end{figure}

The comparison between the non-oscillating and oscillating configurations reveals three key observations. 
First, the imposed horizontal oscillations influence the vertical settling motion, leading to temporally oscillatory (jittering) settling velocity profiles for both particles. 
Second, the presence of oscillations slightly reduces the magnitude of the peak settling velocity during kissing. 
Third, the settling velocity following particle contact is likewise diminished under oscillatory conditions.
In particular, the latter two effects provide a direct explanation for the reduced settling distances.

\subsection{Impact of oscillation parameters}\label{sec:ImpactOscParameters}

The results presented up to this point demonstrate the potential impact of external oscillations on the particle-particle interaction.
To systematically assess the dependence of this interaction on the oscillatory conditions, we vary the parameters $A_f$ and $S$, and compare the resulting evolution of $\zeta_n$ as well as the individual particle trajectories to those of the non-oscillating reference case. 
In addition, we examine the effect of varying oscillation parameters on the evolution of the pressure field in the vicinity of the particles.

\subsubsection{Interparticle distance}

In order to examine the impact of the oscillation parameters on $\zeta_n$, we confine the analysis to the drafting and kissing phases. 
In particular, we assess the extent to which the external oscillations prolong or reduce the duration of contact. 
In Fig.~\ref{fig:TotalDistance_Comparison}, we compare $\zeta_n$ of the non-oscillating reference setup (black solid line) with multiple exemplary oscillation setups (colored dashed lines).
To this end, we choose three exemplary amplitudes, where Fig.~\ref{fig:TotalDistance_Comparison}(a) depicts $A_f = 0.1$, (b)~$A_f = 0.5$, and (c)~$A_f = 1.0$.
For each amplitude, we select eight representative oscillation frequencies to span the entire range of frequencies considered, namely $S=\{0.21, 0.35, 0.49, 0.70, 0.84, 1.05, 1.19, 1.40\}$.
In terms of particle Reynolds number, we obtain $Re_p = \{0.12, 0.27, 0.44, 0.73, 0.93, 1.24, 1.45, 1.78\}$ for Fig.~\ref{fig:TotalDistance_Comparison}(a), $Re_p = \{0.59, 1.34, 2.21,\allowbreak3.64, 4.65, 6.21, 7.27, 8.90\}$ for Fig.~\ref{fig:TotalDistance_Comparison}(b), and $Re_p = \{1.17, 2.68, 4.43, 7.29, 9.30,\allowbreak12.41,\allowbreak14.55, 17.80\}$ for Fig.~\ref{fig:TotalDistance_Comparison}(c).

The illustrations in Fig.~\ref{fig:TotalDistance_Comparison} demonstrate that the deviations in $\zeta_n$ induced by the oscillations become progressively more pronounced with increasing $A_f$. 
For the smallest amplitude, $A_f=0.1$ (Fig.~\ref{fig:TotalDistance_Comparison}a), the particle trajectories closely follow the non-oscillating reference case, with only minor differences emerging after the kissing phase. 
At intermediate amplitude, $A_f=0.5$ (Fig.~\ref{fig:TotalDistance_Comparison}b), deviations are already evident.
For $S \geq 1.05$, the drafting phase exhibits steeper gradients, resulting in an earlier onset of kissing. 
In contrast, smaller values of $S$ produce nearly identical results for $\zeta_n$ during drafting and similar times for the onset of kissing compared to the reference case. 
The end of the kissing phase also depends on $S$, which occurs earlier for $S \geq 1.19$ but later for $S \leq 0.70$.
At the highest amplitude, $A_f=1.0$ (Fig.~\ref{fig:TotalDistance_Comparison}c), these modifications are considerably amplified. 
Steeper gradients of $\zeta_n$ during the drafting phase are observed for $S \geq 0.84$, although only the case $S=1.40$ leads to an earlier entry into the kissing phase relative to the reference. 
For intermediate values of $S$ ($0.84$, $1.05$, and $1.19$), the approach of the particles is strongly damped immediately before contact, delaying the onset of kissing despite the initially accelerated approach of the particles. 
The time that marks the end of the kissing phase also exhibits a systematic dependence.
For $S \geq 0.70$, the time at which the kissing phase ends is drastically reduced, while for the lowest frequencies ($S=0.21$ and $0.35$) it ends slightly later than in the reference case.

\begin{figure}
    \def\stackalignment{l}
    \centering
    \captionsetup[subfigure]{labelformat=empty}
    \begin{subfigure}[b]{\textwidth}
         \centering 
         \topinset{{\footnotesize (a)}}{\includegraphics[trim=3.5cm 0.3cm 4.7cm 2.3cm, clip,width=0.75\textwidth]{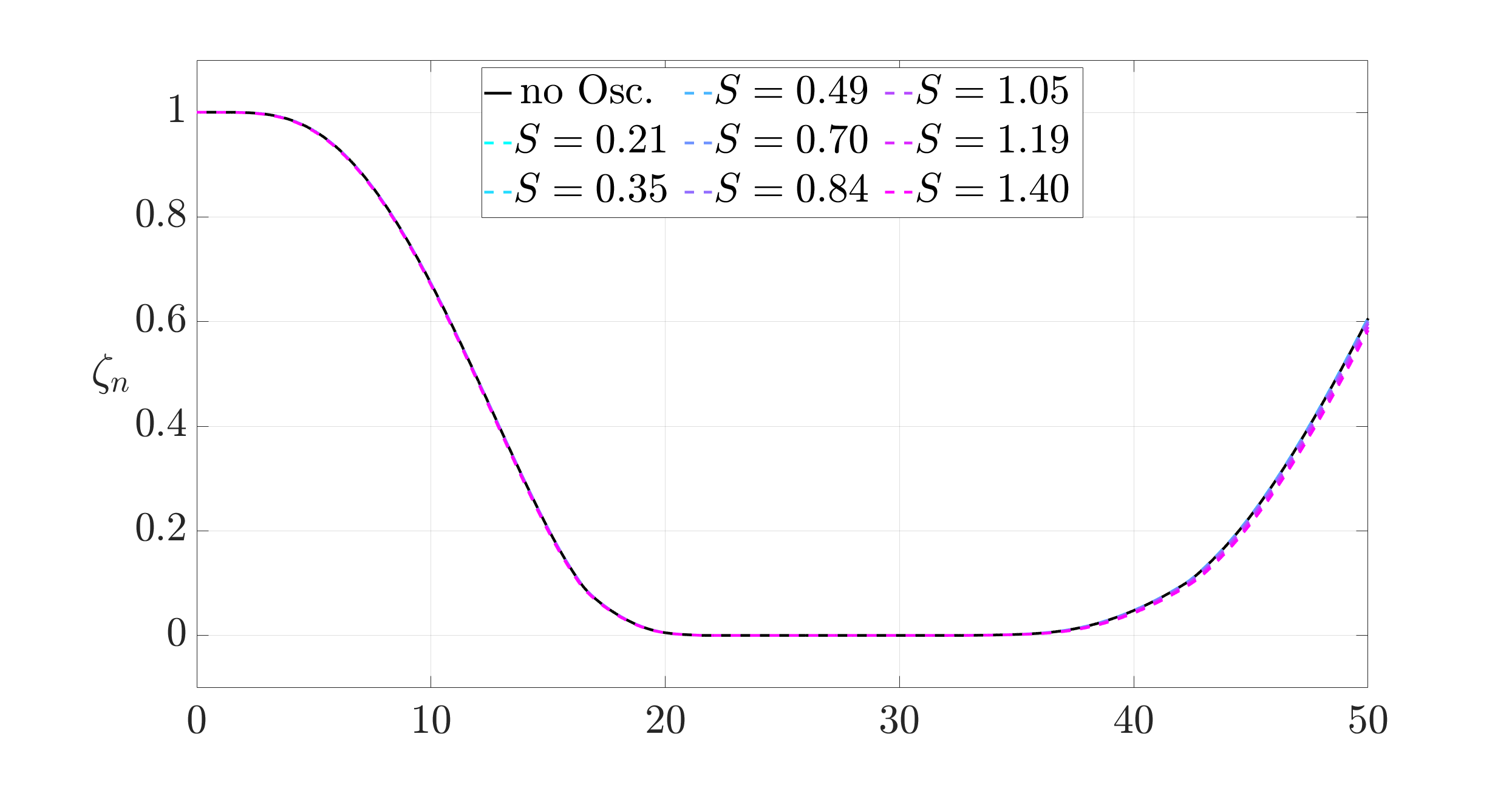}}{0cm}{0cm}
         \topinset{{\footnotesize (b)}}{\includegraphics[trim=3.5cm 0.3cm 4.7cm 2.3cm, clip,width=0.75\textwidth]{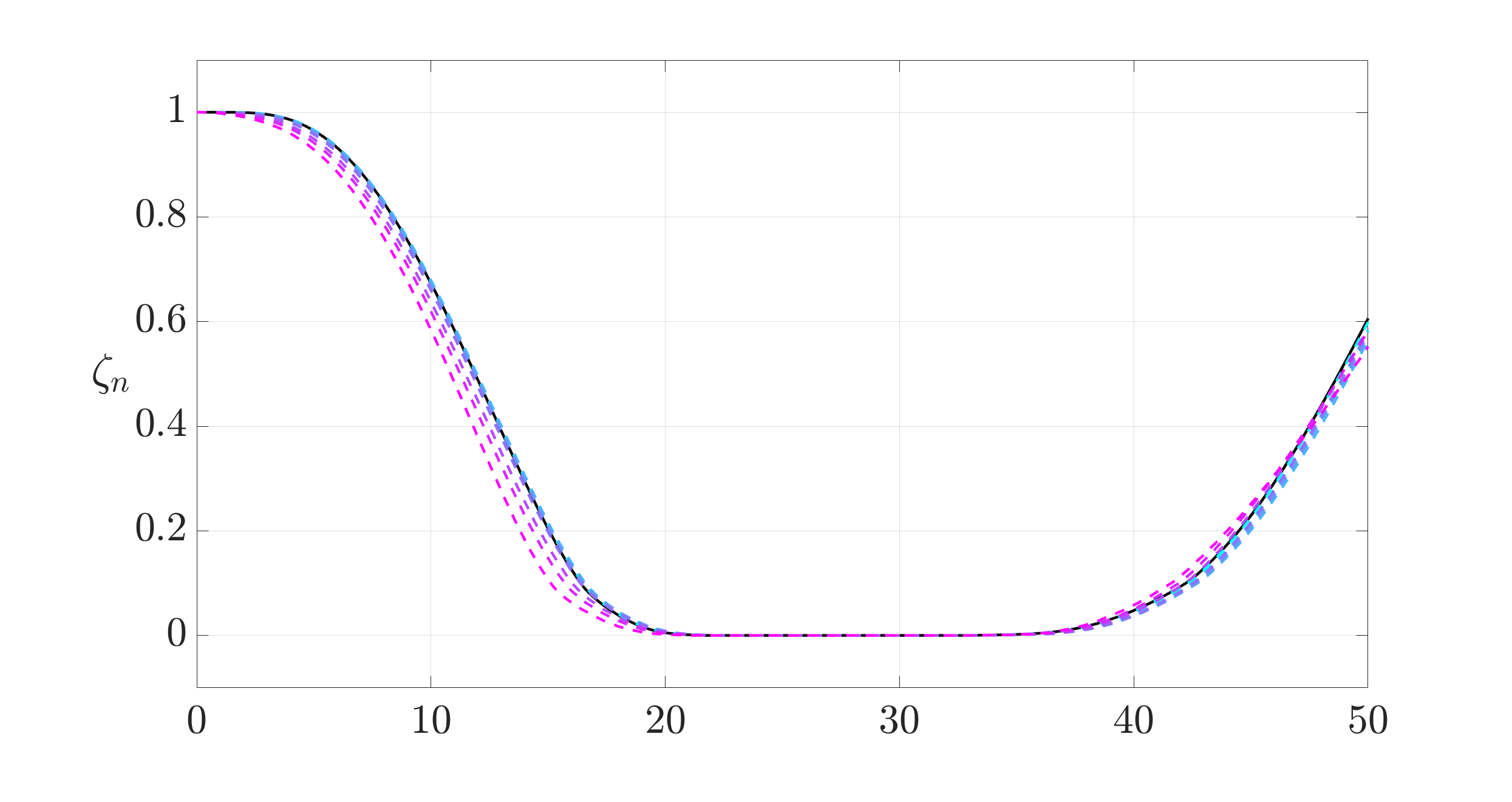}}{0cm}{0cm}
         \topinset{{\footnotesize (c)}}{\includegraphics[trim=3.5cm 0.3cm 4.7cm 2.3cm, clip,width=0.75\textwidth]{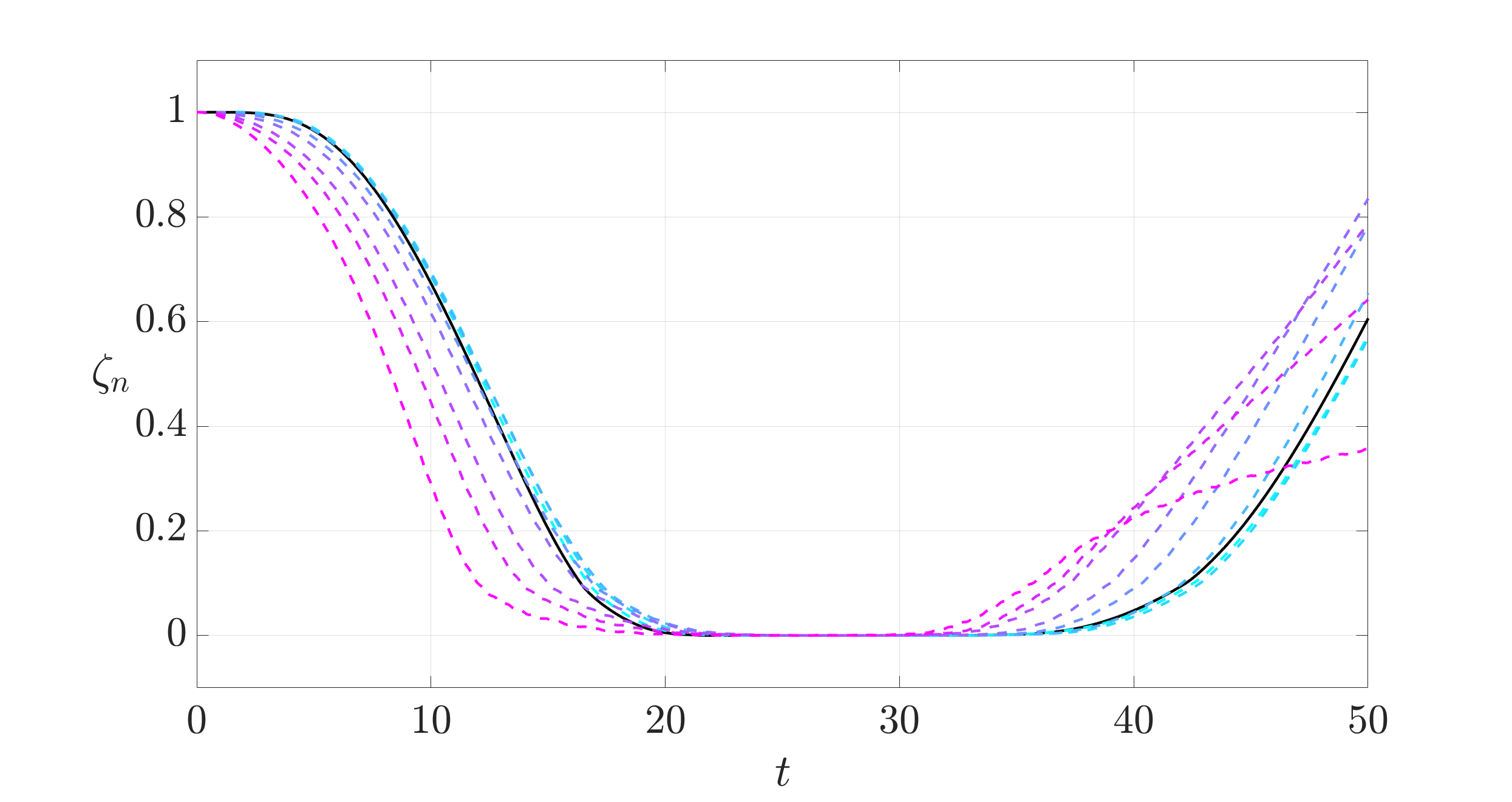}}{0cm}{0cm}
    \end{subfigure}
    \caption{\label{fig:TotalDistance_Comparison}%
    Comparison of surface-to-surface distances $\zeta_n$ between non-oscillating case (solid black line) and oscillating setups with $S=\{0.21, 0.35, 0.49, 0.70, 0.84, 1.05, 1.19, 1.40\}$ (dashed colored lines) for different amplitudes: (a)~$A_f = 0.1$, (b)~$0.5$, and (c)~$1.0$.
    }
\end{figure}

The evolution of $\zeta_n$ in Fig.~\ref{fig:TotalDistance_Comparison} reveals that the imposed oscillations modify the approach and subsequent interaction of the two particles. 
Increasing $S$ and $A_f$ leads to progressively larger deviations from the non-oscillating reference case, with the approach dynamics being modified prior to contact and the effects becoming particularly pronounced once the particles enter the kissing phase. 
While the modified drafting dynamics affect the onset of kissing, the subsequent evolution of $\zeta_n$ indicates that the oscillations primarily alter the duration and stability of the kissing phase, including the occurrence of damping effects that can delay or modify particle contact.

\subsubsection{Modification of particle trajectories}\label{sec:mod_particle_trajectories}

The analysis of $\zeta_n$ as a function of time (Fig. \ref{fig:TotalDistance_Comparison}) shows that both the amplitude and frequency of the oscillation influence the dynamics of the interaction between the two particles. 
To gain a deeper understanding of the underlying mechanisms, we examine the impact of $Re_p$ on the individual particle trajectories.

We focus on three different particle Reynolds numbers by varying $A_f$ and $S$, which are $Re_p = 0.59$ ($A_f = 0.5$, $S=0.21$), $Re_p = 4.67$ ($A_f = 0.5$, $S=0.84$) and $Re_p = 9.35$ ($A_f = 1.0$, $S=0.84$).
To this end, we analyze the individual trajectories along all coordinate directions in Fig.~\ref{fig:Trajectories_EffectLambda}.
We present the different setups by the respective color scheme, where we distinguish between the two particles with solid lines for the leading particle $P_1$ and dashed lines for the trailing particle $P_2$.
The motion of the particles is represented in the non-inertial frame of reference.

Along the $x$-axis (Fig.~\ref{fig:Trajectories_EffectLambda}a), the motion of $P_2$  exhibits a consistent trend for all $Re_p$ considered with only minor deviations occurring before and after the transient stagnation ($t \approx 17–27$), during which its position remains nearly constant.
Pronounced differences appear along the $y$-axis (Fig.~\ref{fig:Trajectories_EffectLambda}b). 
The oscillatory characteristics of the individual particles are governed by $A_f$ and $S$.
The relative motion in the $y$-direction, however, reveals a dependence on~$Re_p$:
for $Re_p < 1$, the particles progressively deviate from one another, whereas for $Re_p > 1$, their motion becomes synchronized, oscillating about a common lateral position.
In the $z$-direction (Fig.~\ref{fig:Trajectories_EffectLambda}c), the trajectories show that the separation between the two particles increases with increasing $Re_p$.

For $Re_p<1$, where viscous effects dominate over oscillatory inertia, the particles exhibit dynamics comparable to the non-oscillating reference case, progressively separating in both lateral directions, $y$ and $z$. 
In contrast, for $Re_p>1$, oscillatory inertia becomes increasingly important and particle motions start to synchronize along the oscillation direction, while their separation in the transverse $z$-direction grows.

\begin{figure}
    \def\stackalignment{l}
    \centering
    \captionsetup[subfigure]{labelformat=empty}
    \begin{subfigure}[b]{\textwidth}
         \centering 
         \topinset{{\footnotesize (a)}}{\includegraphics[trim=4cm 0.3cm 4.6cm 1.5cm, clip,width=0.65\textwidth]{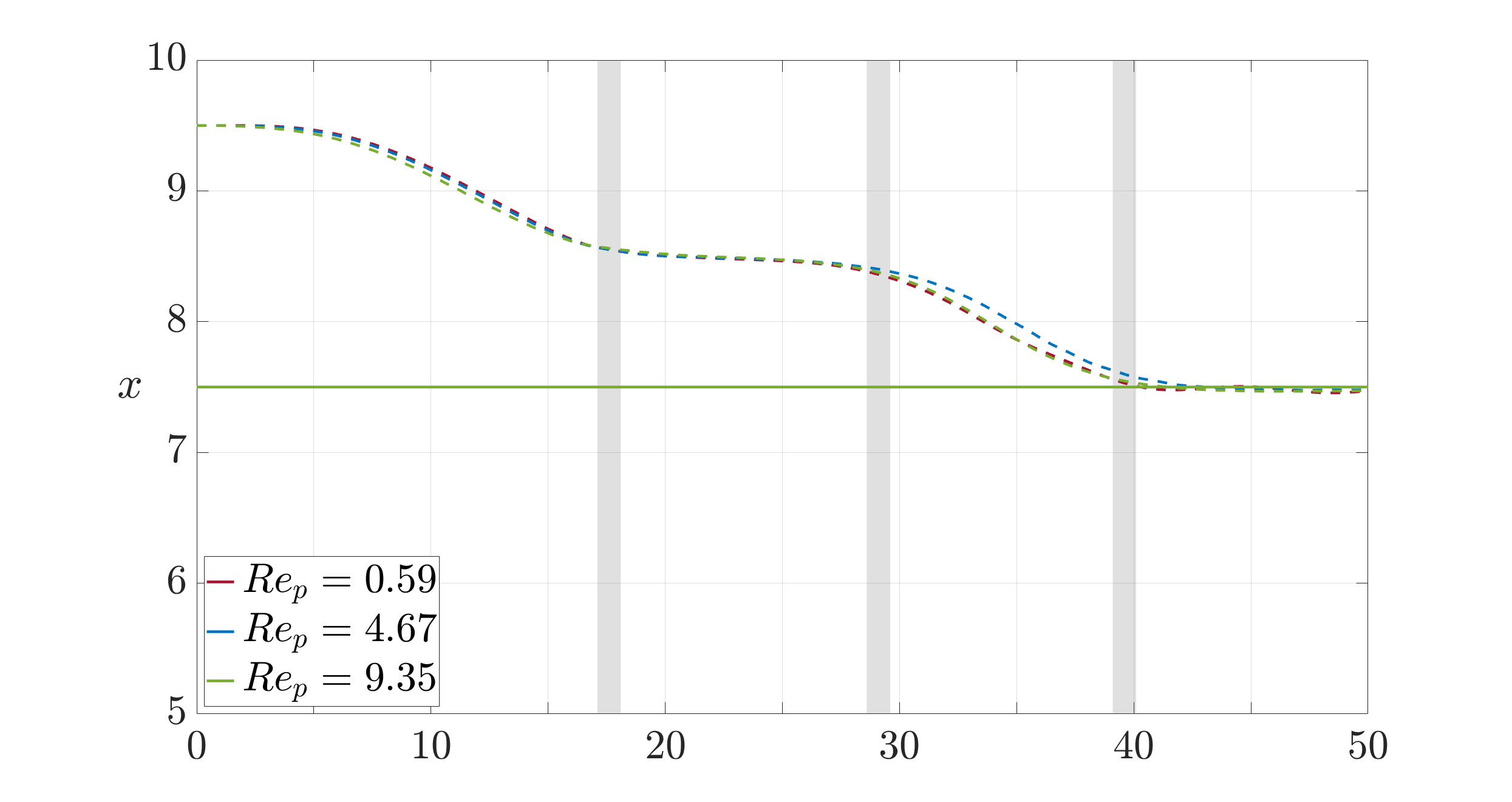}}{0cm}{-0.15cm}
         \put(-210,30){{\footnotesize \RomanNumeralCaps{1}}}
         \put(-144,30){{\footnotesize \RomanNumeralCaps{2}}}
         \put(-84,30){{\footnotesize \RomanNumeralCaps{3}}}
    \end{subfigure}
    \begin{subfigure}[b]{\textwidth}
         \centering
         \topinset{{\footnotesize (b)}}{\includegraphics[trim=4cm 0.3cm 4.6cm 1.5cm, clip,width=0.65\textwidth]{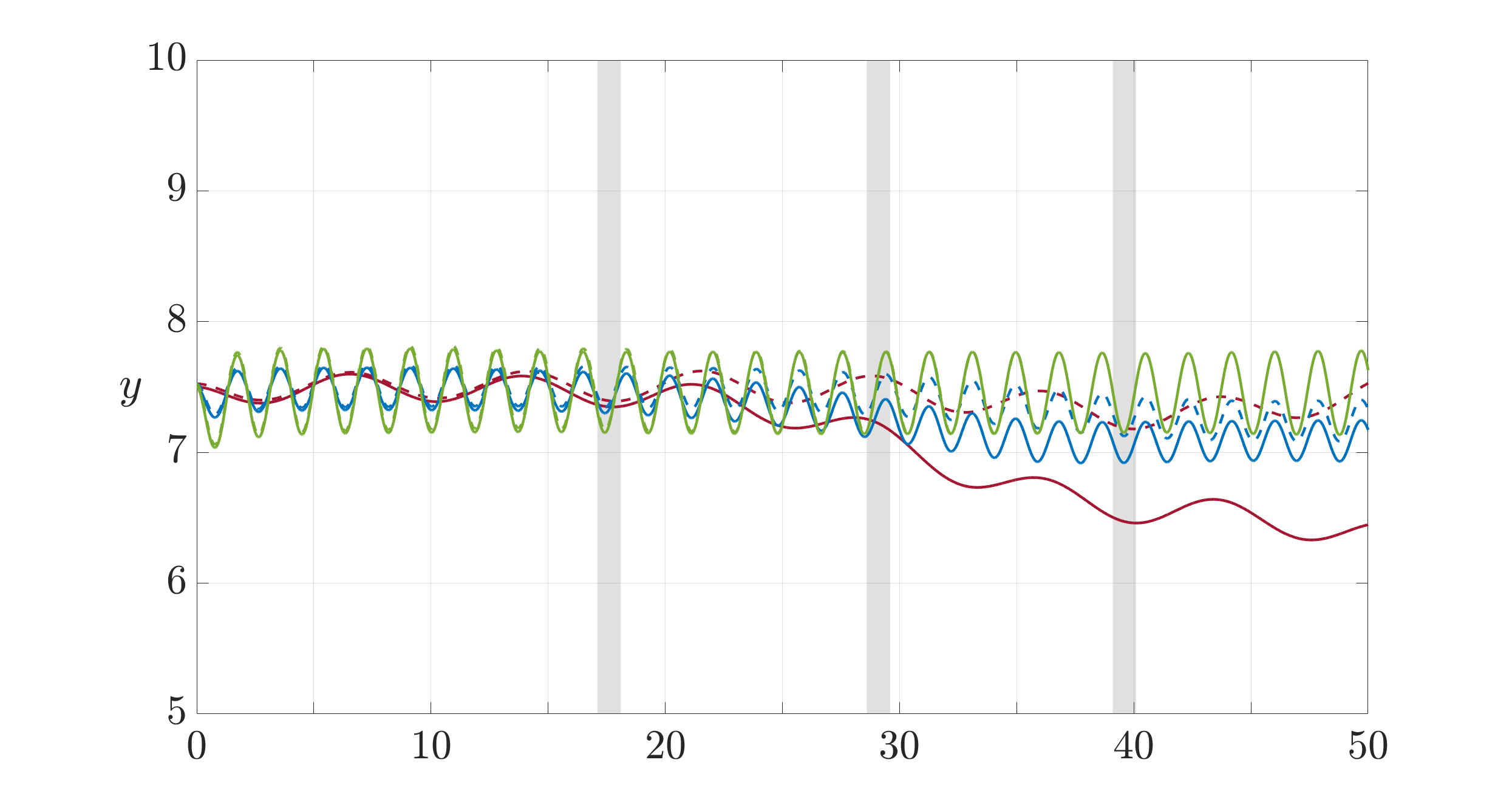}}{0cm}{-0.15cm}
    \end{subfigure}
    \begin{subfigure}[b]{\textwidth}
         \centering
         \topinset{{\footnotesize (c)}}{\includegraphics[trim=4cm 0.3cm 4.6cm 1.5cm, clip,width=0.65\textwidth]{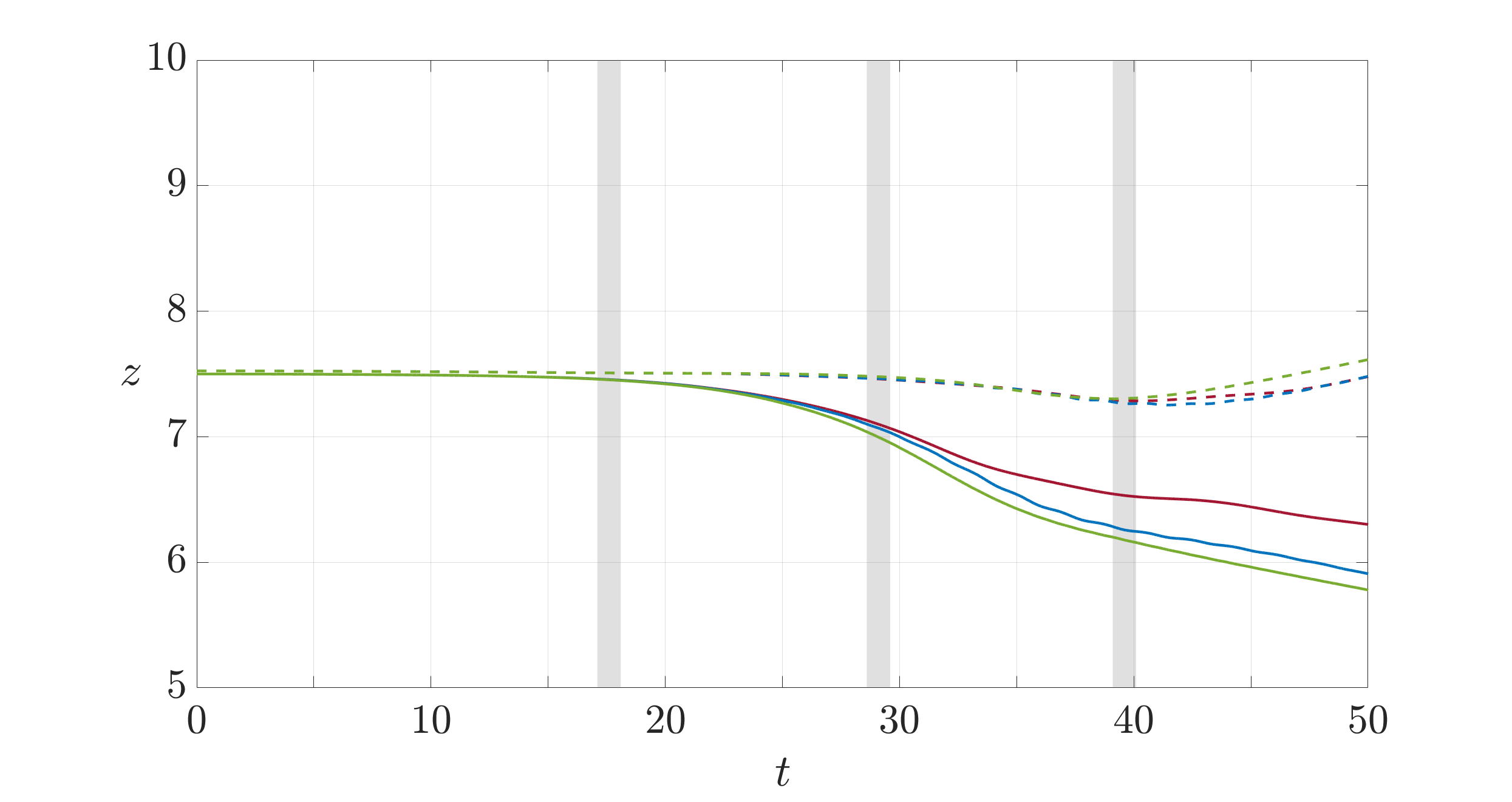}}{0cm}{-0.15cm}
    \end{subfigure}
    \caption{\label{fig:Trajectories_EffectLambda}%
    Analysis of the effect of $Re_p$ on the particle-particle interaction along the (a) $x$-, (b) $y$-, and (c) $z$-direction.
    Solid lines are applied for the leading particle $P_1$ and dashed lines for the trailing particle $P_2$.
    Vertical gray bars labeled by the Roman numerals indicate the time instants that correspond to the visualizations in Fig.~\ref{fig:Visualization_VaryingLambda}.}
\end{figure}

\subsection{Particle Interaction}\label{sec:InteractionPhases} 

The evolution of the inter-particle distance $\zeta_n$ indicates that the imposed oscillations can substantially modify the particle--particle interaction once the particles approach each other. 
To quantify these effects and elucidate their physical origin, we characterize the duration of the kissing phase and examine the oscillation-induced pressure distribution, first for an isolated particle to identify the underlying flow features and then for the two-particle configuration to assess their relevance to the particle--particle interaction.
We note that we also examined the influence of the oscillations on the length of the drafting phase, but found only minor deviations of less than $15\%$ compared to the non-oscillating reference case. 
In addition, these variances might be partly influenced by the choice of initial conditions. Nevertheless, the initial conditions were chosen such that a well developed state was obtained at the time when particles enter the kissing-phase. 
As a consequence, we solely focus on these interactions in the following sections.

\subsubsection{Kissing time} \label{sec:TimeRatio_KissingPhase}

To quantify the impact of the oscillations on the kissing phase, we compare the kissing duration of each oscillating setup with that of the corresponding non-oscillating reference case.
In this regard, we compute the time ratio $\tau_k = t_{k,A_f,S} / t_{k,0}$, where $t_{k,A_f,S}$ denotes the duration of the kissing phase for an oscillating setup with amplitude $A_f$ and frequency $S$, while $t_{k,0}$ denotes the corresponding duration in the non-oscillating reference case.
Thus, $\tau_k>1$ indicates an extended kissing phase, whereas $\tau_k<1$ denotes a reduced kissing duration relative to the reference case.

As described in \S \ref{sec:OscillationSetups}, we restrict the investigated parameter space to $Re_p = [0.01, 17.79]$ and $S = [0.14, 1.4]$. 
This selection ensures that the results remain independent of spatial and temporal resolution, while effectively eliminating the influence of confinement and wall-induced effects.

In Fig.~\ref{fig:timeRatio_kissing}, we present $\tau_k$ in dependence of $Re_p$ with the reference values $\tau_k = 1$ and $Re_{p} = 1$ indicated by the horizontal and vertical dashed lines, respectively.
The results reveal distinct trends in $\tau_k$ with increasing $Re_p$ for the different oscillation amplitudes $A_f$.
For $Re_{p} < 1$, the oscillation amplitude $A_f$ generally has little effect, yielding $\tau_k \approx 1$.
Beyond the threshold of $Re_{p} = 1$, clear deviations from the non-oscillating reference emerge across all oscillating configurations. 
For $A_f = 0.1$, $0.25$, and $0.5$, the oscillations tend to extend the kissing phase, with the most pronounced effect at $A_f = 0.25$.
In contrast, at larger amplitudes ($A_f = 0.75$ and $1.0$), the oscillations lead to a reduction of the duration of the kissing phase. 
Both cases exhibit a similar trend, with $\tau_k$ decreasing until a minimum is reached (which occurs for both cases at $S = 1.12$), followed by a slight increase while remaining below unity. 
The stronger amplitude $A_f = 1.0$ produces a more pronounced reduction. 
Consequently, we find that moderate amplitudes can prolong the duration of the kissing phase, whereas stronger amplitudes destabilize the configuration and shorten the interaction, potentially altering the subsequent onset and characteristics of the tumbling phase.

\begin{figure}[h]
    \centering  
    \includegraphics[width=0.85\linewidth]{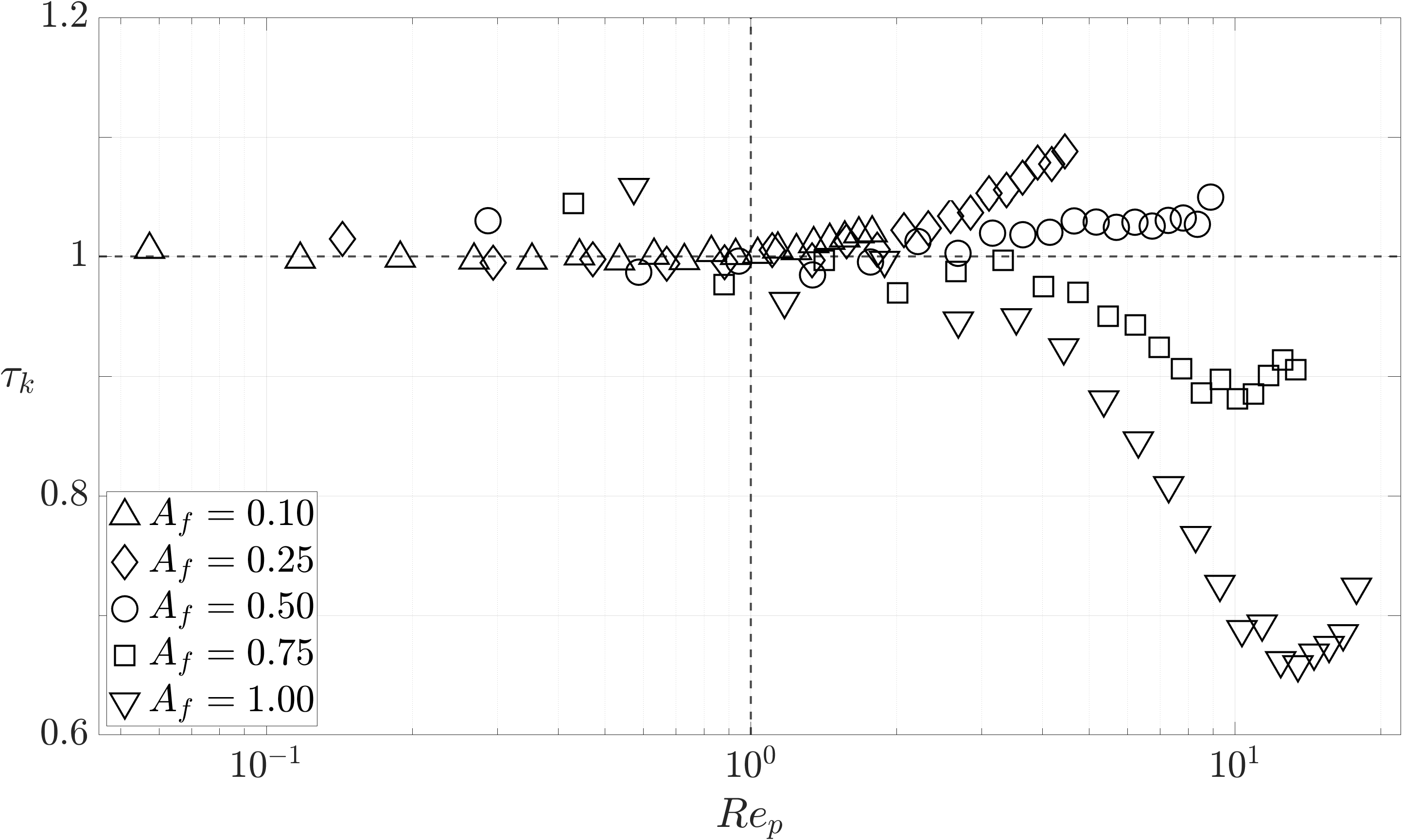}
    \caption{\label{fig:timeRatio_kissing}%
    Time ratio of the kissing phase $\tau_k$ over $Re_{p}$.
    The different symbols indicate the varying amplitudes $A_f$.
    $\tau_k = 1$ is highlighted by the horizontal dashed line and $Re_{p} = 1$ by the vertical dashed line.}
\end{figure}

Our analysis therefore shows that oscillatory effects become significant only for $Re_p > 1$, i.e., when the inertial contributions associated with the oscillations exceed the viscous effects. 
Below this threshold, the dynamics remain largely unaffected by the imposed oscillations. 

\subsubsection{Isolated-particle pressure field}

The observed dependence of the kissing duration on the oscillatory forcing motivates an examination of the underlying flow modification responsible for these changes. 
To isolate the effect of the imposed lateral oscillation, we consider a single settling particle and analyze its surrounding pressure field. 
This single-particle configuration removes the additional complexity associated with particle--particle interactions, thereby allowing us to identify the oscillation-induced pressure structures that may contribute to the observed changes in the kissing dynamics. 
To this end, we employ the same numerical domain described in \S\ref{sec:DKT_setup}, with the sole modification of removing the trailing particle $P_2$. 
The pressure field is evaluated only after the settling velocity has reached a quasi-steady state ($t>20$). 
Furthermore, we integrate the instantaneous pressure fields over one oscillation period, analogous to common approaches in the analysis of steady streaming \cite{2024_Kleischmann_etal,2025_Kleischmann_etal}, to suppress the instantaneous oscillatory fluctuations and isolate the underlying mean pressure distribution.

In Fig.~\ref{fig:PressureFieldSingleParticle}, we visualize the pressure contours in a top-down view.
Fig.~\ref{fig:PressureFieldSingleParticle}a presents $Re_p = 0.59$ ($A_f=0.5$, $S=0.21$) and Fig.~\ref{fig:PressureFieldSingleParticle}b $Re_p = 8.90$ ($A_f=0.5$, $S=1.40$).
The pressure contours range from $-0.1$ (blue) to $0.1$ (red). 
The double-headed arrow indicates the oscillation direction with velocity~$u_f$.
The magnified inset focuses on the near-surface region, highlighting the repulsive high-pressure region indicated in red. 
The extent of this lateral repulsive region is defined as the distance from the particle surface to the transition from repulsive to attractive pressure and is denoted by $\lambda$.
Although determined from the isolated-particle configuration, $\lambda$ captures the essential characteristics of the pressure field and provides a simplified metric for interpreting the two-particle dynamics.

\begin{figure}[h!]
    \centering
    \includegraphics[trim=0.3cm 22cm 0cm 1cm, clip, width=\linewidth]{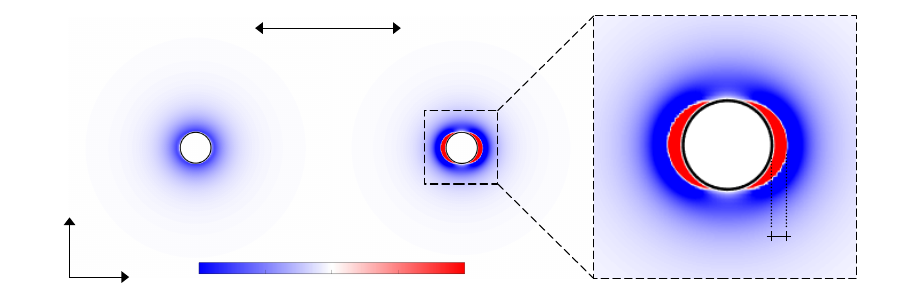}
    \put(-446,15){{$y$}}
    \put(-468,33){{$z$}}
    \put(-328,22){{$p$}}
    \put(-328.5,1){{$0$}}
    \put(-260,1){{$0.1$}}
    \put(-408,1){{$-0.1$}}
    \put(-488,115){{(a)}}
    \put(-320,115){{(b)}}
    \put(-328,155){{$u_f$}}
    \put(-78,20){{$\lambda$}}
    \caption{\label{fig:PressureFieldSingleParticle}%
    Top view of the pressure field surrounding a single particle. 
    The individual cases represent the equivalents to the two particle setups, where (a) presents $Re_p = 0.59$ and (b)~$Re_p = 8.90$.
    The magnified inset highlights the pressure contours in the immediate vicinity of the particle surface. 
    The double-headed arrow indicates the oscillation direction with velocity $u_f$. 
    $\lambda$ denotes the characteristic length of the repulsive pressure.}
\end{figure}
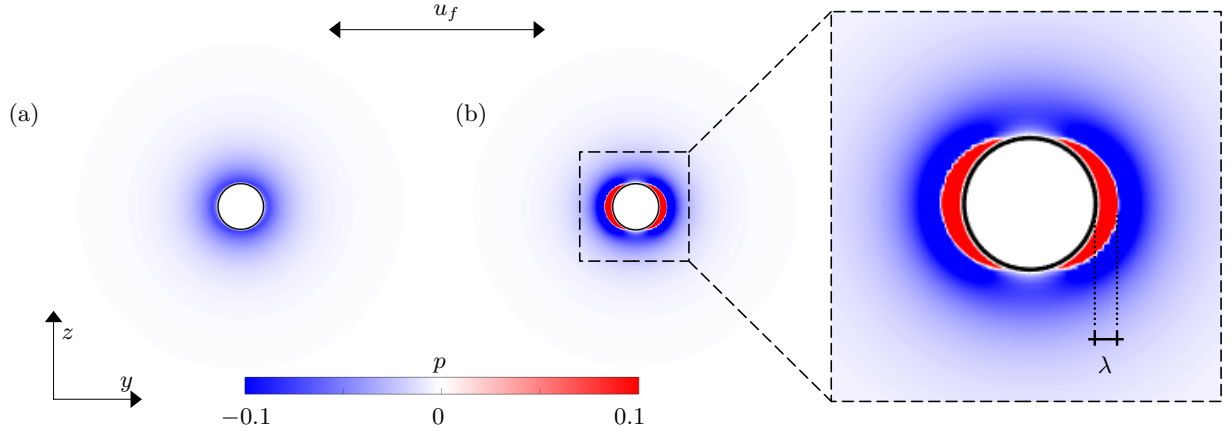

The comparison of the two configurations in Fig.~\ref{fig:PressureFieldSingleParticle} reveals a qualitative change in the pressure distribution. 
For $Re_p=0.59$ (Fig.~\ref{fig:PressureFieldSingleParticle}a), the repulsive pressure region is either absent or negligible, indicating that viscous diffusion suppresses the development of oscillation-induced pressure gradients. 
As a result, the pressure field remains relatively uniform around the particle and does not exhibit a pronounced lateral extension. 
Consequently, the oscillatory motion is expected to have only a limited influence on the surrounding pressure field. 
In contrast, for $Re_p=8.90$ (Fig.~\ref{fig:PressureFieldSingleParticle}b), a distinct laterally extended region of repulsive pressure develops.
As $Re_p$ increases, inertia due to the oscillation increasingly dominates over viscous diffusion, allowing the momentum introduced by the oscillatory motion to persist over larger distances instead of being rapidly dissipated.

Since the imposed oscillatory motion is directed laterally, these pressure disturbances develop preferentially in that direction, causing the repulsive pressure region to expand in the same direction. 
This anisotropic pressure distribution is evident in the magnified view of the near-particle region, where the repulsive pressure field is concentrated along the lateral sides of the particle surface while remaining negligible in the longitudinal regions. 
In the presence of a neighboring particle, this laterally extended pressure region suggests that the hydrodynamic interaction depends on the relative particle position. 
A particle located within the lateral high-pressure region is therefore expected to experience a different hydrodynamic environment than one located in the longitudinal direction. 
In addition, it suggests that the associated pressure forces tend to displace the particles away from the laterally confined high-pressure region towards the longitudinal regions of lower pressure.
These observations demonstrate the potential that the laterally extended pressure field increasingly modifies the hydrodynamic interaction between neighboring particles in a direction-dependent manner.

We further relate the influence of $\lambda$ on the kissing time by plotting this quantity with respect to $\tau_k$ in Fig.~\ref{fig:singleParticle_PressureWidth}.
We determine $\tau_k$ from the two-body interaction addressed in \S~\ref{sec:TimeRatio_KissingPhase}, with $\tau_k > 1$ denoting an extension and $\tau_k < 1$ a reduction of the kissing time. 
The symbols reflect $A_f$ while the color scheme represents $Re_p$.
For small values of~$\lambda$, the oscillation-induced pressure disturbance remains confined to the immediate vicinity of the particle surface. 
Consequently, neighboring particles are expected to interact only weakly with this modified pressure field prior to contact, resulting in kissing dynamics that remain similar to the non-oscillating reference case.
As $\lambda$ increases, the oscillation-induced pressure disturbance extends farther into the surrounding fluid. 
Beyond $\lambda\approx0.2$, $\tau_k$ decreases below unity, particularly for the cases with larger $Re_p$, suggesting that the increasingly extended oscillation-induced pressure disturbance begins to substantially modify the particle--particle interaction. 
Thus, $\lambda\approx0.2$ marks an approximate transition beyond which the spatial extent of the pressure disturbance becomes relevant to the kissing dynamics.

\begin{figure}[h!]
    \centering
    \includegraphics[trim=0cm 0cm 0cm 0cm, clip,width=0.85\textwidth]{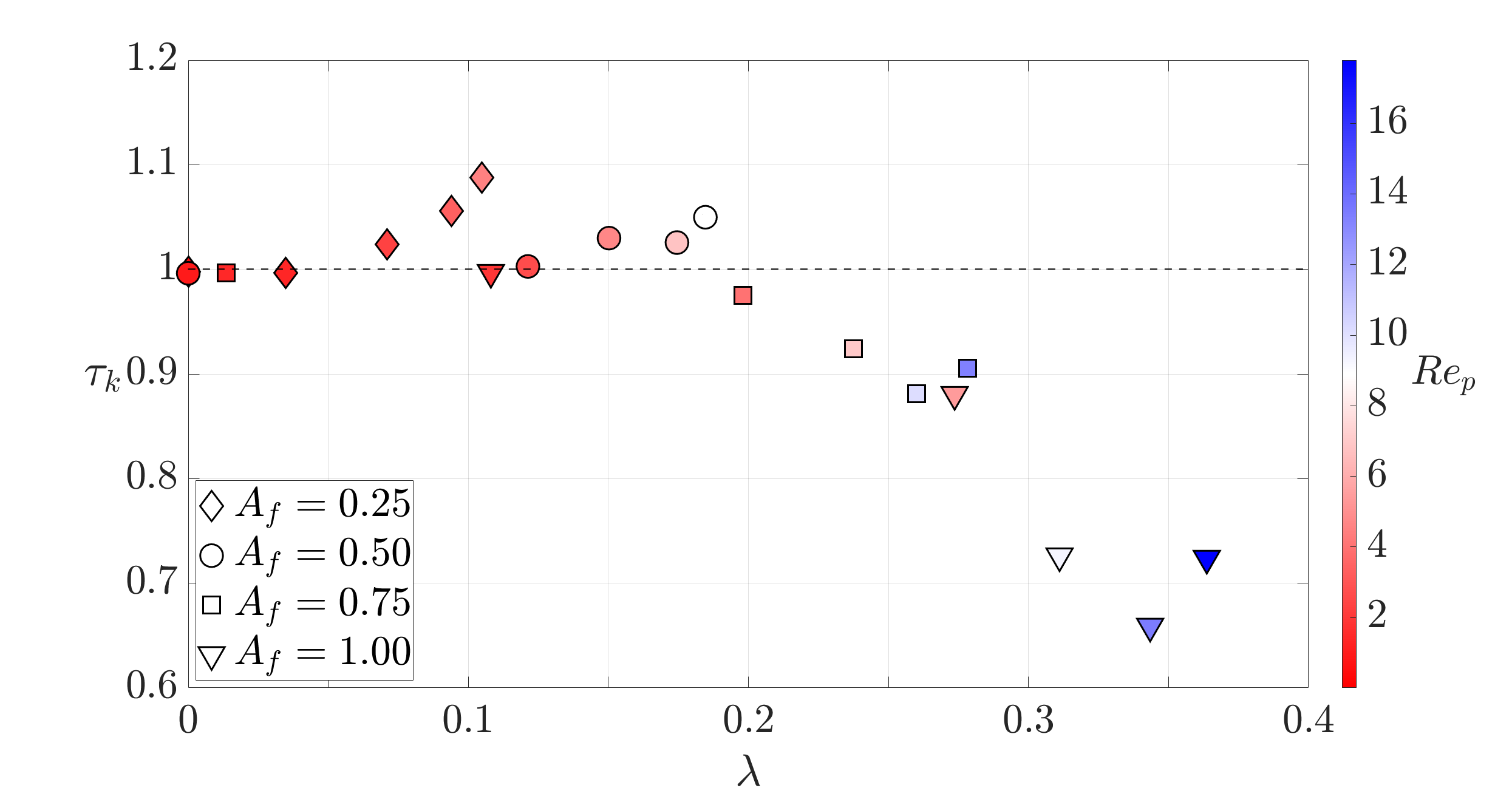}
    \caption{\label{fig:singleParticle_PressureWidth}%
    Correlation between the time ratio during kissing $\tau_k$ of the associated two-particle configurations (cf. Fig~\ref{fig:timeRatio_kissing}) and the width of the lateral repulsive pressure field $\lambda$.
    The symbols represent the amplitude $A_f$ and the color scheme the respective particle Reynolds number $Re_p$.
    }
\end{figure}

\subsubsection{Two-particle pressure distribution}

The single-particle analysis shows that the oscillation-induced pressure disturbance influences the particle--particle interaction.
To assess this effect, Fig.~\ref{fig:Visualization_VaryingLambda} illustrates the pressure field around both particles at the instants marked by the gray vertical bars and Roman numerals in Fig.~\ref{fig:Trajectories_EffectLambda}.

To ensure comparability across cases, we align the selected time intervals with equivalent phases of the particle trajectories, such that each interval represents a comparable dynamical state, rather than the same absolute simulation time.
Within these intervals, we then choose the respective state (e.g., a local peak or trough) for visualization.
Accordingly, \RomanNumeralCaps{1} illustrates the drafting, \RomanNumeralCaps{2} the kissing, and \RomanNumeralCaps{3} the tumbling phase.
As indicated, each column represents a characteristic $Re_p$ with the pressure contours ranging from $-1$ (blue) to $1$ (red) according to the legend.
In general, the pressure field attains its maximum values on the surface of the particles, where the direct fluid-particle interaction occurs, and gradually decays into the surrounding flow. 
This decay is evident in the pressure gradients visible in the contour plots.
For $Re_p<1$, where oscillatory inertia remains weak relative to viscous effects, the pressure distribution retains the characteristic asymmetry associated with gravitational settling, with pressure concentrations confined to the lower side of the leading particle (Figs.~\ref{fig:Visualization_VaryingLambda}a,d) and, during tumbling, to the lower regions of both particles (Fig.~\ref{fig:Visualization_VaryingLambda}g).

In contrast, for $Re_p > 1$, the pressure field is predominantly concentrated along the lateral sides of the particles, with its spatial extent into the surrounding fluid field increasing as $Re_p$ increases. 
Although gravitational settling remains non-negligible in the cases considered, no distinct pressure accumulation is observed on the lower sides, indicating the dominant influence of the horizontal oscillatory forcing.
The differences in the pressure distributions observed for $Re_p > 1$ across the individual rows arise from the specific time instants chosen for visualization and the corresponding instantaneous particle accelerations at those times.
In rows \RomanNumeralCaps{1} and \RomanNumeralCaps{3}, the particle trajectories exhibit local minima in the $y$-direction (Fig.~\ref{fig:Trajectories_EffectLambda}b). 
Owing to the sinusoidal nature of the oscillatory motion, the velocity of the particles vanishes at these positions, whereas the acceleration reaches its maximum immediately before the particles reverse their direction of motion. 
This is reflected by negative pressure contours on the sides of the particles facing in the $y$-direction (e.g. Figs.~\ref{fig:Visualization_VaryingLambda}b,c). 
In contrast, in row \RomanNumeralCaps{2}, the trajectories exhibit local maxima. 
Here, the velocity again vanishes while the acceleration attains its lowest value.
As a consequence, the associated pressure distributions change sign (e.g. Figs.~\ref{fig:Visualization_VaryingLambda}e,f).

In addition to the pressure contours, Fig.~\ref{fig:Visualization_VaryingLambda} also illustrates the particle orientations previously indicated in Fig.~\ref{fig:Trajectories_EffectLambda}. 
During the drafting phase (\RomanNumeralCaps{1}), the particles exhibit similar orientations across all cases, close to their initial configuration. 
While kissing (\RomanNumeralCaps{2}), the particles begin to reorient perpendicular to the direction of oscillation for $Re_p > 1$. 
After contact (\RomanNumeralCaps{3}), the particles are either nearly perpendicular ($Re_p = 4.67$) or fully perpendicular ($Re_p = 9.35$) to the oscillation direction, whereas for $Re_p = 0.59$, they largely retain their initial orientation.

\begin{figure}[h]
    \centering
    \includegraphics[trim=0cm 12.5cm 0cm 1.2cm, clip,width=0.9\linewidth]{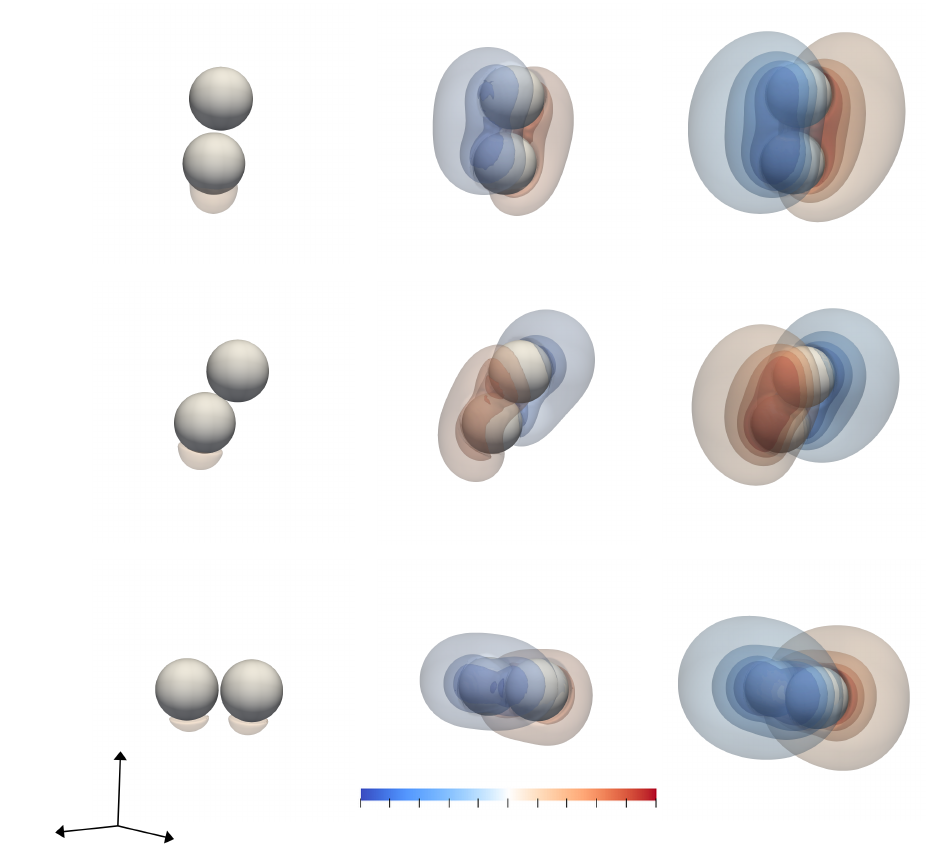}
    \put(-376,410){{$Re_p = 0.59$}}
    %
    \put(-236,410){{$Re_p = 4.67$}}
    %
    \put(-88,410){{$Re_p = 9.35$}}
    \put(-420,352){\makebox(0,0){\shortstack{\RomanNumeralCaps{1}\\Drafting}}}
    \put(-420,220){\makebox(0,0){\shortstack{\RomanNumeralCaps{2}\\Kissing}}}
    \put(-420,77){\makebox(0,0){\shortstack{\RomanNumeralCaps{3}\\Tumbling}}}
    \put(-400,390){{(a)}}
    \put(-260,390){{(b)}}
    \put(-120,390){{(c)}}
    \put(-400,270){{(d)}}
    \put(-260,270){{(e)}}
    \put(-120,270){{(f)}}
    \put(-400,120){{(g)}}
    \put(-260,120){{(h)}}
    \put(-120,120){{(i)}}
    \put(-410,38){{$x$}}
    \put(-430,12){{$y$}}
    \put(-380,9){{$z$}}
    \put(-210,31){{$p$}}
    \put(-210,7){{0}}
    \put(-138,7){{1}}
    \put(-286,7){{-1}}
    \caption{\label{fig:Visualization_VaryingLambda}%
    Visualization of the particles and the surrounding pressure field shown as contour plots.
    The left column (a,d,g) corresponds to $Re_p = 0.59$ ($A_f = 0.5$, $S=0.21$), the middle column (b,e,h) to $Re_p = 4.67$ ($A_f = 0.5$, $S=0.84$), and the right column (c,f,i) to $Re_p = 9.35$ ($A_f = 1.0$, $S=0.84$).
    The rows represent different phases of the interaction: \RomanNumeralCaps{1}~(a–c)~drafting, \RomanNumeralCaps{2} (d–f) kissing, and \RomanNumeralCaps{3} (g–i) tumbling, as indicated by the gray bars in Fig.~\ref{fig:Trajectories_EffectLambda}. 
    Pressure contours are colored from $-1$ (blue) to $1$ (red).}
\end{figure}

The combined trajectory and pressure-field analyses reveal two distinct interaction regimes. 
For $Re_p<1$, the pressure distribution retains the asymmetry associated with gravitational settling, and the particles progressively separate in both lateral directions, $y$ and $z$. 
For $Re_p>1$, the pressure field becomes increasingly concentrated along the lateral sides of the particles, coinciding with synchronized motion in $y$ and enhanced separation in~$z$. 
Thus, increasing $Re_p$ changes not only the magnitude of the oscillation-induced pressure perturbations but also the spatial orientation of the particle pair relative to the oscillation direction.

\subsection{Particle orientation}\label{sec:ParticleOrientation}

The representations and analyses of the particle trajectories provided in the previous sections demonstrate that the mutual orientation of the particles and their alignment relative to the direction of the oscillation evolve over time due to the particle-particle interactions.
This is the case for the non-oscillating reference and the oscillating configurations.
To quantify these changes and examine the impact of the oscillations on the orientation of the particles, we measure the polar and azimuthal angles, $\vartheta$ and $\varphi$ respectively (\emph{c.f.} Fig.~\ref{fig:NumericalSetup}), between the two particles at the beginning and end of the kissing phase. 
The polar angle $\vartheta$ denotes the orientation of the particle pair relative to the vertical settling direction, with $\vartheta = 0 \degree$ corresponding to vertical alignment and $\vartheta = 90 \degree$ to horizontal alignment.
The azimuthal angle~$\varphi$ characterizes the orientation of the particles relative to the imposed oscillation, where $\varphi = 0 \degree$ indicates alignment perpendicular to the direction of the oscillation and $\varphi = 90 \degree$ alignment parallel to it.

\subsubsection{Settling orientation}

Fig.~\ref{fig:ContactAngleTheta} presents the results for the polar angle $\vartheta$ as a function of $Re_p$ at the beginning and end of the kissing phase, where the vertical dashed line denotes $Re_{p} = 1$.
The horizontal dashed line marks the initial particle orientation of all setups, representing an almost vertical alignment, with $\vartheta = 1.01^\circ$.
The dotted horizontal lines show the orientations of the non-oscillating reference case, with $\vartheta = 7.25^\circ$ at the beginning of the kissing phase (\emph{Kissing start}) and $\vartheta = 72.69^\circ$ at the end (\emph{Kissing end}).
At the beginning of kissing, the particle orientations of most oscillatory configurations align with the reference case.
This indicates that, at the onset of contact, the particles are nearly vertically aligned with the settling direction across all cases investigated.
At the end of the contact, the oscillatory cases exhibit particle orientations comparable to the non-oscillating configuration for $Re_{p} \lesssim 2$, however, smaller polar angles are observed for $Re_{p} > 2$, with reductions of up to $20^\circ$. 
This shows that $P_2$ remains above $P_1$ at separation for all configurations, noting that orientations closer to $90^\circ$ correspond to a more horizontal alignment of the particle pair.
Hence, for $Re_{p} > 2$, $P_2$ departs at progressively higher vertical positions, reflected in decreasing values of $\vartheta$.
The results further reveal that the oscillation amplitude does not affect the orientation of the particles with respect to the settling direction.

\begin{figure}[h]
    \centering
    \includegraphics[trim=2.8cm 0.1cm 5.5cm 2.2cm, clip,width=0.85\textwidth]{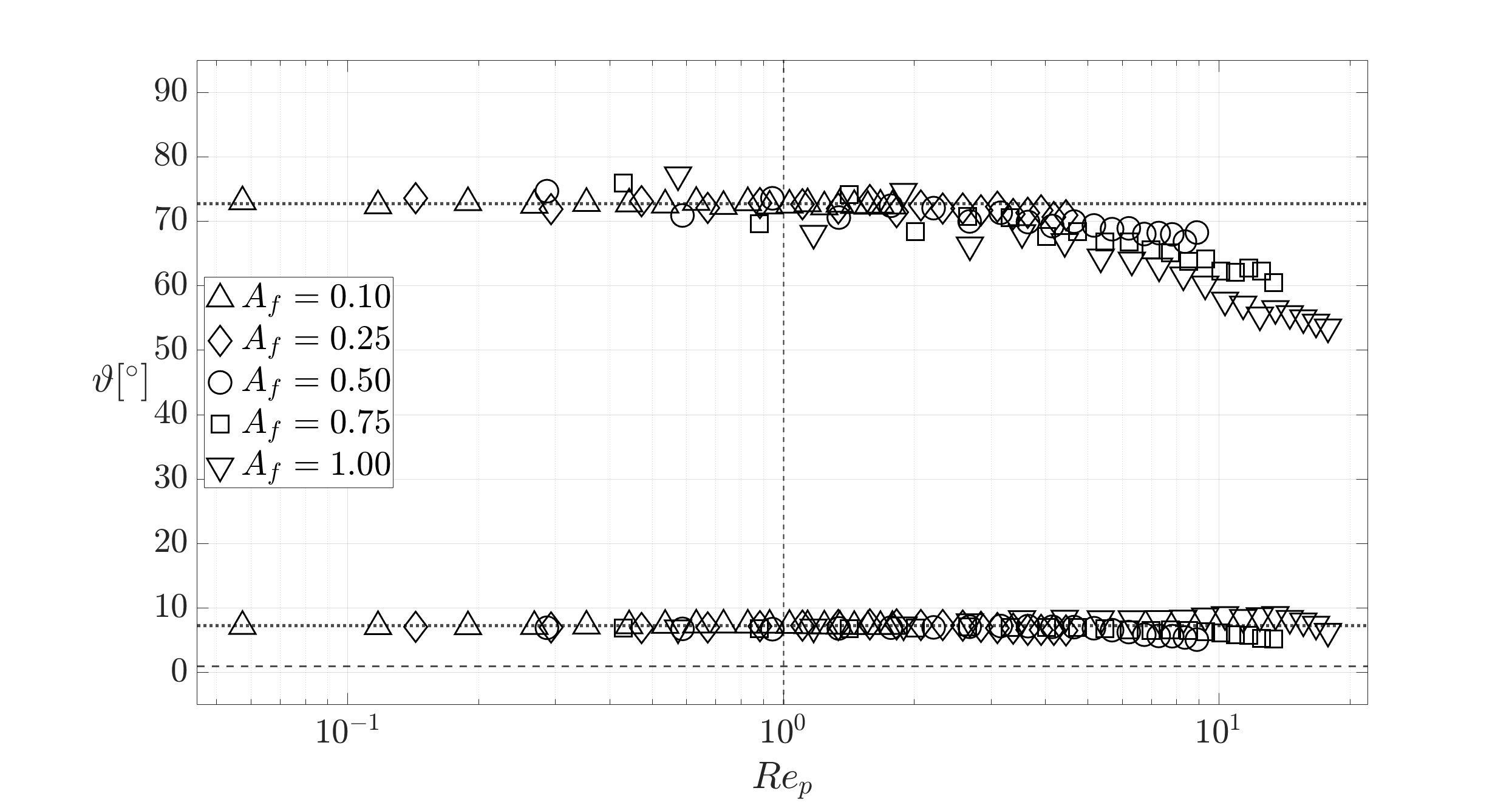}
    \put(-385,37){\footnotesize Initial}
    \put(-385,69){\footnotesize Kissing start}
    \put(-385,209){\footnotesize Kissing end}
    \caption{\label{fig:ContactAngleTheta}%
    Polar angle $\vartheta$ as a function of $Re_p$.
    The vertical dashed line denotes $Re_{p} = 1$.
    The horizontal dashed line marks the initial particle orientation, while the dotted lines show the current orientation of the non-oscillating reference case.
    The lower dotted line with corresponding symbols denotes the start of the kissing phase, and the upper line and symbols indicate its end. 
    The $x$-axis is presented on a logarithmic scale.}
\end{figure}

\subsubsection{Oscillation orientation}

Fig.~\ref{fig:ContactAnglePhi} shows the azimuthal angle $\varphi$ as a function of $Re_p$, evaluated at the end of the kissing phase.
All configurations initially exhibit an orientation of $\varphi = 45 \degree$ as prescribed by the initial condition. 
In the reference case (horizontal dotted line), this orientation remains unchanged, consistent with the nearly symmetric relative motion of the particles in the $y$- and $z$-directions observed in Figs.~\ref{fig:Trajectories_Settling}(a,c) and \ref{fig:Trajectories_IndividualComparison}(c,e). 
In contrast, Fig.~\ref{fig:ContactAnglePhi} demonstrates that the oscillating configurations show orientations similar to the reference case at small $Re_{p}$, but undergo a distinct transition toward a perpendicular alignment close to $\varphi = 0 \degree$ as $Re_{p}$ increases.

In order to further analyze the dependence of $\varphi$ on $Re_{p}$, we apply a shifted and scaled complementary error function, shown by the red curve in Fig.~\ref{fig:ContactAnglePhi} and defined as: 

\begin{equation}\label{eq:erfc}
    \varphi (Re_{p}) = A_{e} \, \text{erfc}\left( \frac{Re_{p} - Re_{p,0}}{\sigma_\varphi} \right) + \varphi_\infty \, ,
\end{equation}
where \(A_e\) denotes the amplitude of the transition, corresponding to half the vertical distance between the asymptotic values of $\varphi$ at low and high $Re_p$, that is, the values that are approached before and after the transition.
The parameter $Re_{p,0}$ defines the horizontal midpoint of the transition and $\sigma_\varphi$ sets its width, thereby stating the range of $Re_{p}$ in which the transition occurs.
The constant $\varphi_\infty$ represents the lower asymptotic value in the limit $Re_{p} \rightarrow \infty$, whereas the upper asymptote in $Re_{p} \rightarrow 0$ is given by $2 A_e + \varphi_\infty$. 
For the oscillatory setups, the coefficients obtained from the curve fit yield $A_e = 22.49\degree$, $Re_{p,0} = 3.83$, $\sigma_\varphi = 2.49$, and $\varphi_\infty = 0.46\degree$.

\begin{figure}[h]
    \centering
    \includegraphics[trim=2.8cm 0.1cm 5.5cm 2.2cm, clip,width=0.85\textwidth]{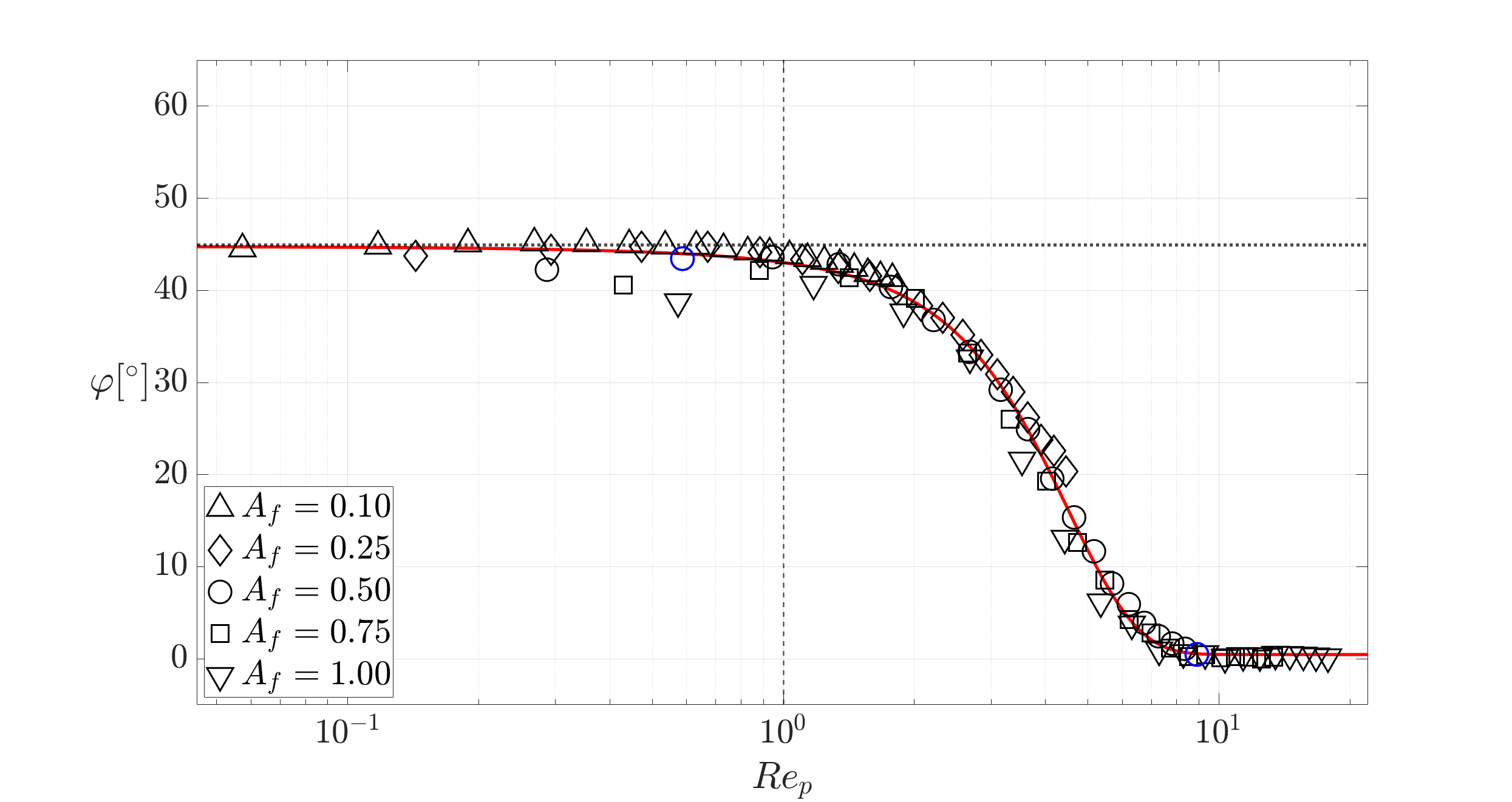}
    \caption{\label{fig:ContactAnglePhi}%
    Azimuthal angle $\varphi$ as a function of $Re_p$ at the end of the kissing phase.
    The $x$-axis is presented on a logarithmic scale.
    The vertical dashed line denotes $Re_{p} = 1$. 
    The horizontal dotted line marks the initial orientation and simultaneously the current orientation of the non-oscillating reference case that does not change over time.
    The red curve represents the fitted complementary error function \eqref{eq:erfc}.}
\end{figure}

An analysis of the coefficients fitted for the end of the kissing phase indicates that the transition occurs between an upper asymptote of $\varphi = 45.44 \degree$ and a lower asymptote of $\varphi = 0.46 \degree$, thus quantifying the gradual reorientation induced by oscillatory forcing.
By calculating the range of $Re_p$ in which the reorientation occurs ($Re_{p,0} \, \pm \, \sigma_\varphi$), we obtain $Re_{p} = [1.34, 6.32]$.
These results show that for $Re_{p} < 1$, the orientations of the particles remain close to the initial arrangement ($\varphi \approx 45 \degree$), which may be attributed to a mechanism similar to that in the reference case, namely persistent separation of the particles in the $y$-~and~$z$-directions.  
In contrast, for $Re_{p} > 6$, the particles align perpendicular to the oscillation direction ($\varphi \approx 0 \degree$), highlighting the strong reorienting influence of the oscillatory forcing at higher $Re_{p}$.
This systematic shift in orientation underscores the decisive role of oscillations in controlling particle alignment, with stronger oscillatory forcing accelerating the reorientation, and thereby potentially influencing the subsequent separation and tumbling dynamics.

The analysis shows that $Re_p$ represents the decisive parameter in determining the orientation of the particles relative to the directions of settling and oscillation.
However, in contrast to the findings for the interaction phases (\S \ref{sec:InteractionPhases}), the oscillation amplitude $A_f$ has no discernible effect on particle orientation.
Moreover, the observation that the particles reach an orientation that is not perpendicular to the oscillation direction ($\varphi > 0\degree$) constitutes a previously unreported phenomenon.
In microgravity environments, \textcite{2017_Fabre_etal} hypothesized and \textcite{2024_Kleischmann_etal} demonstrated that two particles subjected to horizontal oscillations tend to reorient perpendicularly to the oscillation direction once they deviate from a perfect alignment. 
In the absence of gravity, this perpendicular arrangement arises solely from the inertial response of the particles to the oscillatory forcing.

In summary, the reorientation behavior in this study is influenced not only by gravity but also through the horizontal oscillation. Particles align perpendicular to the oscillation direction only when the particle Reynolds number, $Re_p$, exceeds unity. $Re_p$ quantifies the balance between oscillatory inertia and viscous diffusion, incorporating both oscillation amplitude and frequency. For $Re_p<1$, viscous forces dominate, resulting in reversible flow and minimal impact on particle orientation. When $Re_p>1$, inertial effects alter hydrodynamic interactions, driving particles toward a perpendicular configuration.

\subsection{Hydrodynamic mechanisms underlying particle reorientation}\label{sec:SingleParticle}

To identify the physical origin of the observed particle reorientation, we consider two representative configurations resulting in two distinctively different particle orientations, highlighted as blue hollow circles in Fig.~\ref{fig:ContactAnglePhi}: 
(i) $Re_p=0.59$ ($A_f=0.5$, $S=0.21$), which results in $\varphi\approx45^\circ$ at the end of the kissing phase, and (ii) $Re_p=8.90$ ($A_f=0.5$, $S=1.40$), which yields $\varphi\approx0^\circ$.
In the following, we progressively link observed dynamics to hydrodynamic processes to uncover the mechanism driving particle reorientation.
First, we analyze the temporal evolution of particle orientation and trajectories to identify geometric reorientation. 
Next, we examine relative velocities to explain trajectory changes and assess hydrodynamic force imbalances to reveal the dynamic mechanism.

\subsubsection{Evolution of the particle orientation}

The reorientation presented in Fig.~\ref{fig:ContactAnglePhi} corresponds to the particle orientation angles at the end of the kissing phase. 
To elucidate the instantaneous evolution of the particle orientation throughout the DKT process, we present the temporal evolution of $\varphi$ for the two representative cases $Re_p=0.59$ in~(a) and $Re_p=8.90$ in~(b) of Fig.~\ref{fig:ReorientationEvolution}.
The vertical dashed lines indicate the beginning and end of the kissing phase, while the horizontal dashed line denotes the initial particle orientation, which also corresponds to the orientation of the non-oscillating reference case ($\varphi = 45^\circ$).

\begin{figure}[h!]
    \def\stackalignment{l}
    \centering
    \captionsetup[subfigure]{labelformat=empty}
    \begin{subfigure}[b]{\textwidth}
         \topinset{{\footnotesize (a)}}{\includegraphics[trim=7cm 1.2cm 4cm 2.9cm, clip,width=0.48\textwidth]{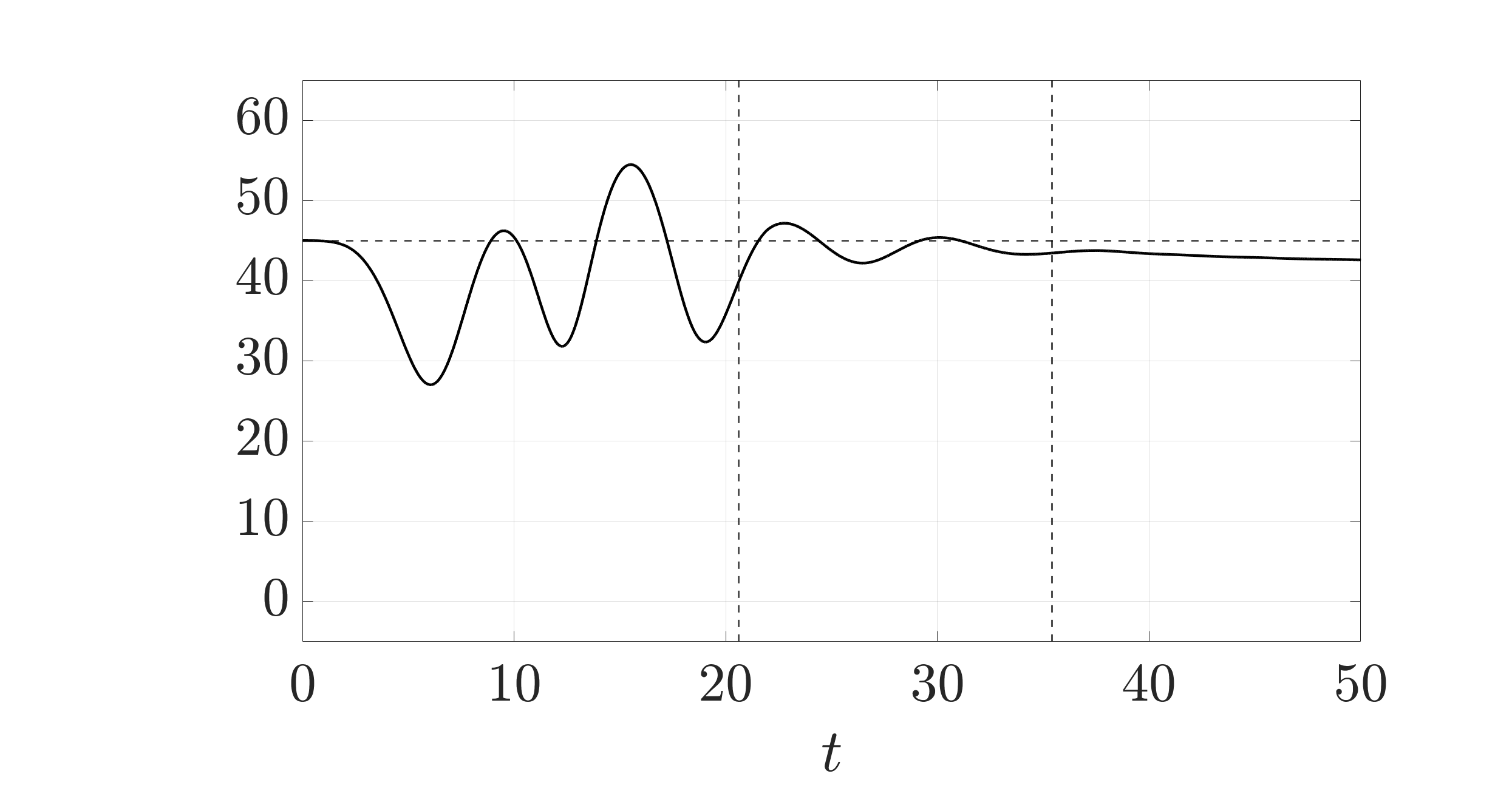}}{0cm}{-0.15cm}
         \topinset{{\footnotesize (b)}}{\includegraphics[trim=7cm 1.2cm 4cm 2.9cm, clip,width=0.48\textwidth]{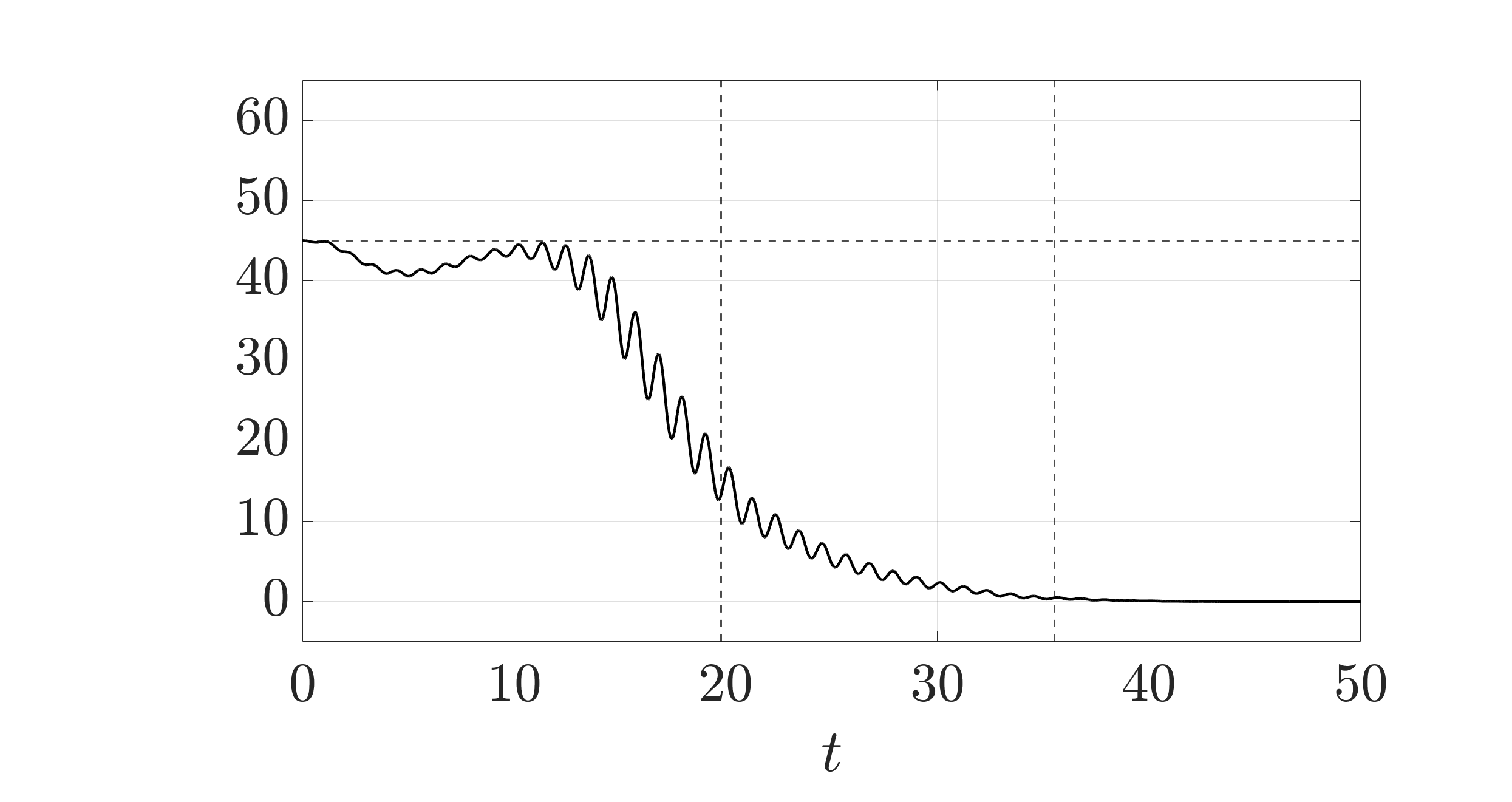}}{0cm}{-0.15cm}    
         \put(-500,80){\small $\varphi [^\circ]$}
    \end{subfigure}
    \caption{\label{fig:ReorientationEvolution}%
    Lateral orientation angle $\varphi$ over time.
    (a) represents $Re_p = 0.59$ and (b)~$Re_p = 8.90$.
    The vertical dashed lines indicate the beginning and end of the kissing phase.
    The horizontal dashed line marks the initial orientation and simultaneously the orientation of the non-oscillating reference case at $\varphi = 45 ^\circ$.}
\end{figure}

For $Re_p=0.59$ (Fig.~\ref{fig:ReorientationEvolution}a), the orientation exhibits pronounced long-wave oscillations during drafting, with successive extrema increasingly deviating from the initial $\varphi=45^\circ$. 
Despite these excursions, the mean orientation remains close to $45^\circ$, and the particles return to approximately their initial orientation at the onset of kissing. 
During kissing, the oscillation gradually decreases and the pair remains near its initial orientation, followed by only a slight reduction in $\varphi$ during tumbling, resulting in a final orientation close to the non-oscillating reference case.

In contrast, the orientation for $Re_p=8.90$ (Fig.~\ref{fig:ReorientationEvolution}b) follows a markedly different evolution. 
During early drafting, $\varphi$ initially decreases slightly before recovering towards its initial value, while exhibiting only small-amplitude, short-wavelength oscillations associated with the larger $S$. 
As the particles approach, the mean orientation decreases steadily before the onset of kissing. 
This trend intensifies during kissing, where $\varphi$ continuously decreases towards $0^\circ$ with periodic oscillations superimposed on the decreasing mean. 
After separation, the oscillations decay, and the particle pair retains an orientation of approximately $\varphi=0^\circ$ throughout tumbling.

\subsubsection{Lateral motion of the particles}\label{sec:lateral_particle_motion}

The particle orientation, $\varphi$, is determined by the relative positions of the two particles in the lateral plane. 
We therefore examine the evolution of the particle trajectories in the lateral directions, i.e. the oscillation direction ($y$) and the transverse direction ($z$). 
The trajectories for the lower forcing case, $Re_p=0.59$, are already shown in Figs.~\ref{fig:Trajectories_EffectLambda}b and \ref{fig:Trajectories_EffectLambda}c as the red curves. 
The trajectories for $Re_p=8.90$ are not shown again, as they exhibit the same qualitative behavior as $Re_p = 9.35$, shown by the green curves in Figs.~\ref{fig:Trajectories_EffectLambda}b and \ref{fig:Trajectories_EffectLambda}c. 
We therefore focus here on the systematic changes in particle motion with increasing $Re_p$.

For low forcing ($Re_p=0.59$), both particles initially undergo periodic oscillations in the $y$-direction with similar amplitudes and phases. 
During the kissing phase, the trajectories begin to diverge, resulting in a gradual increase in the relative particle separation along the oscillation direction. 
In the transverse direction ($z$), the particles likewise separate at a comparable rate. 
Consequently, the relative particle separation develops nearly proportionally in both lateral directions, leading to only a small net change in the particle orientation, as shown in Fig.~\ref{fig:ReorientationEvolution}(a).

With increasing forcing, the particle motion changes systematically. 
Both particles remain nearly synchronized in the oscillation direction ($y$), performing almost identical periodic motions, while continuously separating in the transverse direction ($z$) as they approach one another. 
This transverse migration begins shortly before the onset of the kissing phase and continues throughout the subsequent evolution. 
Thus, the growing transverse separation dominates the evolution of the relative particle position, while the particles remain synchronized in $y$. 
Consequently, the ratio between the two relative separations changes progressively, causing the line connecting the particle centers to rotate towards an orientation perpendicular to the oscillation direction. 
This behavior gives rise to the continuous decrease of the particle orientation angle from $45^\circ$ to approximately $0^\circ$ for $Re_p=8.90$ observed in Fig.~\ref{fig:ReorientationEvolution}(b).

\subsubsection{Relative particle velocities}

In order to understand the dynamics during DKT,
we now examine the corresponding particle velocity differences in the lateral directions: 
\begin{equation}
    v_{rel} = v_2 - v_1 \; , \qquad \; w_{rel} = w_2 - w_1 
\end{equation}}
We show these two relative velocity components as a two-dimensional vectorial plot in Fig.~\ref{fig:RelativeVelocityTrajectories}, where the temporal evolution is indicated by the color scale. 
In this figure, we compare the non-oscillating case with $Re_p=0.59$ and  $Re_p=8.90$ in their respective sub-panels.

\begin{figure}[h!]
    \centering
    \includegraphics[trim=2.7cm 5.1cm 1.7cm 3.2cm, clip,width=\textwidth]{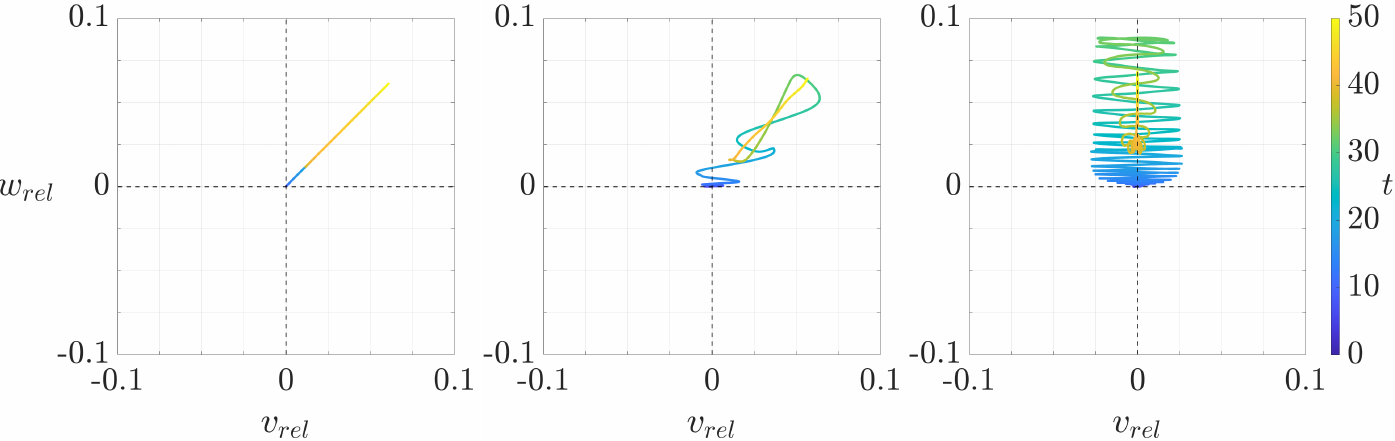}
    \put(-495,165){\footnotesize (a)}
    \put(-340,165){\footnotesize (b)}
    \put(-185,165){\footnotesize (c)}
    \caption{\label{fig:RelativeVelocityTrajectories}%
    Relative particle velocity components in the ($v_{rel}$, $w_{rel}$)-plane.
    The temporal evolution is represented by the color scale and the dashed lines highlight $v_{rel}=0$ and $w_{rel}=0$.
    (a)~non-oscillating reference case, (b) $Re_p=0.59$ and (c)~$Re_p=8.90$. }
\end{figure}

For the non-oscillating reference case (Fig.~\ref{fig:RelativeVelocityTrajectories}a), the relative velocity vector components show a linear behavior. This shows that the relative motion develops proportionally in both lateral directions. 
A similar behavior is observed for $Re_p=0.59$ (Fig.~\ref{fig:RelativeVelocityTrajectories}b). 
However, the relative horizontal velocity vector components no longer follow a perfect linear dependency, but they  exhibit oscillatory deviations induced by the imposed background oscillation. 
Nevertheless, the relative velocities undulate around the same linear trend throughout the DKT process.
Consequently, the relative particle separation develops similar to the non-oscillating case. 

This proportionality is no longer observed for $Re_p=8.90$ (Fig.~\ref{fig:RelativeVelocityTrajectories}c). 
The relative velocity components attain substantially larger values, particularly in the transverse direction, resulting in a markedly more anisotropic evolution. 
Following a brief initial transient, the relative velocity in the oscillation direction exhibits periodic oscillations about an approximately constant value. 
The transverse relative velocity remains positive throughout the DKT process.
The relative velocity vector evolution becomes increasingly elongated in the transverse direction and the relative particle motion is progressively dominated by the transverse component. 
This trend becomes particularly pronounced towards the end of the simulation time, where $v_{rel}$ approaches zero while $w_{rel}$ remains positive.

\subsubsection{Hydrodynamic forces}

While the previous subsection established the relative particle velocity as the kinematic quantity governing the particle reorientation, the origin of this relative motion must be sought in the hydrodynamic forces acting on the particles.
As described in \S\ref{sec:DKT_governingEqs}, the hydrodynamic forces are obtained by integrating the fluid stress over the surface of each particle at every time step.
While these forces characterize the individual fluid--particle interactions, the particle reorientation is governed by the two lateral hydrodynamic force components acting on the two particles. 
Analogous to the relative velocity introduced in the previous subsection, we therefore consider the relative hydrodynamic force, defined as
\begin{equation}
    F_{\mathrm{rel},y} = F_{h,y,P_2} - F_{h,y,P_1} \; , \qquad \; F_{\mathrm{rel},z} = F_{h,z,P_2} - F_{h,z,P_1}\, ,
\end{equation}
where $F_{h,i,P_{1}}$ and $F_{h,i,P_{2}}$ denote the hydrodynamic forces acting on the particles $P_1$ and $P_2$, respectively, in the coordinate direction $i={y,z}$.

\begin{figure}[h!]
    \centering
    \includegraphics[trim=2.5cm 4.8cm 1.5cm 3.5cm, clip,width=\textwidth]{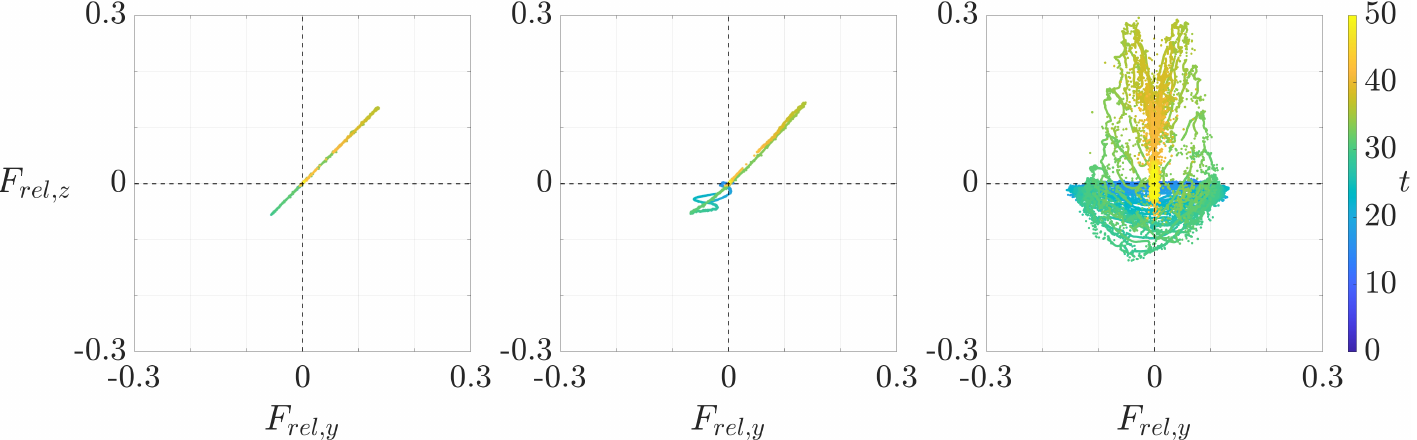}
    \put(-495,165){\footnotesize (a)}
    \put(-340,165){\footnotesize (b)}
    \put(-185,165){\footnotesize (c)}
    \caption{\label{fig:ForceTrajectories}%
    Relative hydrodynamic force paths in the $(F_{\mathrm{rel},y},F_{\mathrm{rel},z})$-plane.
    The temporal evolution is represented by the color scale and the dashed lines highlight $F_{\mathrm{rel},y}=0$ and $F_{\mathrm{rel},z}=0$.
    (a) illustrates the non-oscillating reference case, (b) $Re_p=0.59$ and (c)~$Re_p=8.90$.}
\end{figure}

Fig.~\ref{fig:ForceTrajectories} illustrates the relative hydrodynamic force path in the $(F_{\mathrm{rel},y},F_{\mathrm{rel},z})$-plane, with the temporal evolution represented by the color scale.
We compare the non-oscillating reference case, $Re_p=0.59$, and $Re_p=8.90$ in their respective sub-panels. 
For the non-oscillating reference case (Fig.~\ref{fig:ForceTrajectories}a), the relative hydrodynamic force follows an approximately linear path, indicating that the two lateral force components remain strongly correlated throughout the DKT process. 
This behavior directly reflects the path analyzed for the transverse velocities in Fig.~\ref{fig:RelativeVelocityTrajectories}a.
A similar behavior is observed for $Re_p=0.59$ (Fig.~\ref{fig:ForceTrajectories}b), where the imposed oscillation introduces only small loops around the reference path. 
These periodic deviations indicate that the oscillatory forcing merely modulates the lateral hydrodynamic forcing without altering its mean settling orientation. 
Accordingly, the relative particle velocity exhibits only weak oscillatory deviations from the reference behavior (Fig.~\ref{fig:RelativeVelocityTrajectories}b).
At the stronger forcing of $Re_p=8.90$, the force trajectory changes qualitatively (Fig.~\ref{fig:ForceTrajectories}c). 
Rather than following a well-defined linear path, the relative hydrodynamic force occupies a broad region in the $(F_{rel,y},F_{rel,z})$-plane, demonstrating that both the magnitude and direction of the lateral hydrodynamic forcing evolves substantially throughout the DKT process. 
Consequently, the proportionality between the lateral force components observed at low $Re_p$ no longer persists. 
Instead, the lateral hydrodynamic forcing exhibits a pronounced directional redistribution, which is consistent with the anisotropic relative particle velocities observed in Fig.~\ref{fig:RelativeVelocityTrajectories}c.

\begin{figure}[h!]
    \centering
    \includegraphics[trim=1.5cm 4.8cm 1.5cm 3.25cm, clip,width=\textwidth]{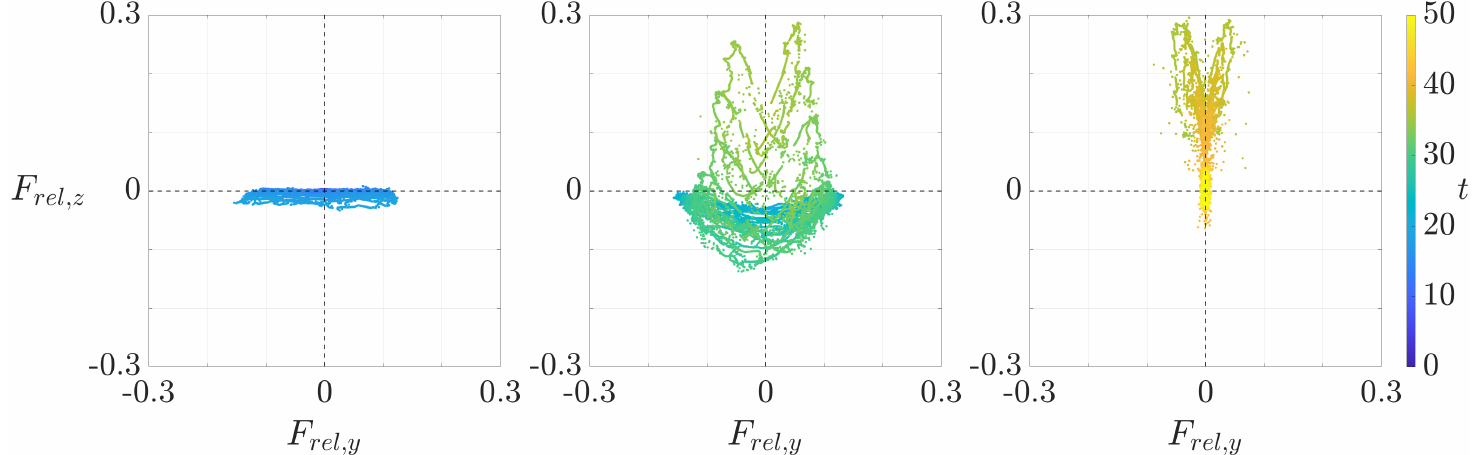}
    \put(-495,165){\footnotesize (a)}
    \put(-340,165){\footnotesize (b)}
    \put(-185,165){\footnotesize (c)}
    \caption{\label{fig:ForceTrajectories_Phases}%
    Relative hydrodynamic force paths in the $(F_{\mathrm{rel},y},F_{\mathrm{rel},z})$-plane for $Re_p=8.90$, decomposed into the individual phases of the DKT process.
    The temporal evolution is represented by the color scale. 
    The dashed lines inside the plots indicate $F_{\mathrm{rel},y}=0$ and $F_{\mathrm{rel},z}=0$. 
    (a) Drafting, (b) kissing, and (c) tumbling. }
\end{figure}

Having identified a qualitative transition in the relative hydrodynamic force path, we now examine how the corresponding lateral forcing  evolves throughout the individual phases of the DKT process for the representative reorienting case $Re_p=8.90$. 
To this end, we decompose its force path shown in Fig.~\ref{fig:ForceTrajectories}c  into its constituent DKT phases and present the individual respective paths in Fig.~\ref{fig:ForceTrajectories_Phases}.
We apply the same color scheme for the drafting, kissing and tumbling phase in the respective sub-panels of this figure.

During the drafting phase, the transverse force component $F_{rel,z}$ remains considerably smaller than the component $F_{rel,y}$ aligned with the oscillation-direction. 
As the particles enter the kissing phase, the force path expands substantially in both lateral directions. 
In particular, the emergence of increasingly positive $F_{rel,z}$ coincides with the onset of the continuous particle reorientation.
Following particle separation, the force path collapses towards $F_{rel,y}\approx0$ while remaining predominantly positive in the $z$-direction ($F_{rel,z}$), indicating a dominant hydrodynamic forcing transverse to the oscillation direction. 
This transverse forcing drives the continued separation of the particles along $z$ while their motion in the $y$-direction becomes synchronized, thereby maintaining the reoriented configuration.

The analysis laid out above revealed the change in force path in the evolution of the DKT process as the oscillatory forcing increases.
To examine how the relative importance of the force components in the transverse and oscillation-direction changes over time, we now consider the temporal variation of the normalized root-mean-square (RMS) ratio,
\begin{equation}
    R_{\mathrm{RMS}}
    =
    \frac{\mathrm{RMS}(F_{rel,z})}
         {\mathrm{RMS}(F_{rel,y})+\mathrm{RMS}(F_{rel,z})} \; ,
    \label{eq:RMS_ratio}
\end{equation}
with
\begin{equation}
    \mathrm{RMS}(F_{\mathrm{rel},i})
    =
    \sqrt{
    \frac{1}{T}
    \int_{t_0}^{t_0+T}
    F_{\mathrm{rel},i}^2(t)\,\mathrm{d}t
    },
    \qquad i\in\{y,z\}.
    \label{eq:RMS}
\end{equation}
Here, we evaluate $RMS$ over each oscillation period $T$ by integrating over the corresponding time interval. This corresponds to $7.38$ and $1.11$ non-dimensional time units $t$ per $T$ for $Re_p = 0.59$ and $Re_p = 8.90$, respectively.
$R_{RMS}$ provides a compact measure of the relative contributions of the transverse and oscillation-direction force components throughout the DKT process.
Thus, $R_{RMS}<0.5$ indicates that the oscillation-direction component $F_{rel,y}$ dominates, whereas $R_{RMS}>0.5$ indicates a dominant transverse component $F_{rel,z}$. 
A value of $R_{RMS}=0.5$ corresponds to equal contributions from both components.

\begin{figure}[h!]
    \centering
    \includegraphics[trim=0.8cm 0.1cm 2.5cm 0.8cm, clip,width=0.85\textwidth]{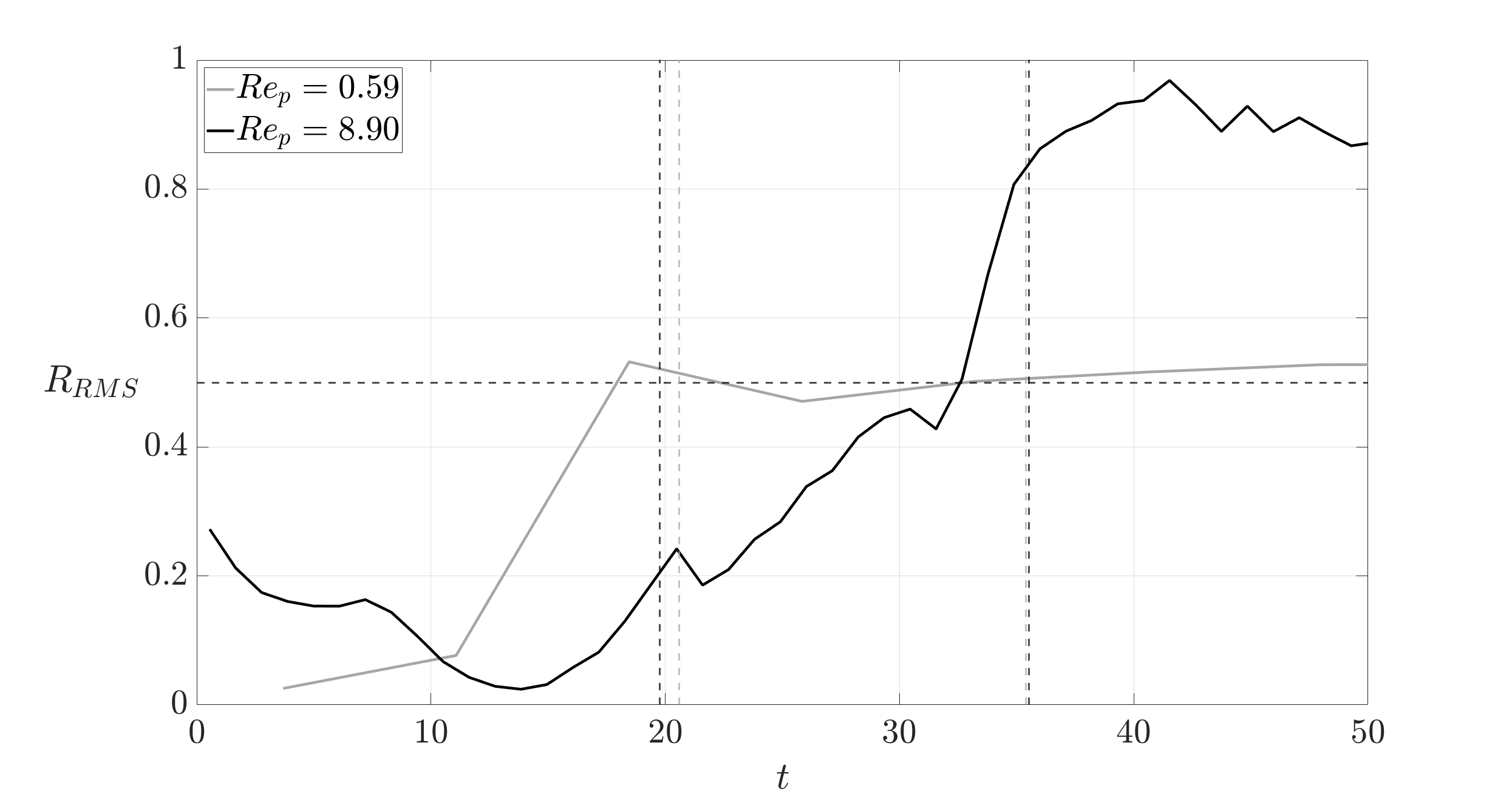}
    \caption{\label{fig:RMS_over_time}%
    Temporal evolution of the lateral hydrodynamic forcing, quantified by the normalized RMS ratio defined in \eqref{eq:RMS_ratio}, for the representative cases $Re_p = 0.59$ (gray) and $Re_p = 8.90$ (black). 
    The vertical dashed lines indicate the beginning and end of the kissing phase for the respective cases applying the corresponding color scheme. 
    The horizontal dashed line denotes $R_{\mathrm{RMS}} = 0.5$.
    The resulting data points of each curve correspond to the midpoints of the respective oscillation periods $T$ used for \eqref{eq:RMS}. 
    As the lower oscillation frequency for $Re_p = 0.59$ results in a longer oscillation period, the corresponding curve begins at a later time.}
\end{figure}

Fig.~\ref{fig:RMS_over_time} compares the temporal evolution of $R_{RMS}$ for the two representative cases \mbox{$Re_p=0.59$} (gray) and $Re_p=8.90$ (black). 
For the non-reorienting case ($Re_p=0.59$), the lateral hydrodynamic forcing is initially dominated by $F_{rel,y}$, ($R_{RMS}<0.5$). 
During the late drafting phase, $R_{RMS}$ increases gradually towards $0.5$ and subsequently remains nearly constant throughout the remainder of the DKT process.
This behavior of the balanced force components is consistent with the approximately proportional force trajectories shown in Fig.~\ref{fig:ForceTrajectories}(b).
For the reorienting case ($Re_p=8.90$), the lateral hydrodynamic forcing is likewise initially dominated by $F_{rel,z}$. 
However, close before contact and throughout the kissing phase, $R_{RMS}$ increases continuously and the transverse force component $F_{rel,z}$ becomes dominant near the end of the kissing phase. 
During the subsequent tumbling phase, $R_{RMS}$ approaches values close to unity.

The two selected representative cases analyzed in this section provided detailed insight into the temporal evolution of the hydrodynamic force balance. 
We now investigate whether the same trends persist throughout the entire parameter space of $Re_p$ and $A_f$ explored in the present study.
To this end, we evaluate the phase-averaged hydrodynamic force balance $\langle R_{RMS} \rangle$ for the drafting, kissing, and tumbling phases to characterize the ratio during each stage of the DKT process.

\begin{figure}[h!]
    \centering
    \includegraphics[trim=0.3cm 0.8cm 2.6cm 0.8cm, clip,width=0.85\textwidth]{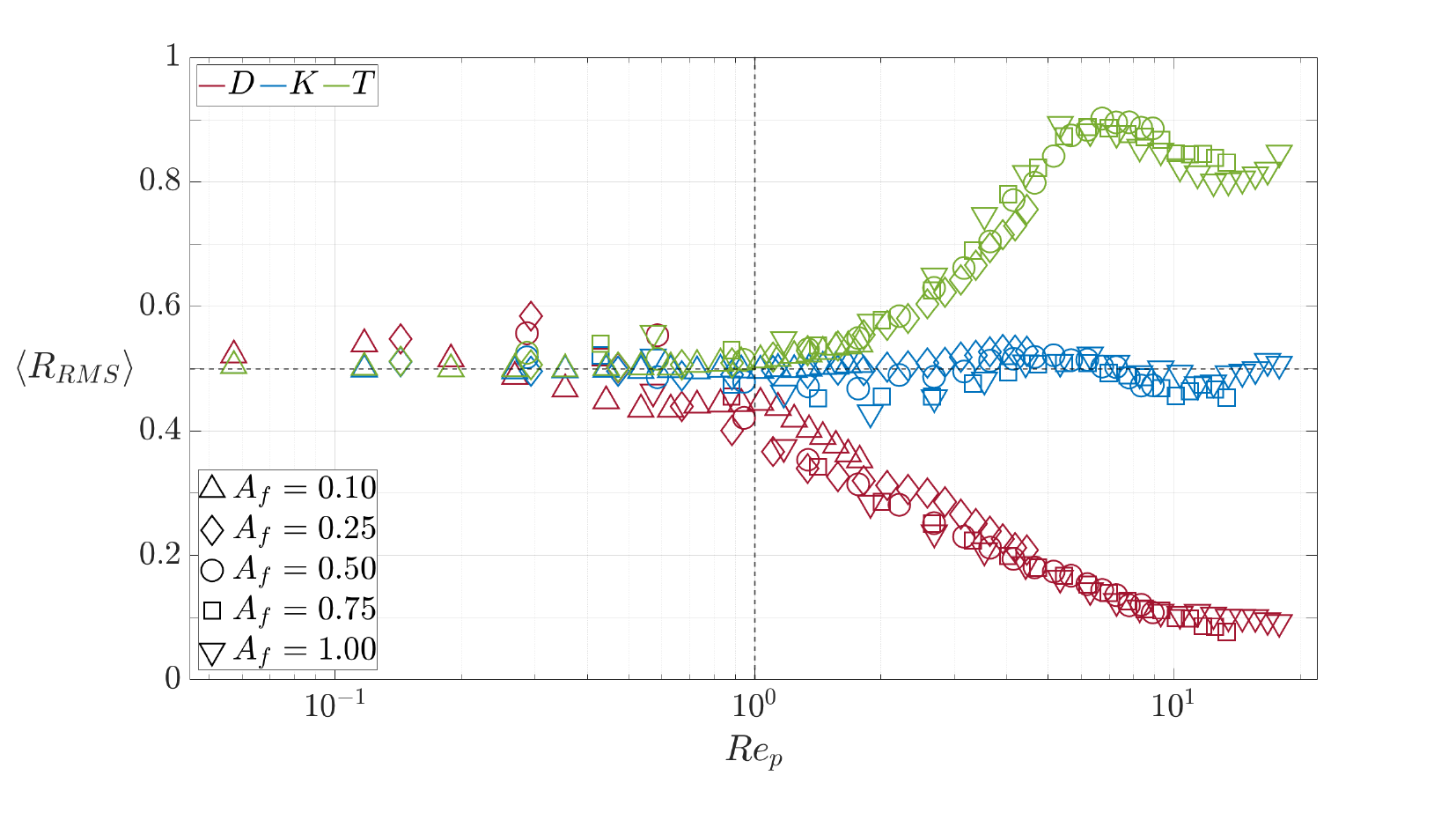}
    \caption{\label{fig:RMS_DKT}%
    Phase-averaged hydrodynamic force balance, $\langle R_{RMS} \rangle$, as a function of the oscillatory particle Reynolds number, $Re_p$, during the drafting (D, red), kissing (K, blue), and tumbling (T, green) phases. 
    Different symbols denote the imposed oscillation amplitudes, $A_f$, as indicated in the legend. 
    The horizontal and vertical dashed lines indicate $\langle R_{RMS} \rangle = 0.5$ and $Re_p = 1$, respectively.}
\end{figure}

Fig.~\ref{fig:RMS_DKT} shows $\langle R_{RMS} \rangle$ as a function of $Re_p$. 
The drafting phase (D) is presented in red, the kissing phase (K) in blue, and the tumbling phase (T) in green.
During all three phases, $\langle R_{RMS} \rangle$ remains close to  $0.5$ for $Re_p \lesssim 1$, indicating nearly equal contributions of the oscillation-direction and transverse force components. 
Beyond this threshold, the results of the averaged DKT phases differ significantly.
While drafting, $\langle R_{RMS} \rangle$ gradually decreases, indicating an increasing dominance of $F_{rel,y}$.
During the kissing phase, $\langle R_{RMS} \rangle$ remains close to $0.5$ throughout the investigated parameter range. 
This is consistent with the time-resolved analysis of the representative cases shown in Fig.~\ref{fig:RMS_over_time}: 
while the non-reorienting case ($Re_p=0.59$) maintained a nearly constant value of $R_{\mathrm{RMS}}\approx0.5$, the reorienting case ($Re_p=8.90$) exhibited a transition from $R_{\mathrm{RMS}}<0.5$ to $R_{\mathrm{RMS}}>0.5$. 
Conversely, during tumbling $\langle R_{RMS} \rangle$ increases rapidly and approaches values close to $0.9$, demonstrating that $F_{rel,z}$ becomes dominant.

The results demonstrate that the systematic redistribution of the hydrodynamic force imbalance drives the particle reorientation, with the dominant contribution shifting from the oscillation direction during drafting to the transverse direction during tumbling. 
This redistribution emerges at $Re_p\approx1$, coinciding with the onset of particle reorientation shown in Fig.~\ref{fig:ContactAnglePhi}.

\section{Conclusions} \label{sec:Conclusion}

In the present study, we investigated the impact of horizontal oscillations on the drafting-kissing-tumbling process of two monodisperse particles that settle under gravity.
By conducting particle-resolved direct numerical simulations, we analyzed the effects of varying oscillation frequencies and amplitudes on the particle-particle interaction.
To this end, we systematically examined the influence of the oscillations on the individual stages of the DKT process by analyzing the particle orientation, particle trajectories, relative settling velocities, and particle-particle separation. 
Building upon these observations, we investigated the underlying hydrodynamic mechanisms through a detailed analysis of the relative hydrodynamic forces acting on the particles and the associated pressure distributions surrounding individual particles. 
While this mechanistic investigation is based on a simplified two-particle configuration with a single initial arrangement and particle size as well as density combination, it provides fundamental insight into the physical processes governing the observed reorientation.

The results demonstrate that the imposed oscillations significantly influence the particle motion in the direction of oscillation, as the particles closely follow the prescribed oscillatory forcing. 
This lateral motion modifies the overall hydrodynamic conditions and, consequently, affects the vertical settling behavior. 
As a result, the effective settling velocity is reduced in the oscillating configurations compared to the non-oscillating reference case.
However, the kissing phase is only affected for particle Reynolds numbers $Re_p > 1$, indicating that the inertial contributions associated with the oscillations must surpass viscous forces to affect the dynamics of the particles. 
The influence of the oscillations is dependent on the oscillation amplitude $A_f$, with low to moderate amplitudes tending to prolong the interaction, while higher amplitudes destabilize the configuration and shorten the phase.
The results indicate that $Re_p$ governs the threshold at which the system becomes responsive to oscillatory forcing, whereas $A_f$ controls both the magnitude and the nature of this response. 
The analysis of the pressure distribution around an individual particle as well as around both particles further show that the magnitude and spatial extent of the repulsive pressure increase with increasing $Re_p$.
This enhanced repulsive interaction promotes an earlier separation of the particles and thereby contributes to a reduction of the kissing phase.

The analysis of the particle orientation with respect to the direction of oscillation shows a clear dependence on $Re_p$.
For $Re_p < 1 $, the simulations remain similar to the non-oscillating reference, where the particles remain in their initial orientation of $\varphi \approx 45^\circ$.
For $Re_p > 6$, the particles approach an  alignment that is perpendicular to the oscillation direction ($\varphi \approx 0^\circ$) while setups with intermediate values of $Re_p$ undergo a transitional regime.
This stepwise transition between the initial lateral arrangement and the perpendicular orientation constitutes a previously unreported observation. 
It unifies two well-established limiting behaviors:
(i)~the preservation of the lateral particle orientation throughout the drafting, kissing, and tumbling phases in the classical DKT process \cite{2015_Cao_etal, 2022_Li_etal}; 
and (ii)~the reorientation of initially misaligned particles towards an orientation perpendicular to the oscillation direction under horizontal oscillations in the absence of settling \cite{2001_Lyubimov_etal, 2002_Voth_etal, 2002_Wunenburger_etal, 2007_Klotsa_etal, 2009_Klotsa_etal, 2017_Fabre_etal, 2024_Kleischmann_etal}. 
The present results demonstrate that the coupled action of gravitational settling and horizontal oscillations gives rise to a transition between these two regimes, with $Re_p \approx 1$ marking the critical threshold.

To elucidate the physical origin of this transition, we further investigated representative oscillation setups exhibiting the distinct reorientation regimes by successively analyzing the particle trajectories, the relative particle velocities, and the lateral hydrodynamic forcing acting on the particles.
The path analysis reveals the emergence of an asymmetric particle motion for $Re_p>1$, which is accompanied by the development of a sustained transverse relative velocity responsible for the progressive reorientation of the particle arrangement. 
The hydrodynamic force analysis further demonstrates that this behavior is associated with a shift in the lateral hydrodynamic force components, with oscillation-induced forces increasingly dominating the particle dynamics as $Re_p$ exceeds unity.  
These analyses establish a consistent mechanistic framework linking the transition at $Re_p\approx1$ to the onset of oscillation-induced inertial effects, which overcome viscous damping, generate a persistent hydrodynamic force imbalance, and ultimately drive the particle arrangement towards the oscillation-perpendicular orientation.

The findings of the present study indicate that $Re_p$ governs the onset of oscillation-induced effects, while the amplitude of the oscillation primarily determines the magnitude and qualitative characteristics of the particle-particle interactions. 
These results provide a mechanistic understanding of how oscillatory flows modulate DKT dynamics and lay the groundwork for further investigations into more details of the distinct roles of oscillation amplitude and frequency. 
Extending the study to include variations in the particle arrangement and the particle properties would broaden the applicability of these findings and would support a deeper understanding of particle–particle interactions under horizontal oscillatory forcing during gravitational settling.

\section{Acknowledgments} 
The authors gratefully acknowledge support through the German Research Foundation (DFG) grant VO2413/2-1. The authors also acknowledge the GCS Supercomputer SUPERMUC-NG at Leibniz Supercomputing Centre and ZIH at TUD Dresden University of Technology for providing computing facilities.

\section{Data Availability} 
All the data used for the analyses is available on Zenodo at  \url{https://zenodo.org/records/20554387} \cite{supplemental_material}.

\appendix

\section{}\label{appendix:ParticleAmplitude}

We compute the particle amplitude $A_p$ by considering the oscillations of the particles throughout the entire time of each simulation.
We disregard the varying phases, that is, drafting, kissing, and tumbling, to account for as many data points as possible, because the smaller $S$, the smaller is the total number of oscillations during the time period considered ($t=100$).

\begin{figure}[h!]
    \centering  
    \includegraphics[width=0.8\linewidth]{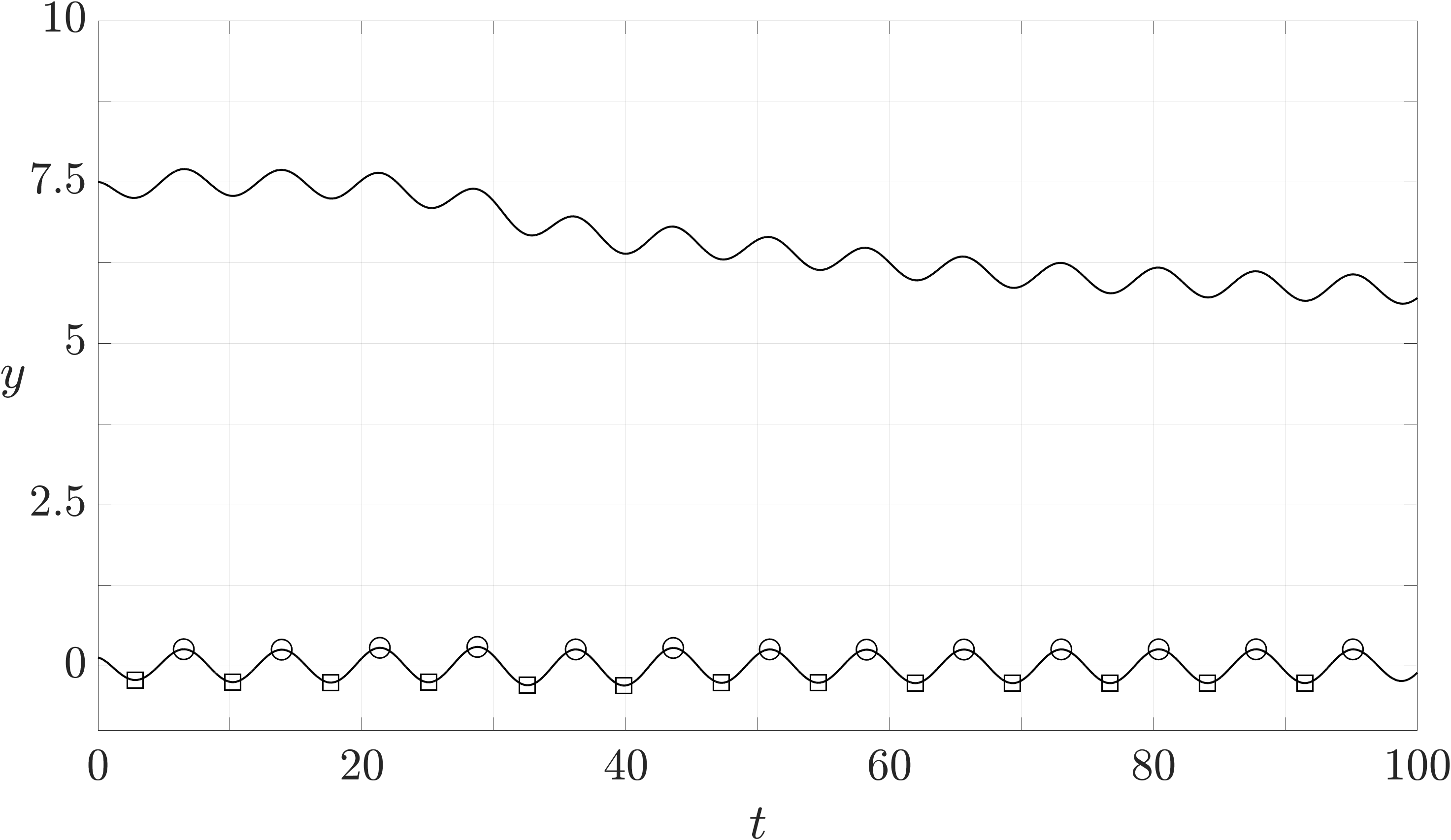}
    \caption{\label{fig:amplitudeAnalysis}%
    Analysis of the particle amplitude exemplary for the leading particle $P_1$ of setup $A_f = 1.0$ and $S = 0.21$. 
    Upper graph represents the trajectory of the $y$-axis and the lower graph the amplitude after removing the particle drift.
    Circles indicate the maximum and squares the minimum of the amplitude.}
\end{figure}

To compute the oscillation amplitude of each particle in the direction of oscillation ($y$-axis), we first remove the underlying trend from the trajectory of the respective particle to extract the oscillatory component (detrending), before identifying the local maxima and minima.
This is exemplary illustrated in Fig.~\ref{fig:amplitudeAnalysis} for the leading particle of the setup $A_f = 1.0$ and $S = 0.21$, where the upper graph is the actual trajectory in the $y$-direction.
The lower graph represents the result of detrending with circles highlighting the local maxima and squares the local minima.
The actual amplitude of a respective oscillation period is half of the distance between the associated maximum and minimum.
The amplitude for a considered setup is then computed by
\begin{equation}
    A_p = \frac{1}{2} \left( \frac{1}{N_T} \sum_{i=1}^{N_T} A_{1,i} \, + \, \frac{1}{N_T} \sum_{j=1}^{N_T} A_{2,j} \right) ,
\end{equation}
with $N_T$ the number of oscillation periods $T$, as well as $A_{1}$ and $A_{2}$ the individual amplitudes of $P_1$ and $P_2$.

\vspace{-0.5cm}

\section{}\label{appendix:KissingDistance}

Following the definition in \S\ref{sec:ParticleDynamics}, we consider two particles as kissing when their surface-to-surface distance $\zeta_n \leq \zeta_k$.
For the analyses presented in \S\ref{sec:DKT_results}, we choose the distance at which kissing begins as $\zeta_k =  2 \zeta_{min}$, with $\zeta_{min}$ the chosen roughness of a particle surface defined as $\zeta_{min} = 3 \cdot 10^{-3} \, R_m$ \cite{2017_Biegert_etal}.
To analyze the impact of this choice on the drafting and kissing phases, we compare the present value $\zeta_k =  0.003 \, d_p$ with $\zeta_{k} = 0.013 \, d_p$ following the work of \textcite{1994_Feng_etal}, and $\zeta_{k} = 0.04 \, d_p$ based on the study of \textcite{2014_Wang_etal}.
The values considered are summarized and ordered by decreasing size in Table~\ref{tab:ValuesKissingDistance}.

\begin{table}[h!]
    \caption{\label{tab:ValuesKissingDistance}%
    Distances at which the particles start to enter the kissing phase, $\zeta_k$, arranged in descending order.}
    \begin{ruledtabular}
        \begin{tabular}{ccccc}
            Reference & $\zeta_k \, [d_p]$ \\
            \colrule
            \textcite{2014_Wang_etal} & 0.040 \\
            \textcite{1994_Feng_etal} & 0.013 \\
            \textcite{2017_Biegert_etal} & 0.003 \\
        \end{tabular}
    \end{ruledtabular}
\end{table}

\begin{figure}[h!]
    \def\stackalignment{l}
    \centering
    \captionsetup[subfigure]{labelformat=empty}
    \begin{subfigure}[b]{\textwidth}
         \centering
         \topinset{{\footnotesize (a)}}{\includegraphics[trim=6.5cm 2cm 5cm 2.5cm, clip,width=0.8\textwidth]{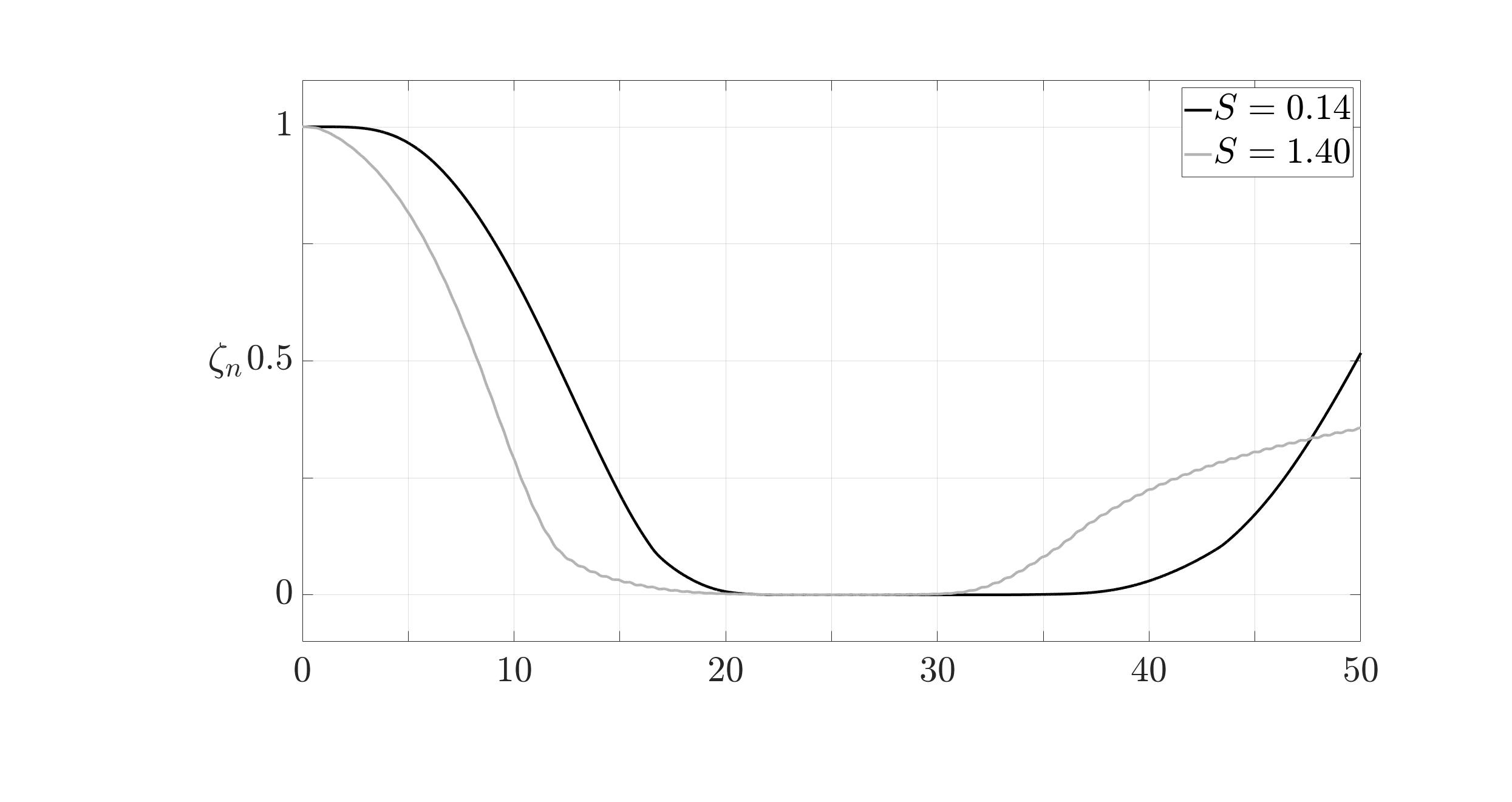}}{0cm}{0cm}
         \topinset{{\footnotesize (b)}}{\includegraphics[trim=6.5cm 2cm 5cm 2.5cm, clip,width=0.8\textwidth]{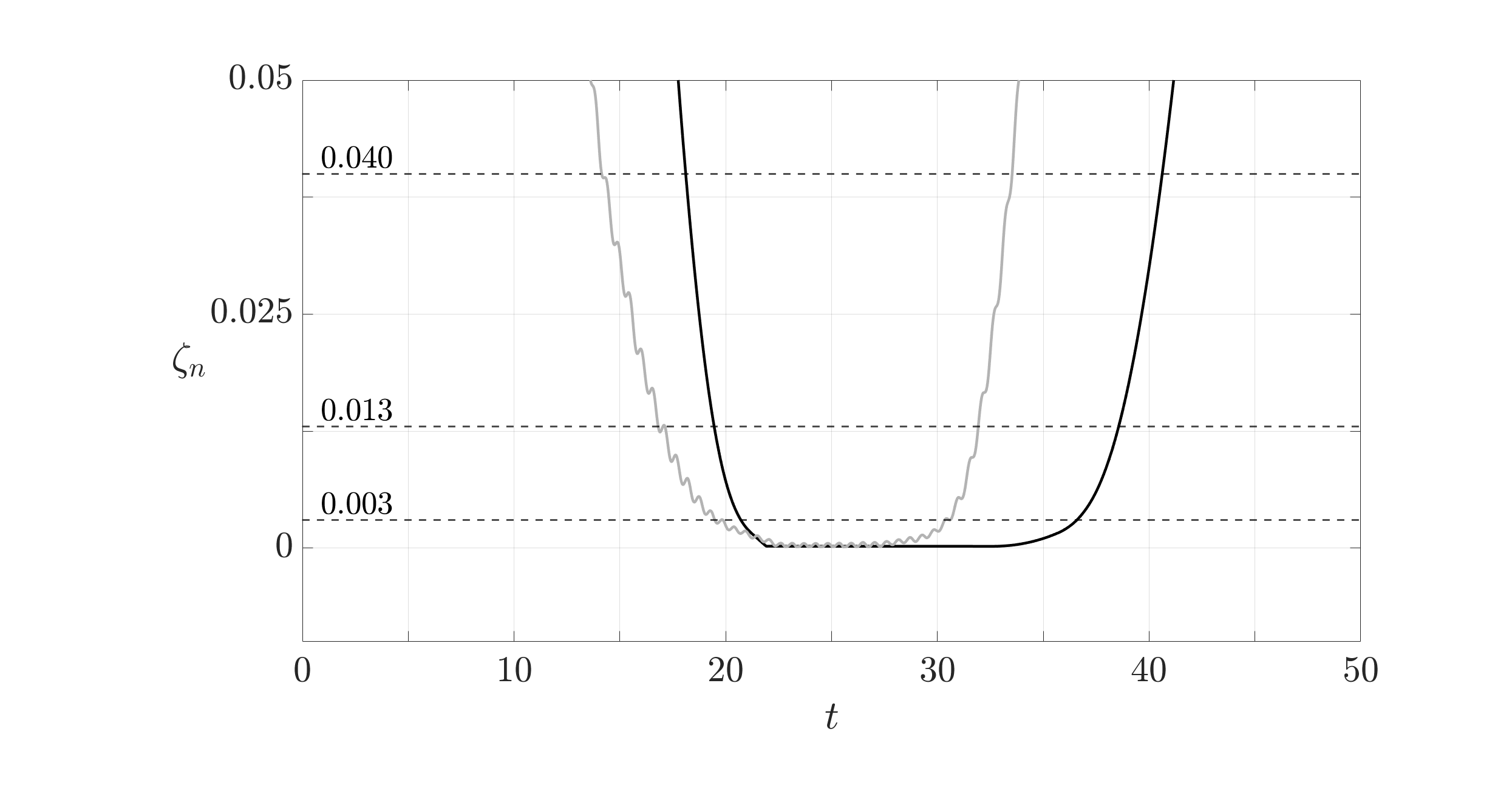}}{0cm}{0cm}
    \end{subfigure}
    \caption{\label{fig:KissingDistance}%
    Surface-to-surface distances $\zeta_n$ over time for $S=0.14$ and $S=1.40$ with $A_f=1.0$. 
    (a) total presentation and (b) zoom-in of the same representation focusing on the kissing phase with kissing distances $\zeta_k$ indicated according to Table ~\ref{tab:ValuesKissingDistance}.}
\end{figure}

In Fig.~\ref{fig:KissingDistance}, we show the temporal evolution of the distance between the particles $\zeta_n$ for $S=0.14$ and $S=1.40$ with $A_f=1.0$, corresponding to $Re_p = 1.17$ and $17.80$, respectively.
The complete trajectories are presented in Fig.~\ref{fig:KissingDistance}(a), while Fig.~\ref{fig:KissingDistance}(b) provides a magnified view of the kissing phase. 
In this close-up, the different thresholds $\zeta_k$ considered to define the beginning of kissing are indicated.
The comparison shows that varying $\zeta_k$ only alters the definition of the kissing interval: a higher value of $\zeta_k$ shifts the apparent onset of kissing to an earlier time and the offset of kissing to a later time.
Importantly, this influence is relative rather than absolute, since the kissing phase of the reference case without oscillations would be shifted consistently. 
In all cases, the fluctuations of $\zeta_n$ remain small enough that the threshold is not repeatedly crossed, ensuring that the definition of kissing is robust with respect to the specific choice of $\zeta_k$.

For clarity and consistency, we select the smallest value, $\zeta_{k} = 0.003 \, d_p$.
This threshold represents the closest approach to actual particle contact while ensuring that the fluctuations of $\zeta_n$ do not cross this value during the actual kissing phase.
Since the specific choice of $\zeta_k$ affects only the formal definition of the onset and offset of kissing and not the physical outcome, this convention is sufficient for our analysis.

\clearpage


%

\end{document}